\documentclass{raa}
\usepackage{natbib,times}
\usepackage{lscape}
\bibpunct{(}{)}{,}{a}{}{;}

\providecommand{\bm}[1]{\mbox{\boldmath$#1$\unboldmath}}

\begin{document}

\title{Lensing clusters of galaxies in the SDSS-III
 $^*$
\footnotetext{\small $*$ Supported by the National Natural Science Foundation of China.} }

\volnopage{{\bf 2011} Vol.\ {\bf 00}~~ No.\ {\bf 0},~~~ 00 -- 00}
   \setcounter{page}{1}

\author{Z. L. Wen,
        J. L. Han,
        and 
        Y. Y. Jiang}

\institute{National Astronomical Observatories, Chinese Academy of Sciences,
                 Beijing 100012, China;
                 {\it zhonglue@nao.cas.cn}}


\abstract{
We identify new strong lensing clusters of galaxies from the Sloan
Digital Sky Survey III (SDSS DR8) by visually inspecting
color images of a large sample of clusters of galaxies. We find 68 new
clusters showing giant arcs in addition to 30 known lensing
systems. Among 68 cases, 13 clusters are ``{\it almost certain}'' lensing 
systems with tangential giant arcs, 22 clusters are ``{\it probable}''
and 31 clusters are ``{\it possible}'' lensing systems. We also find two exotic
systems with blue rings. The giant arcs have angular separations of
$2.0''-25.7''$ from the bright central galaxies.
We note that the rich clusters are more likely to be lensing systems,
and the separations between arcs and the central galaxies increase
with cluster richness.
\keywords{galaxies: clusters: general --- gravitational lensing} }

\authorrunning{Wen, Han and Jiang}
\titlerunning{Lensing clusters of galaxies in the SDSS-III}

\maketitle

\section{Introduction}

Gravitational lensing is the phenomenon of light deflection by
gravitational fields. Strong lensing occurs when a lens and a
background source lie very close along the line of sight. Thus, it is
a relatively rare phenomenon in the universe. As the largest bound
systems in the universe, galaxy clusters are the most powerful
gravitational lenses to magnify background sources. They distort
background galaxies into giant arcs or multiple images. Strong lensing
by galaxy clusters can be used to study the faint background galaxies
at high redshift \citep[e.g.][]{mf93,mkm+03} and to directly measure
cosmological parameters \citep[e.g.,][]{bb03}. The statistics of giant
lensing arcs help to understand the structure formation and cosmology
paradigm \citep[e.g.][]{bhc+98,lmj+05}. The strong lensing by galaxies
or galaxy clusters has been proposed to test other gravity theories
\citep[e.g., TeVeS theory,][]{cz06}. Moreover, strong lensing is a
unique method to examine the model of mass distribution in halos
\citep[e.g.][]{tc01,szb+08,cm10}. The determined mass from lensing
features is model independent and therefore can be used to calibrate
the mass estimates derived from X-ray observations \citep{wu00}.

Generally, lensed arcs are very faint. They can be found from imaging
surveys and confirmed by later spectroscopic observations. Searches
for arcs are usually performed with high quality image data.  Up to
now, less than two hundred strong lensing clusters have been
discovered by visual inspection of images
\citep[e.g.][]{ghy+03,ste+05} or by automated search of the images
\citep[e.g.,][]{cad+07,lcg+09}.
Systematic searches for lensing systems have been made for massive
clusters due to their powerful magnification. From 38 X-ray luminous
clusters, \citet{lgh+99} found eight clusters with giant arcs, two clusters
with arclets and six lensing candidates. Using the {\it HST} WFPC2 data,
\citet{ste+05} found 104 giant arcs from 54 rich clusters.
\citet{hgo+08} found 16 new lensing clusters with giant arcs, with
12 likely lensing clusters and 9 possible candidates based on 240 rich
clusters from the Sloan Digital Sky Survey (SDSS) and
follow-up deep imaging observations.

The SDSS has provided five broad band ($u$, $g$, $r$, $i$, and $z$)
relatively shallow images with a detection limit of $r=22.5$ and a
seeing of 1.43$''$ \citep{slb+02}. Because it covers one-fifth of the
sky, tens of lensing systems have been recently discovered.
Some lensing systems were serendipitously discovered, e.g., the ``8
O'clock arc'' \citep{atl+07} and the Hall's arc \citep{ead+07}. More
lens systems, e.g, the ``Cosmic horseshoe'' and the ``Cheshire Cat''
\citep{bem+07,beh+09}, were discovered in the systematic searches. The
``CASSOWARY''\footnote{http://www.ast.cam.ac.uk/ioa/research/cassowary}
project searches for wide separation ($>1.5''$) gravitational lensing
systems around massive elliptical galaxies in the SDSS with follow-up
spectroscopic observations \citep{pcd+10,cdp+10}. The ``Sloan Bright
Arcs Survey'' is a similar project to search for lensing candidates around
massive galaxies and make spectroscopic confirmation
\citep{daa+09,kaa+09}.
By visually inspecting color images of 39,668 clusters, \citet{whx+09}
preformed a search for giant arcs and found four new lensing systems and
another nine lensing candidates. Most of them have been spectroscopically
confirmed later by other observers (see Table~\ref{lens.tab}).
Some arcs in the SDSS images, e.g., the ``8 O'clock arc'' and the
``Cosmic horseshoe'', are lensed by individual galaxies, but more
giant arcs are generated by clusters or groups of galaxies.

In this paper, we report the discovery of 68 lensing cluster
candidates from a large galaxy cluster sample identified from the 
SDSS-III.

\section{New Lensing clusters in the SDSS-III}

Visually inspecting images is an efficient method to search for lensed arcs
\citep[e.g.,][]{lgh+99, ste+05}. Following the procedure of
\citet{whx+09}, we first identify a large sample of galaxy clusters 
and then visually inspect their color images for lensing systems.

Using photometric redshifts of galaxies, \citet{whl09} identified
39,668 clusters from $\sim$$8400$ deg$^2$ of the SDSS DR6.  We now
improve the method to identify clusters of galaxies from
$\sim$$14,000$ deg$^2$ of the SDSS-III \citep[DR8,][]{dr8}. A cluster
is identified when the richness reaches $R=12$ within a radius
$r_{200}$ and a photometric redshift gap between $z\pm0.04(1+z)$. 
Here $r_{200}$ is the radius within which the mean density
is 200 times the critical density of the universe. The cluster
richness, $R$, is defined to be the average number of $L^{\ast}$
galaxies, i.e., $R=\sum L_r/L^{\ast}$, where $\sum L_r$ is the
$r$-band total luminosity of cluster galaxies after the background
subtraction. $L^{\ast}$ is the characteristic luminosity in the
Schechter luminosity function \citep{bhb+03}. From the SDSS-III data,
we now identify about 130,000 clusters in the redshift range of $0.05<
z <0.8$ (Wen et al. 2011, in preparation). 
By inspecting color images on the SDSS web page\footnote
{http://skyserver.sdss3.org/dr8/en/tools/chart/list.asp}, we find 68
new clusters with giant arcs, in addition to 30 known lensing
clusters.  They are listed in Table~\ref{lens.tab}.  Among them, 13
clusters are {\it almost certain} lensing systems (see
Fig.~\ref{lens_sure}) which show tangential giant arcs or multiple
tangential giant arcs with respect to the bright central galaxies
with a separation of $S=2.0''-18.5''$. The faint but clear arcs usually
have blue colors compared with the bright central
galaxies. Remarkably, SDSS J005848.9$-$072156 shows the giant arc
almost forming a half circle with a radius of $S=14.4''$. In a
$\Lambda$CDM cosmology ($H_0=$72 ${\rm km~s}^{-1}$ ${\rm Mpc}^{-1}$,
$\Omega_m=0.3$ and $\Omega_{\Lambda}=0.7$, hereafter), we estimate the
mass within this radius by \citet{wu00},
\begin{equation}
M(<S)=\frac{c^2S^2}{4G}\frac{D_{\rm l}D_{\rm s}}{D_{\rm ls}},
\label{lensm}
\end{equation}
where $D_{\rm s}$ and $D_{\rm l}$ are the angular diameter distances
of the source and lens from the observer, and $D_{\rm ls}$ is the
angular diameter distance of the source from the lens. We get the mass
$M(<S)=1.1\times10^{14}~M_{\odot}$ if the source has a
redshift $z_{\rm s}=1$, or $M(<S)=0.61\times10^{14}~M_{\odot}$
if $z_{\rm s}=2$.

\begin{table}[!hpt]
\centering
\begin{minipage}[]{100mm}
\caption[]{Strong lensing clusters in the SDSS-III.} 
\label{lens.tab}\end{minipage}

\vspace{-3mm} \fns


\tabcolsep 1.5mm
\begin{tabular}{lcrrrrl}

\noalign{\smallskip}\hline\noalign{\smallskip}
\multicolumn{1}{c}{Cluster name}&\multicolumn{1}{c}{Cluster $z$}&\multicolumn{1}{c}{Richness}&
\multicolumn{1}{c}{$S$}&\multicolumn{1}{c}{$r$} &\multicolumn{1}{c}{$g-r$}&Reference, Notes\\
    &   &   &\multicolumn{1}{c}{($''$)} &\multicolumn{1}{c}{(mag)}&\multicolumn{1}{c}{(mag)} &  \\
\multicolumn{1}{c}{(1)}&\multicolumn{1}{c}{(2)}&\multicolumn{1}{c}{(3)}&\multicolumn{1}{c}{(4)}
 &\multicolumn{1}{c}{(5)}&\multicolumn{1}{c}{(6)}&\multicolumn{1}{c}{(7)}\\
\hline\noalign{\smallskip}
SDSS J002240.9$+$143110&  0.380&  18.96&  3.3& 21.37$\pm$  0.08& $ 0.19\pm$  0.11& 1   \\
SDSS J014656.0$-$092952&  0.448&  66.11&  9.2& 21.75$\pm$  0.12& $ 3.00\pm$  0.74& 2,3 \\
SDSS J023953.1$-$013455&  0.375& 169.43& 25.7& 22.38$\pm$  0.18& $ 2.04\pm$  0.55& 4   \\
SDSS J024803.4$-$033145&  0.188& 108.18& 16.4& 22.67$\pm$  0.21& $ 0.20\pm$  0.26& 5   \\
SDSS J082728.4$+$223245&  0.335& 108.26&  2.2& 21.12$\pm$  0.06& $-0.20\pm$  0.08& 6   \\
SDSS J090002.6$+$223404&  0.489&  19.37&  7.0& 21.08$\pm$  0.06& $-0.11\pm$  0.08& 7,8 \\
SDSS J095240.2$+$343446&  0.354&  46.10&  6.9& 22.09$\pm$  0.14& $ 0.92\pm$  0.26& 9   \\
SDSS J095739.2$+$050931&  0.429&  16.06&  8.0& 19.90$\pm$  0.07& $ 0.19\pm$  0.09& 7,9,10  \\
SDSS J103843.6$+$484917&  0.426&  19.44&  7.8& 22.06$\pm$  0.14& $ 0.09\pm$  0.17& 10,11,12\\
SDSS J111310.6$+$235639&  0.336&  80.47& 11.5& 23.25$\pm$  0.26& $ 0.07\pm$  0.32& 7,12    \\
SDSS J113313.2$+$500840&  0.370&  18.78& 10.4& 23.07$\pm$  0.24& $ 0.36\pm$  0.33& 5    \\
SDSS J113740.1$+$493635&  0.448&  15.43&  3.7& 20.35$\pm$  0.04& $ 0.05\pm$  0.05& 7,12 \\
SDSS J115200.2$+$331342&  0.357&  63.85&  7.9& 23.61$\pm$  0.51& $-0.56\pm$  0.55& 10   \\
SDSS J120602.1$+$514229&  0.422&  18.96&  3.8& 19.93$\pm$  0.06& $ 0.39\pm$  0.08& 13   \\
SDSS J120735.9$+$525459&  0.282&  24.77&  9.4& 22.54$\pm$  0.18& $-0.06\pm$  0.21& 7,9  \\
SDSS J120923.7$+$264046&  0.559& 110.76& 10.6& 22.98$\pm$  0.30& $-0.11\pm$  0.35& 10,14\\
SDSS J122651.7$+$215225&  0.433& 117.13& 10.9& 22.28$\pm$  0.18& $ 0.26\pm$  0.24& 7,10 \\
SDSS J124032.3$+$450902&  0.251&  19.40&  3.8& 20.16$\pm$  0.04& $-0.27\pm$  0.04& 11   \\
SDSS J131811.5$+$394226&  0.475&  26.81&  8.5& 22.16$\pm$  0.15& $ 0.33\pm$  0.21& 7,9  \\
SDSS J134332.9$+$415503&  0.418&  37.84& 12.7& 22.31$\pm$  0.19& $ 0.20\pm$  0.24& 7,8,10\\
SDSS J141912.2$+$532611&  0.638&  22.80& 10.0& 21.85$\pm$  0.14& $ 0.26\pm$  0.17& 15   \\
SDSS J151118.7$+$471340&  0.451&  31.15&  5.4& 23.26$\pm$  1.18& $-2.56\pm$  1.18& 12   \\
SDSS J152745.8$+$065233&  0.400&  92.84& 17.7& 21.65$\pm$  0.16& $-0.02\pm$  0.17& 3,10,16\\
SDSS J153713.2$+$655621&  0.251&  32.99&  8.1& 22.64$\pm$  0.24& $ 0.33\pm$  0.31& 9    \\
SDSS J162132.4$+$060719&  0.361&  49.20& 16.1& 21.51$\pm$  0.10& $ 0.96\pm$  0.20& 7,10 \\
SDSS J172336.2$+$341158&  0.442&  34.78&  4.7& 20.56$\pm$  0.04& $-0.21\pm$  0.05& 7,9  \\
SDSS J211119.3$-$011423&  0.638&  22.70& 10.9& 21.23$\pm$  0.17& $ 0.04\pm$  0.22& 3    \\
SDSS J223831.3$+$131955&  0.413&  38.35&  9.3& 22.61$\pm$  0.16& $ 0.60\pm$  0.27& 7,10 \\
SDSS J223933.1$-$042917&  0.553&  23.10&  3.4& 22.34$\pm$  0.20& $-0.16\pm$  0.23& 17   \\
SDSS J224712.3$-$020537&  0.329&  88.14&  9.0& 22.44$\pm$  0.19& $ 0.66\pm$  0.30& 5    \\[1mm]
SDSS J004827.2$+$031116&  0.337&  13.64&  8.0& 22.04$\pm$  0.14& $-0.25\pm$  0.16&  almost certain \\
SDSS J005848.9$-$072156&  0.619&  33.35& 14.4& 20.03$\pm$  0.11& $ 0.14\pm$  0.14&  almost certain \\
SDSS J014504.3$-$045551&  0.604&  16.64&  2.0& 21.83$\pm$  0.14& $ 0.08\pm$  0.17&  almost certain \\
SDSS J100859.8$+$193717&  0.306&  43.85&  6.1& 22.99$\pm$  0.28& $ 0.51\pm$  0.44&  almost certain* \\
SDSS J111017.7$+$645948&  0.601&  29.55& 14.9& 21.42$\pm$  0.11& $ 0.40\pm$  0.15&  almost certain \\
SDSS J111504.4$+$164538&  0.588&  15.35&  6.9& 21.89$\pm$  0.12& $ 0.15\pm$  0.15&  almost certain \\
SDSS J113829.2$+$155417&  0.450&  27.84&  6.8& 23.35$\pm$  0.38& $-0.32\pm$  0.43&  almost certain \\
SDSS J114723.3$+$333153&  0.212&  34.47&  5.3& 23.12$\pm$  0.30& $-0.09\pm$  0.35&  almost certain \\
SDSS J115605.5$+$191112&  0.516&  13.05&  5.5& 22.76$\pm$  0.23& $-0.13\pm$  0.27&  almost certain \\
SDSS J122718.7$+$172551&  0.308&  14.73& 11.8& 24.42$\pm$  0.58& $-0.61\pm$  0.65&  almost certain* \\
SDSS J143956.5$+$325024&  0.418&  56.22&  7.7& 22.94$\pm$  0.23& $-0.03\pm$  0.28&  almost certain \\
SDSS J144823.2$+$450021&  0.554&  36.75& 15.3& 22.66$\pm$  0.18& $-0.14\pm$  0.22&  almost certain \\
SDSS J230017.3$+$221329&  0.444&  46.64& 18.5& 21.46$\pm$  0.14& $ 0.41\pm$  0.19&  almost certain \\[1mm]
SDSS J001331.9$+$351220&  0.228&  25.94& 10.0& 20.99$\pm$  0.10& $ 0.25\pm$  0.13&  probable \\
SDSS J002556.3$+$362439&  0.322&  19.55&  9.3& 22.41$\pm$  0.18& $ 1.59\pm$  0.43&  probable \\
SDSS J003237.3$+$073649&  0.481&  24.21&  6.9& 20.94$\pm$  0.07& $ 0.07\pm$  0.09&  probable \\
SDSS J010049.2$+$181827&  0.650&  29.07&  5.5& 22.55$\pm$  0.22& $ 0.52\pm$  0.30&  probable \\
SDSS J014350.1$+$160739&  0.452&  20.13&  2.8& 22.15$\pm$  0.18& $-0.38\pm$  0.20&  probable \\
SDSS J090122.4$+$181432&  0.346&  21.01&  6.5& 22.35$\pm$  0.16& $ 0.92\pm$  0.27&  probable \\
SDSS J100202.5$+$602026&  0.573&  46.01& 14.8& 22.32$\pm$  0.20& $ 1.54\pm$  0.48&  probable \\
SDSS J100226.8$+$203101&  0.321& 135.69& 14.1&      --         &   --            &  probable \\
SDSS J104044.6$+$330520&  0.615&  41.43&  7.9& 22.52$\pm$  0.16& $ 0.03\pm$  0.20&  probable \\
SDSS J104601.1$+$104852&  0.258&  14.06&  5.8& 21.72$\pm$  0.14& $ 0.07\pm$  0.17&  probable \\
SDSS J115120.6$+$645530&  0.561&  36.75& 17.2& 22.17$\pm$  0.14& $ 1.05\pm$  0.28&  probable \\
SDSS J122221.6$+$241909&  0.501&  52.21& 11.7& 22.88$\pm$  0.23& $ 0.21\pm$  0.31&  probable \\
SDSS J122656.8$+$052045&  0.504&  25.84&  8.6& 22.04$\pm$  0.16& $ 0.48\pm$  0.24&  probable \\
SDSS J130137.7$+$551917&  0.606&  24.11&  6.8& 21.37$\pm$  0.09& $-0.16\pm$  0.10&  probable \\
SDSS J140115.1$-$075014&  0.508&  96.27& 12.4& 21.23$\pm$  0.09& $ 0.22\pm$  0.12&  probable \\
SDSS J162028.2$-$010236&  0.370&  13.68&  3.4& 22.03$\pm$  0.14& $ 1.69\pm$  0.43&  probable \\
SDSS J163415.8$+$250843&  0.221&  20.41& 14.2& 21.02$\pm$  0.09& $ 0.84\pm$  0.15&  probable \\
\noalign{\smallskip}\hline\noalign{\smallskip}
\end{tabular}
\end{table}
\begin{table}[!!!!!hpt]
\addtocounter{table}{-1}
\centering
\begin{minipage}[]{100mm}
\caption[]{-- {\it continued}} 
\end{minipage}

\vspace{-3mm} \fns

\tabcolsep 1.5mm
\begin{tabular}{lcrrrrl}

\noalign{\smallskip}\hline\noalign{\smallskip}
\multicolumn{1}{c}{Cluster name}&\multicolumn{1}{c}{Cluster $z$}&\multicolumn{1}{c}{Richness}&
\multicolumn{1}{c}{$S$}&\multicolumn{1}{c}{$r$} &\multicolumn{1}{c}{$g-r$}&Reference, Notes\\
   &   &   &\multicolumn{1}{c}{($''$)}    &\multicolumn{1}{c}{(mag)}&\multicolumn{1}{c}{(mag)}& \\
\multicolumn{1}{c}{(1)}&\multicolumn{1}{c}{(2)}&\multicolumn{1}{c}{(3)}&\multicolumn{1}{c}{(4)}
 &\multicolumn{1}{c}{(5)}&\multicolumn{1}{c}{(6)}&\multicolumn{1}{c}{(7)}\\
\hline\noalign{\smallskip}
SDSS J195835.3$+$595058&  0.231&  74.24&  7.9& 20.37$\pm$  0.04& $ 0.50\pm$  0.07&  probable \\
SDSS J224405.0$+$275915&  0.352&  54.52&  6.4& 22.53$\pm$  0.21& $ 1.29\pm$  0.42&  probable \\
SDSS J224621.2$+$223337&  0.571&  12.72&  2.7& 21.88$\pm$  0.10& $ 0.37\pm$  0.13&  probable \\
SDSS J234924.0$+$053835&  0.692&  54.02&  6.4& 21.53$\pm$  0.11& $ 0.66\pm$  0.18&  probable \\
SDSS J235335.0$+$254257&  0.521&  12.70&  8.5& 21.48$\pm$  0.12& $ 0.30\pm$  0.17&  probable \\[1mm]
SDSS J002547.5$+$285329&  0.342&  19.47&  7.2& 20.77$\pm$  0.05& $ 0.03\pm$  0.06&  possible \\
SDSS J002824.1$+$224802&  0.449&  32.12& 18.8& 20.48$\pm$  0.07& $ 1.07\pm$  0.13&  possible \\
SDSS J005403.9$+$185504&  0.477&  19.27&  7.5& 19.74$\pm$  0.18& $ 4.58\pm$  6.73&  possible \\
SDSS J005529.8$+$074114&  0.366&  22.17& 11.4& 20.79$\pm$  0.10& $ 2.48\pm$  0.58&  possible \\
SDSS J010842.0$+$062443&  0.564&  35.23&  3.1& 20.85$\pm$  0.07& $-0.02\pm$  0.08&  possible \\
SDSS J020638.9$+$044803&  0.268&  38.72&  5.4& 20.25$\pm$  0.08& $ 0.31\pm$  0.10&  possible \\
SDSS J080731.5$+$441048&  0.449&  15.54&  2.9& 22.83$\pm$  0.26& $ 0.29\pm$  0.36&  possible \\
SDSS J082854.3$+$103408&  0.312&  32.38& 12.8& 21.04$\pm$  0.06& $ 0.51\pm$  0.09&  possible \\
SDSS J084647.5$+$044605&  0.242&  38.94&  4.2& 21.91$\pm$  0.14& $-0.38\pm$  0.15&  possible \\
SDSS J085428.7$+$100814&  0.298&  28.59&  4.5& 22.30$\pm$  0.17& $ 0.19\pm$  0.23&  possible \\
SDSS J105559.4$+$155514&  0.284&  16.03&  6.1& 21.74$\pm$  0.15& $ 0.27\pm$  0.19&  possible \\
SDSS J110934.8$+$142953&  0.476&  22.20& 10.2& 22.39$\pm$  0.15& $ 1.12\pm$  0.27&  possible \\
SDSS J112229.5$+$212416&  0.380&  12.00&  4.3& 21.35$\pm$  0.07& $ 0.65\pm$  0.12&  possible \\
SDSS J113912.9$+$165435&  0.579&  28.89& 20.9& 20.33$\pm$  0.05& $ 0.65\pm$  0.08&  possible \\
SDSS J120453.5$+$135753&  0.533&  20.22&  3.2& 22.98$\pm$  0.27& $-0.08\pm$  0.33&  possible \\
SDSS J123736.2$+$553343&  0.410&  24.66&  4.5& 19.96$\pm$  0.03& $-0.04\pm$  0.04& 7*, possible \\
SDSS J133145.3$+$513431&  0.283&  33.34&  3.4& 23.37$\pm$  0.29& $-0.12\pm$  0.35&  possible \\
SDSS J133314.4$+$184637&  0.315&  15.44&  5.3& 22.68$\pm$  0.18& $ 1.27\pm$  0.39&  possible \\
SDSS J140110.4$+$565420&  0.493&  32.84&  3.8& 21.50$\pm$  0.08& $ 0.45\pm$  0.12&  possible \\
SDSS J142040.9$+$400454&  0.634&  23.26& 14.6& 20.46$\pm$  0.07& $ 0.36\pm$  0.09&  possible \\
SDSS J144355.0$+$585302&  0.156&  21.55&  7.3& 23.72$\pm$  0.51& $-1.06\pm$  0.53&  possible \\
SDSS J145836.3$-$002400&  0.570&  25.02&  3.8& 21.69$\pm$  0.11& $ 1.00\pm$  0.21&  possible \\
SDSS J150236.6$+$292053&  0.554&  22.88&  2.2& 21.46$\pm$  0.08& $ 0.72\pm$  0.12&  possible \\
SDSS J150510.8$+$172042&  0.511&  28.24&  4.0& 22.54$\pm$  0.17& $ 2.11\pm$  0.51&  possible \\
SDSS J172535.3$+$654153&  0.519&  20.40&  6.3& 22.69$\pm$  0.24& $ 0.13\pm$  0.31&  possible \\
SDSS J212326.0$+$015312&  0.551&  15.19&  3.6& 22.05$\pm$  0.12& $ 0.75\pm$  0.20&  possible \\
SDSS J213340.5$-$050055&  0.316&  55.44&  9.0& 20.46$\pm$  0.05& $ 0.12\pm$  0.06&  possible \\
SDSS J222208.6$+$274535&  0.488&  37.59& 10.8& 21.66$\pm$  0.11& $ 0.25\pm$  0.14&  possible \\
SDSS J224859.0$+$014711&  0.372&  37.23&  8.8& 22.06$\pm$  0.16& $ 0.41\pm$  0.24&  possible \\
SDSS J234028.5$+$294747&  0.496&  12.05&  2.7& 22.29$\pm$  0.20& $-0.39\pm$  0.22&  possible \\
SDSS J234744.1$+$051324&  0.491&  24.75&  2.4& 21.83$\pm$  0.12& $ 1.51\pm$  0.34&  possible \\[1mm]
SDSS J113843.7$+$120802&  0.177&  13.66&  9.2& 20.80$\pm$  0.06& $ 0.22\pm$  0.09&  exotic ring \\
SDSS J155719.4$+$270416&  0.068&  12.00& 16.8& 22.21$\pm$  0.14& $ 1.03\pm$  0.25&  exotic ring \\
\noalign{\smallskip}\hline\noalign{\smallskip}
\end{tabular}
\tablecomments{0.9\textwidth}{Here we list the name, redshift and
richness of galaxy clusters in columns (1), (2) and (3). The angular separation ($S$) between the arc
and the central galaxy is listed in column (4), the $r$-band magnitude and color ($g-r$) 
of the brightest part of arc are given in columns (5) and (6). References and notes 
on the lensing systems are given in column (7).  
References are: (1) \citet{atl+07}; (2) \citet{ead+07}; (3) \citet{hgo+08}; 
(4) \citet{hr89}; (5) \citet{ste+05}; (6) \citet{sso+08}; 
(7) \citet{whx+09}; (8) \citet{daa+09}; (9) \citet{kad+10}; 
(10) \citet{bhg+11}; (11) \citet{beh+09}; (12) \citet{kaa+09}; 
(13) \citet{lba+09}; (14) \citet{osk08}; (15) \citet{ghy+03}; 
(16) \citet{kgh+10}; (17) \citet{sbs+06}.
*:CASSOWARY candidates, see http://www.ast.cam.ac.uk/research/cassowary}
\end{table}

\newpage

\begin{figure}[hp]
\vs\vs
\resizebox{35mm}{!}{\includegraphics{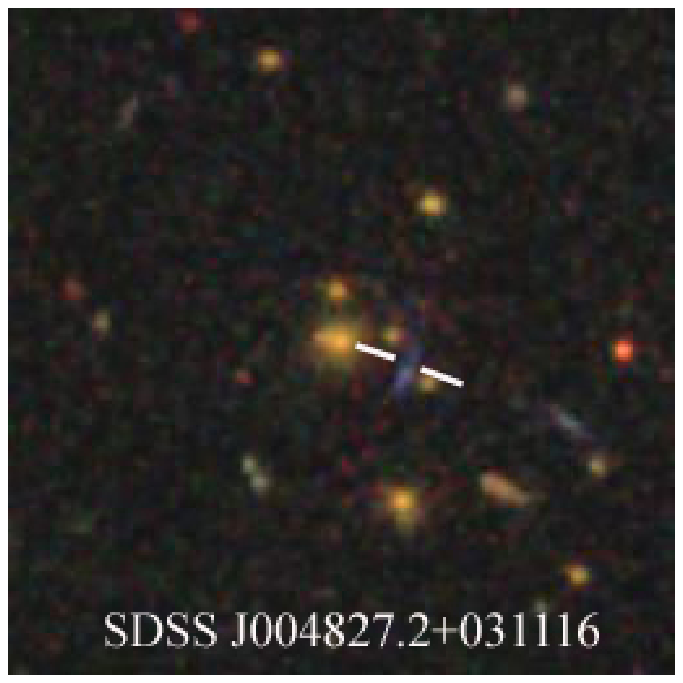}}~%
\resizebox{35mm}{!}{\includegraphics{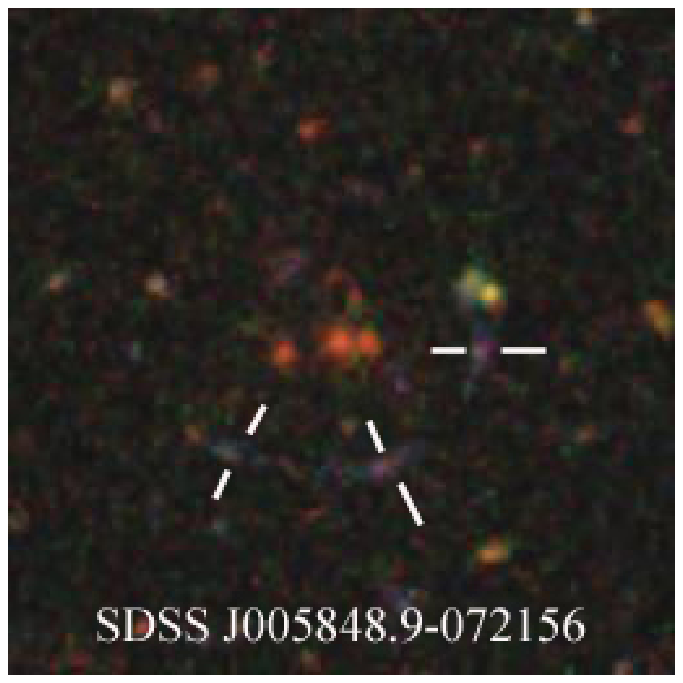}}~%
\resizebox{35mm}{!}{\includegraphics{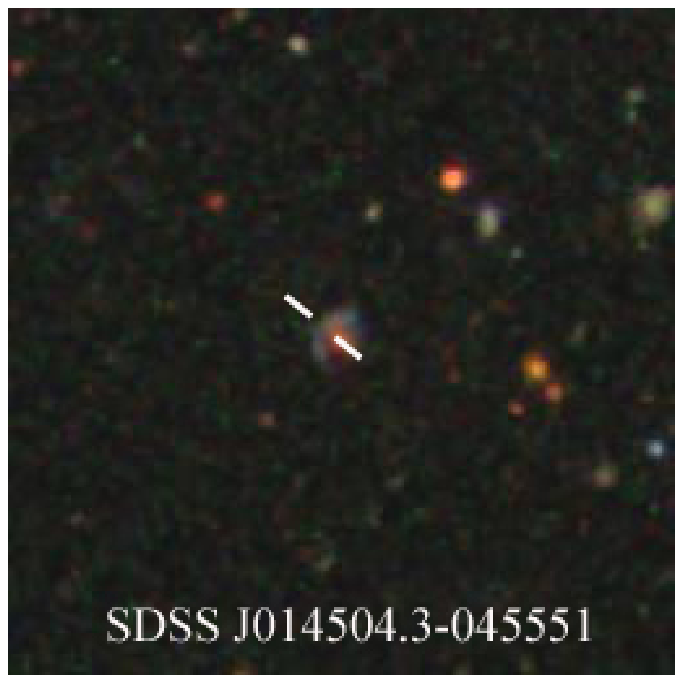}}~%
\resizebox{35mm}{!}{\includegraphics{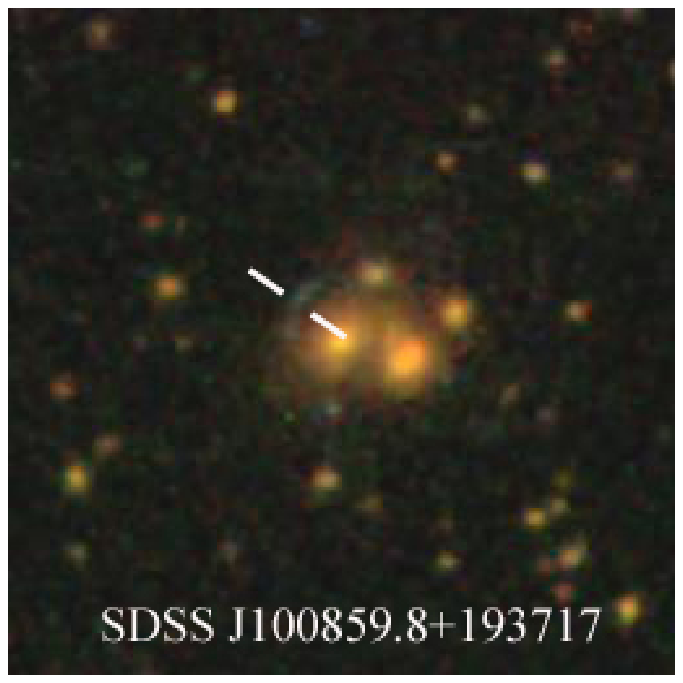}}\\[0.5mm]
\resizebox{35mm}{!}{\includegraphics{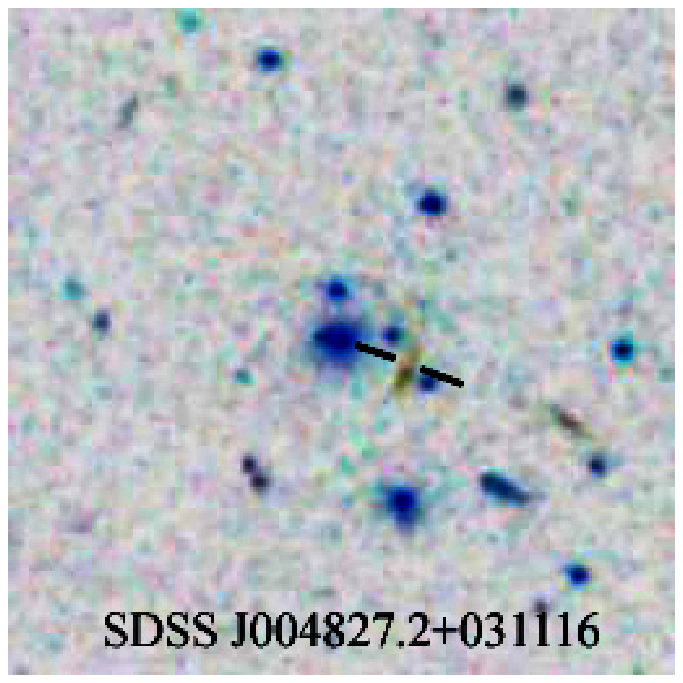}}~%
\resizebox{35mm}{!}{\includegraphics{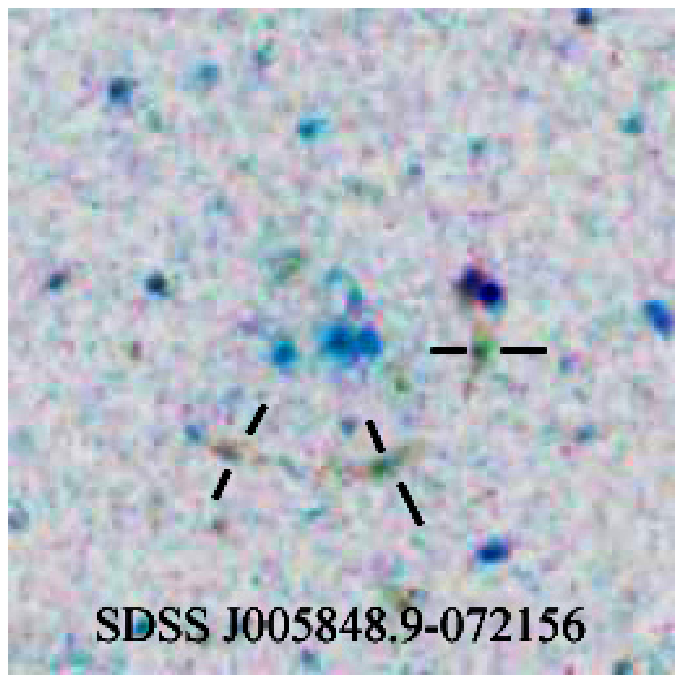}}~%
\resizebox{35mm}{!}{\includegraphics{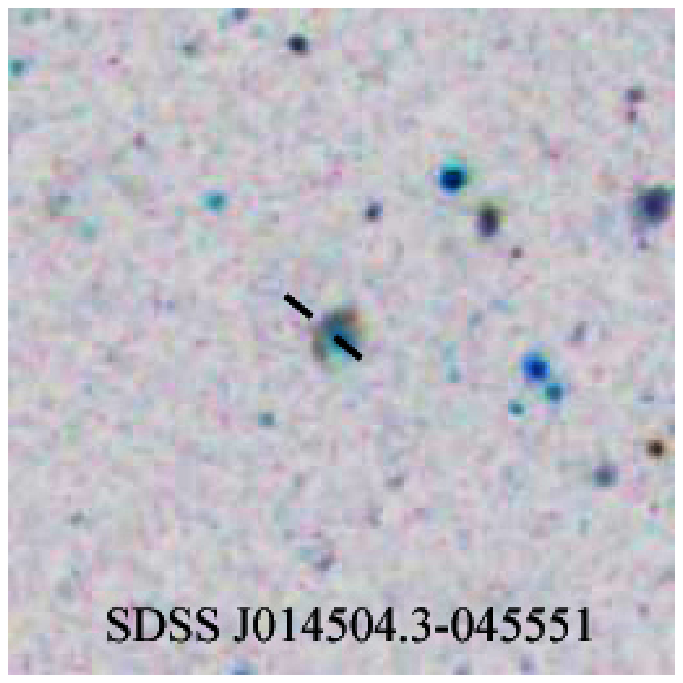}}~%
\resizebox{35mm}{!}{\includegraphics{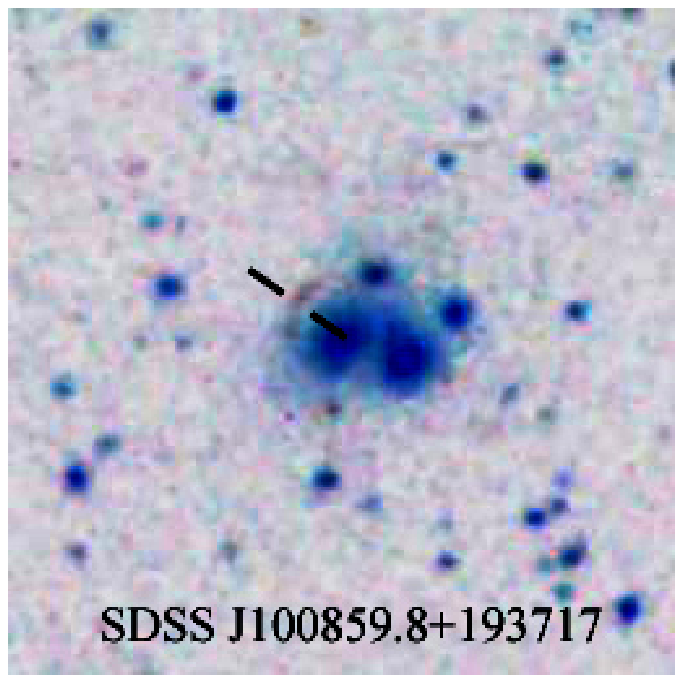}}\\[0.5mm]
\resizebox{35mm}{!}{\includegraphics{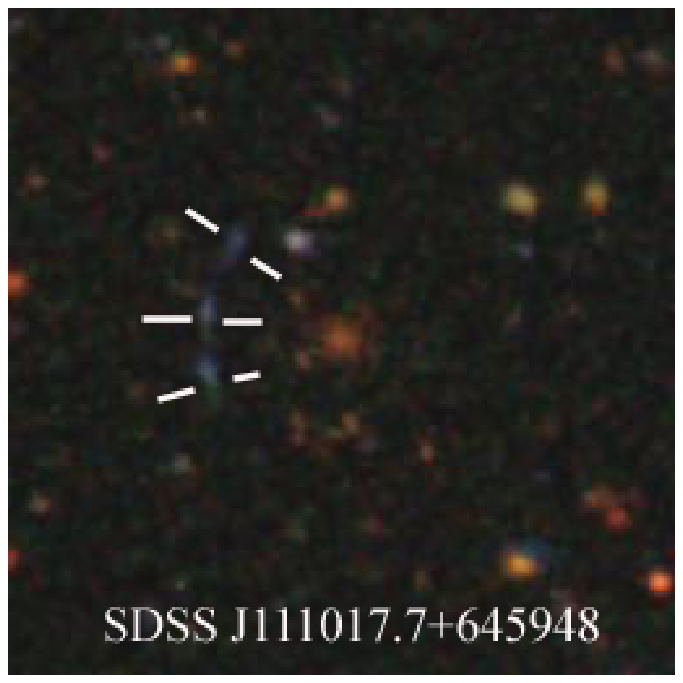}}~%
\resizebox{35mm}{!}{\includegraphics{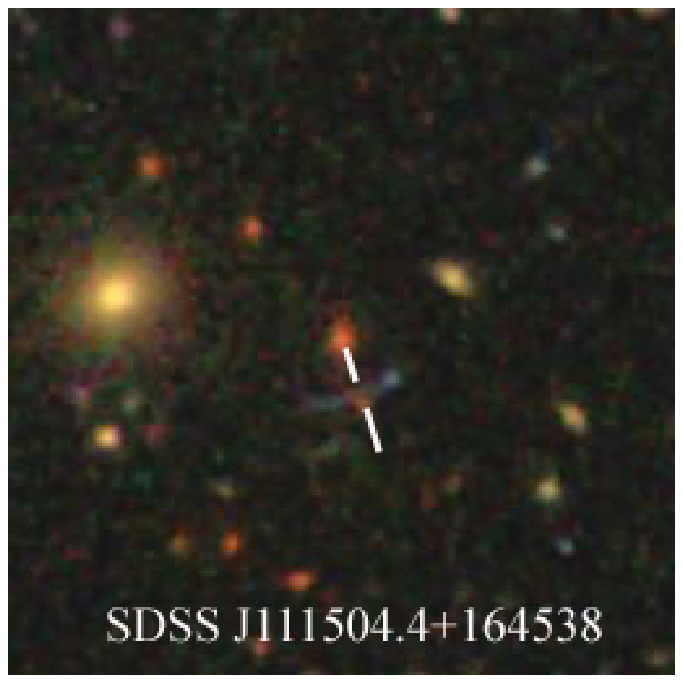}}~%
\resizebox{35mm}{!}{\includegraphics{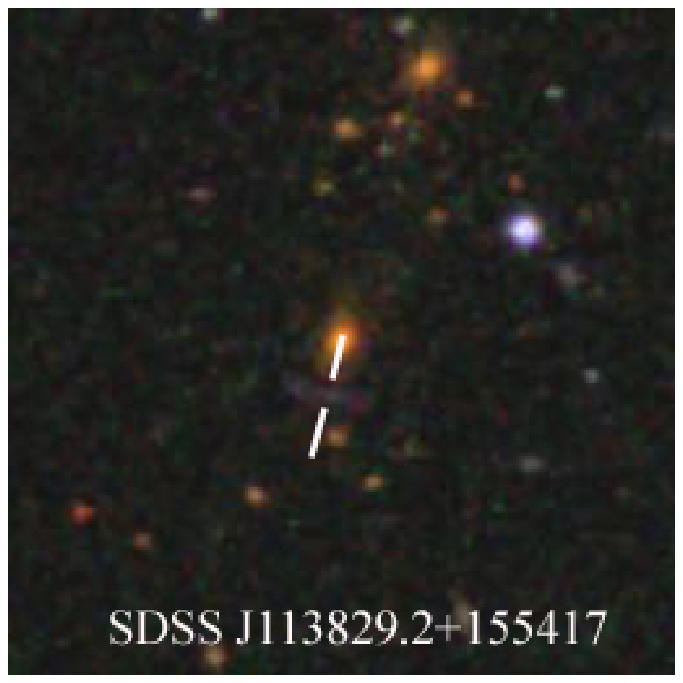}}~%
\resizebox{35mm}{!}{\includegraphics{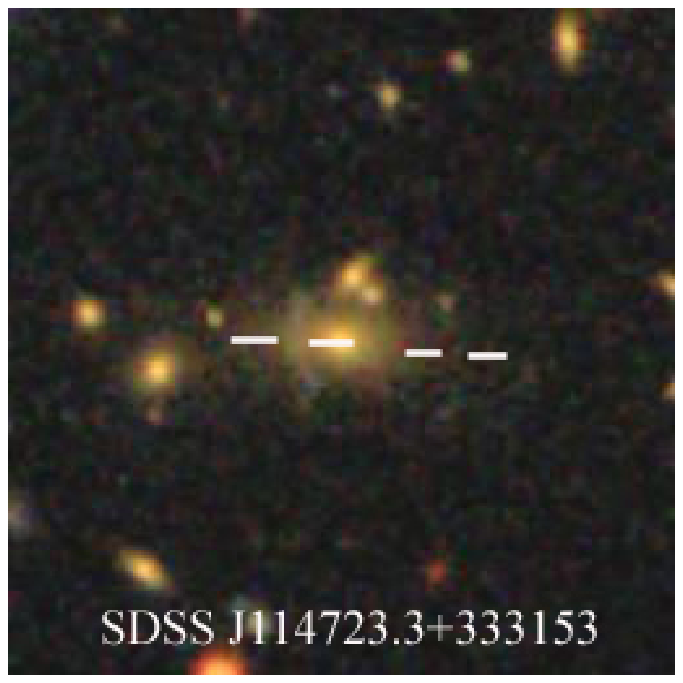}}\\[0.5mm]
\resizebox{35mm}{!}{\includegraphics{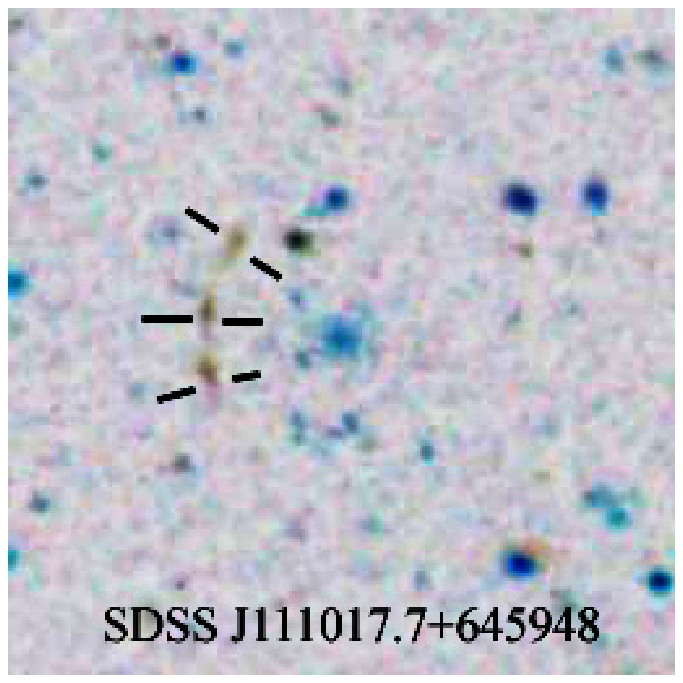}}~%
\resizebox{35mm}{!}{\includegraphics{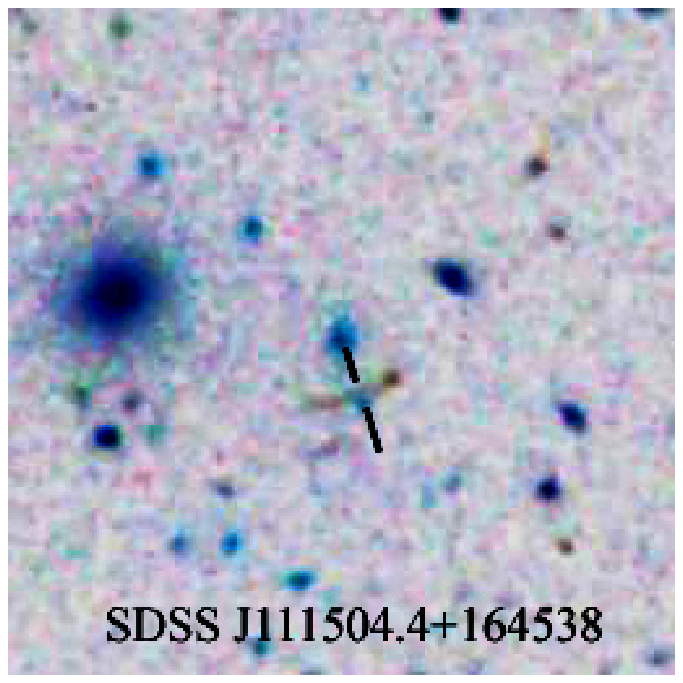}}~%
\resizebox{35mm}{!}{\includegraphics{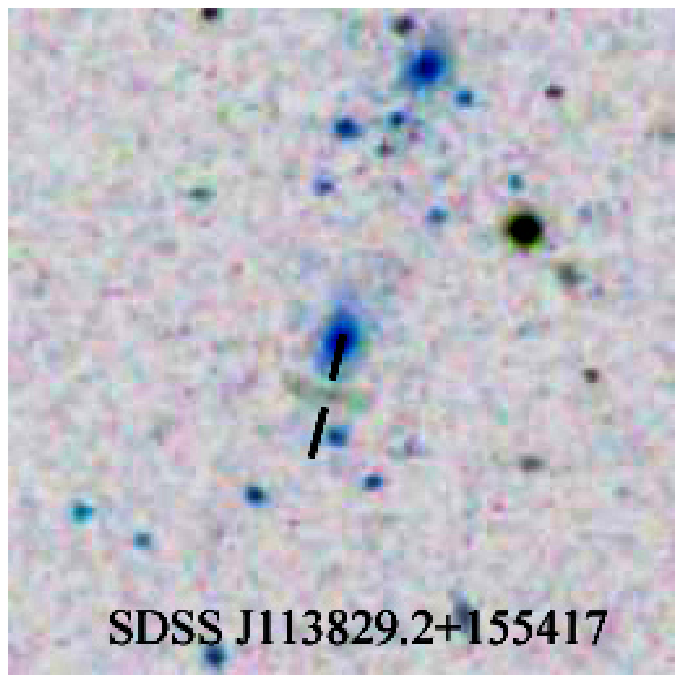}}~%
\resizebox{35mm}{!}{\includegraphics{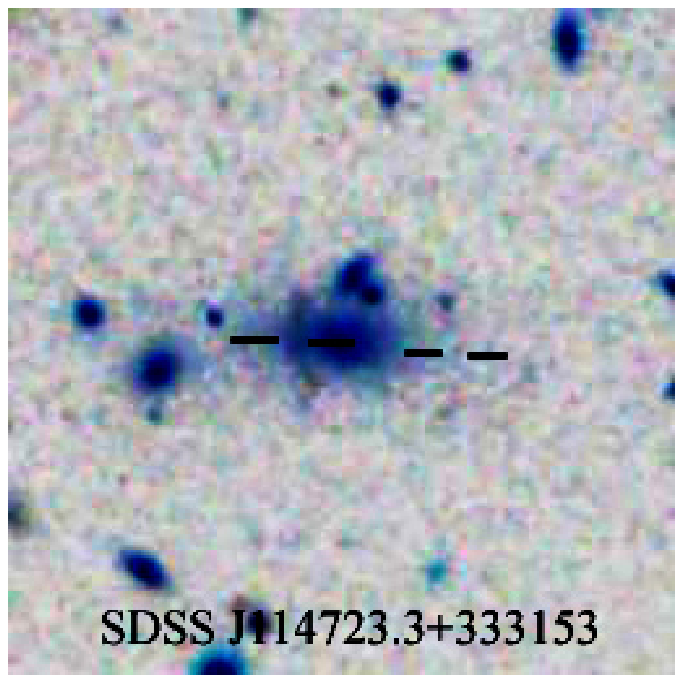}}\\[0.5mm]
\resizebox{35mm}{!}{\includegraphics{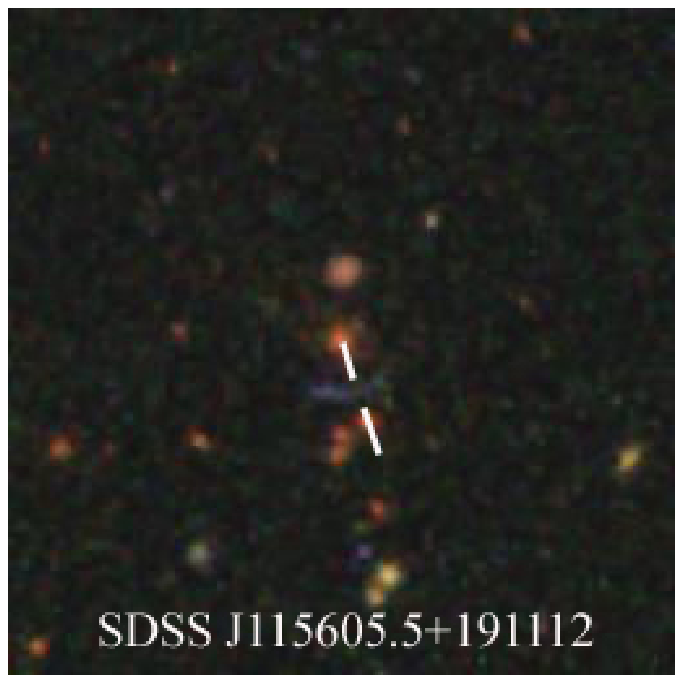}}~%
\resizebox{35mm}{!}{\includegraphics{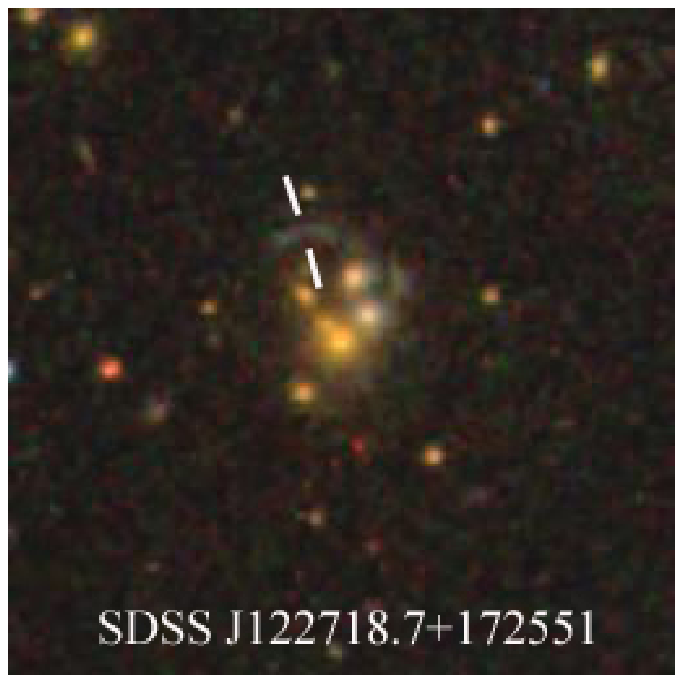}}~%
\resizebox{35mm}{!}{\includegraphics{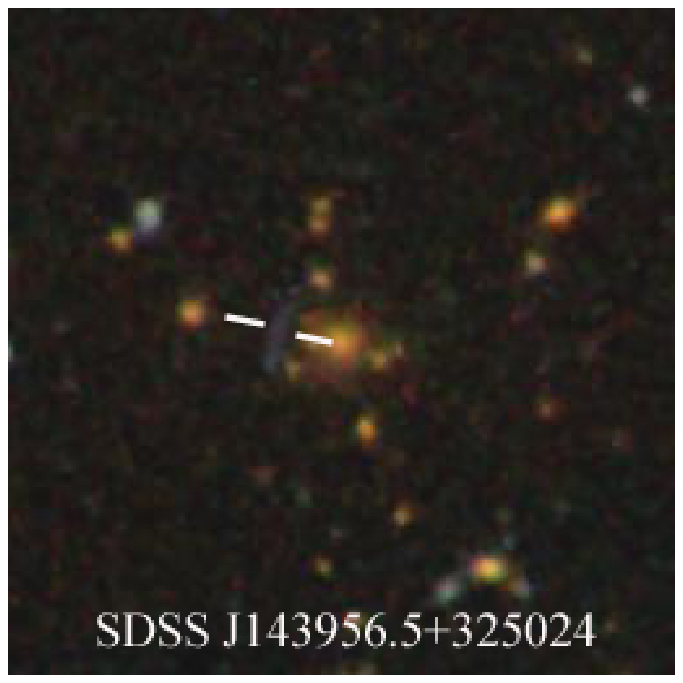}}~%
\resizebox{35mm}{!}{\includegraphics{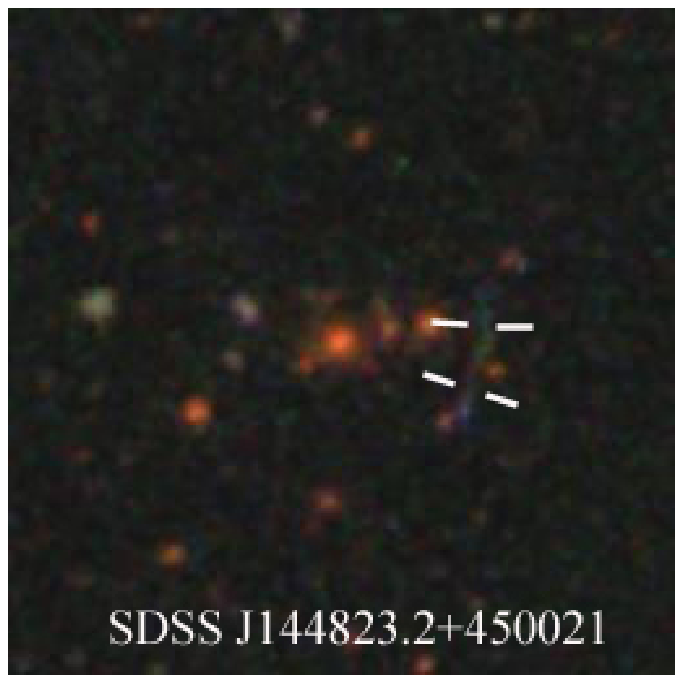}}\\[0.5mm]
\resizebox{35mm}{!}{\includegraphics{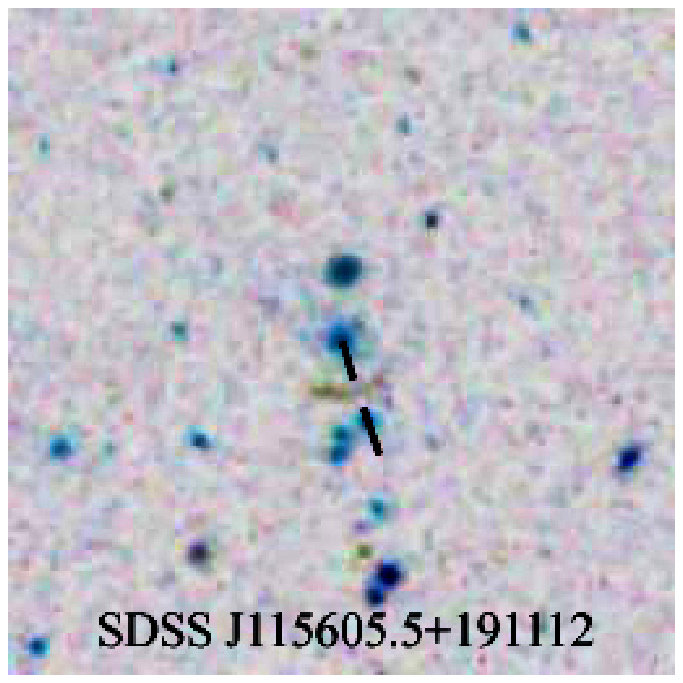}}~%
\resizebox{35mm}{!}{\includegraphics{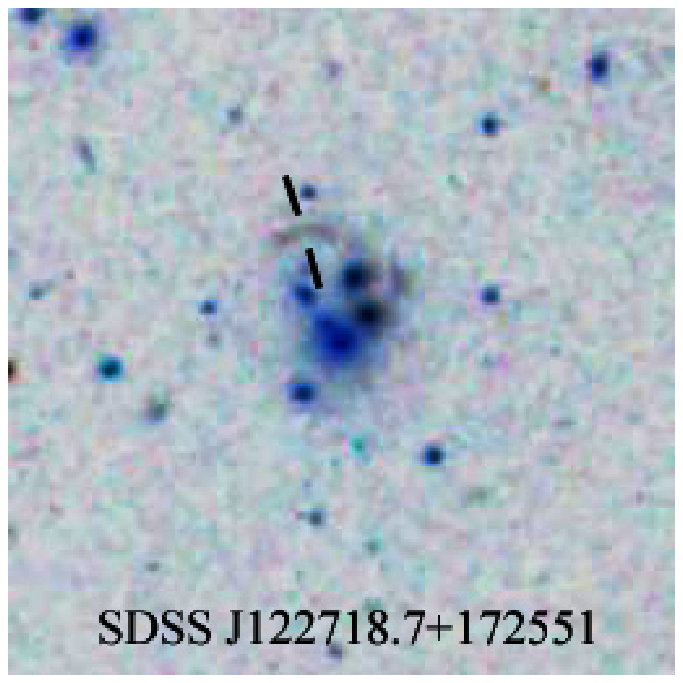}}~%
\resizebox{35mm}{!}{\includegraphics{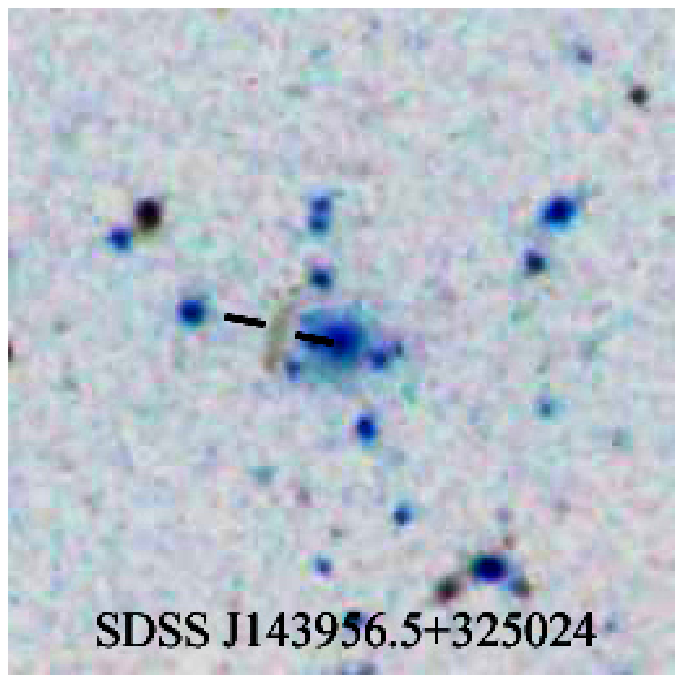}}~%
\resizebox{35mm}{!}{\includegraphics{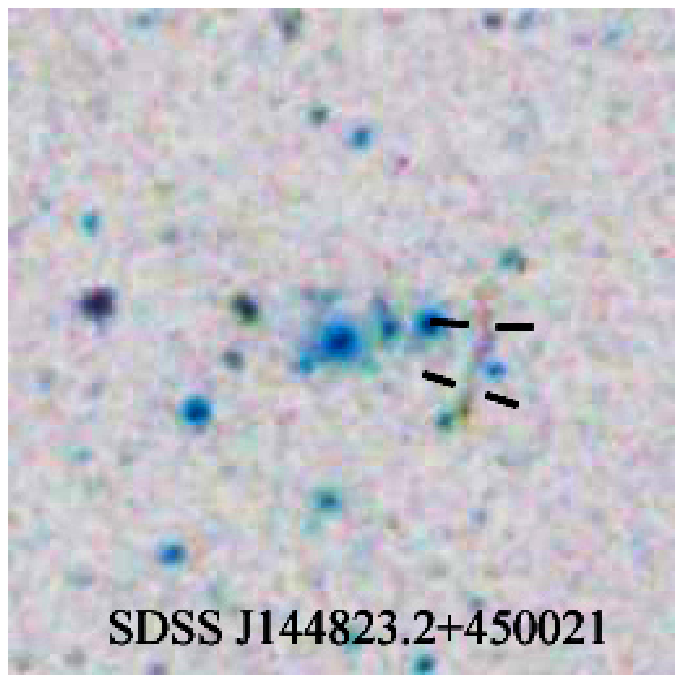}}
\caption{\baselineskip 3.6mm
SDSS composite color images ($g$, $r$ and $i$) of 13
clusters with a field of view of 
1.2$'\times$1.2$'$. They are almost certain gravitational 
lensing systems. The negative images are also shown 
in the second rows to see the lensing features more clearly.
\label{lens_sure}}
\end{figure}

\begin{figure}[!hpt]
\resizebox{35mm}{!}{\includegraphics{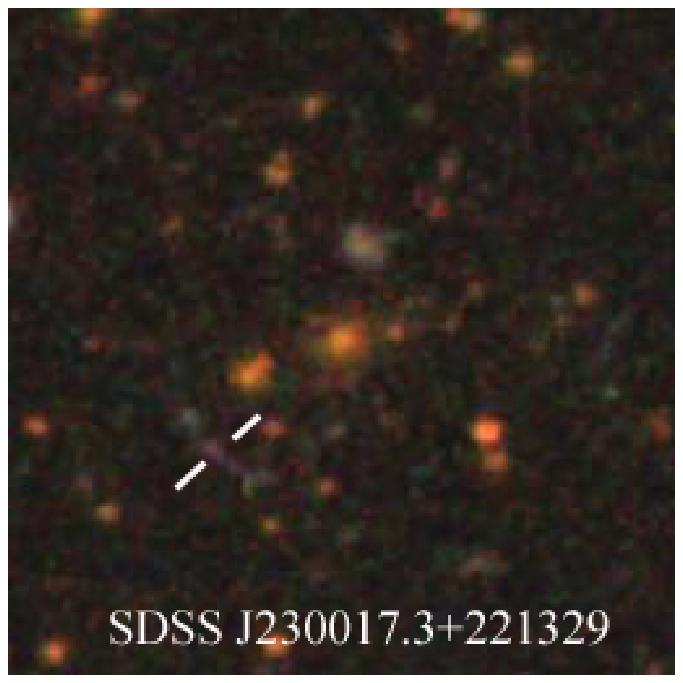}}~%
\resizebox{35mm}{!}{\includegraphics{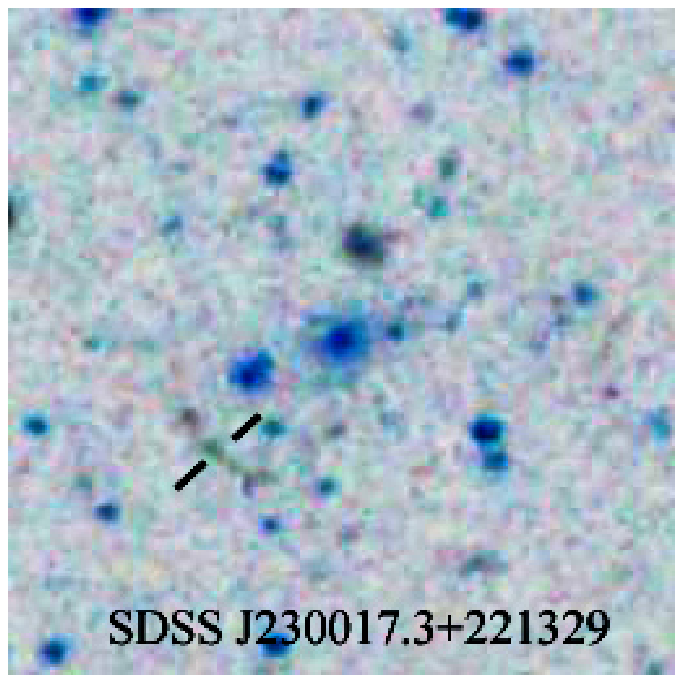}}
\setcounter{figure}{0}
\caption{{\it Continued}}
%
\resizebox{35mm}{!}{\includegraphics{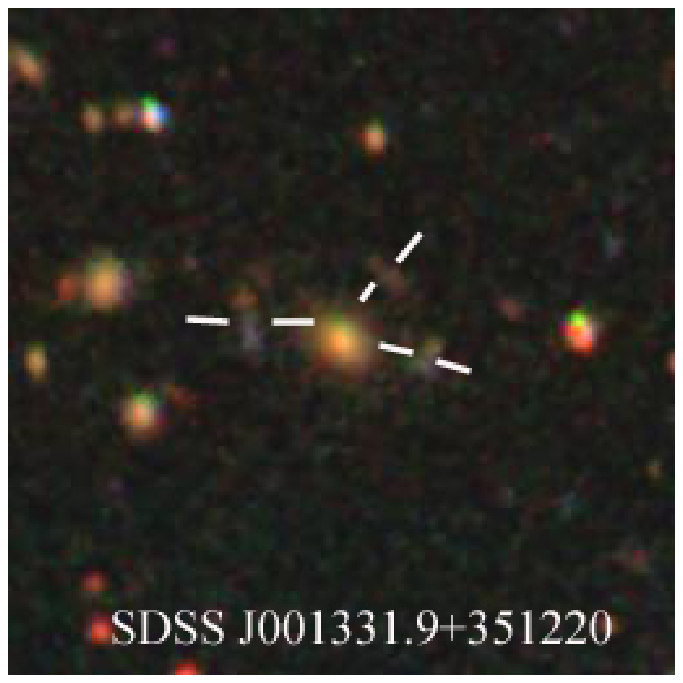}}~%
\resizebox{35mm}{!}{\includegraphics{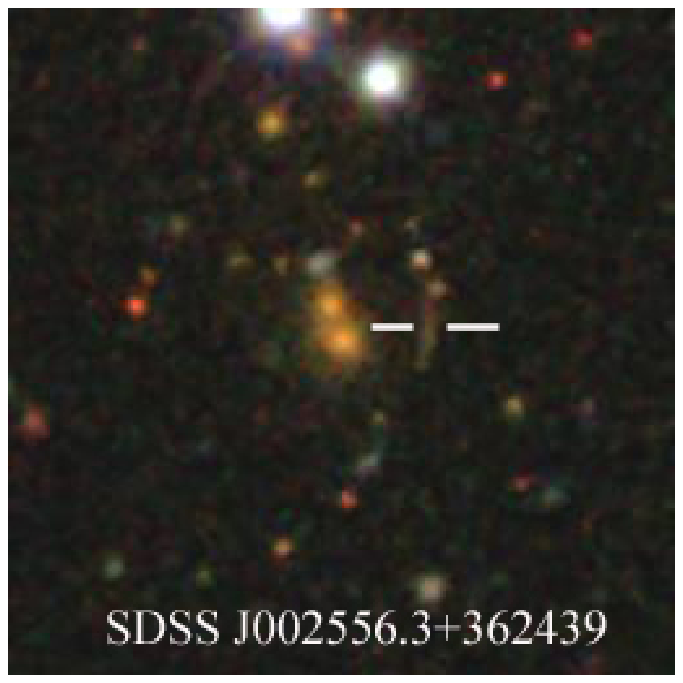}}~%
\resizebox{35mm}{!}{\includegraphics{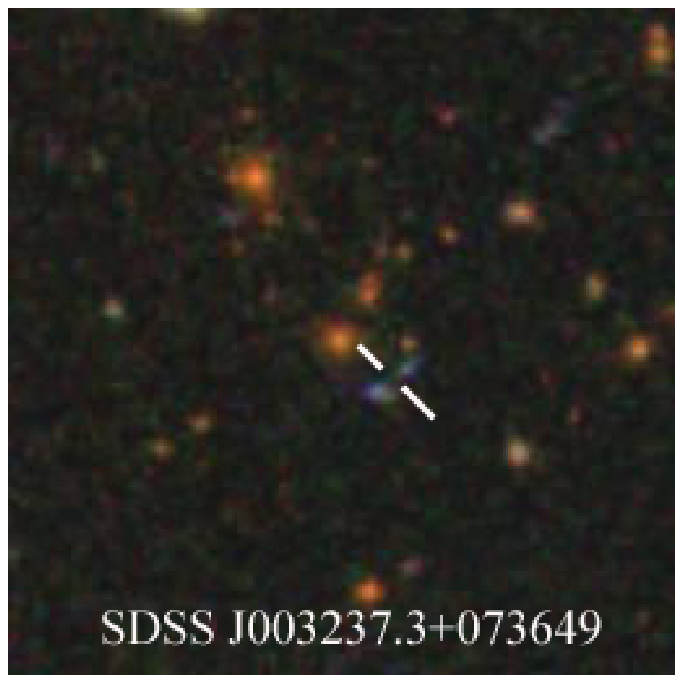}}~%
\resizebox{35mm}{!}{\includegraphics{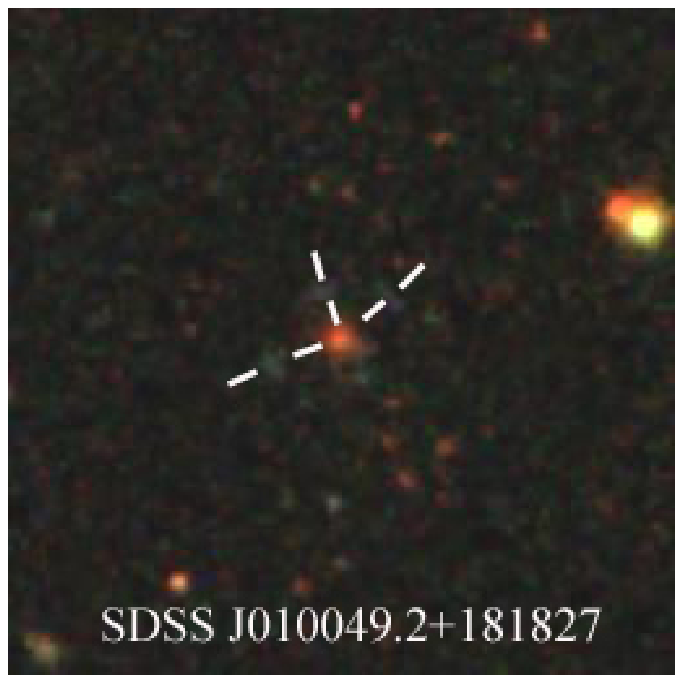}}\\[0.5mm]
\resizebox{35mm}{!}{\includegraphics{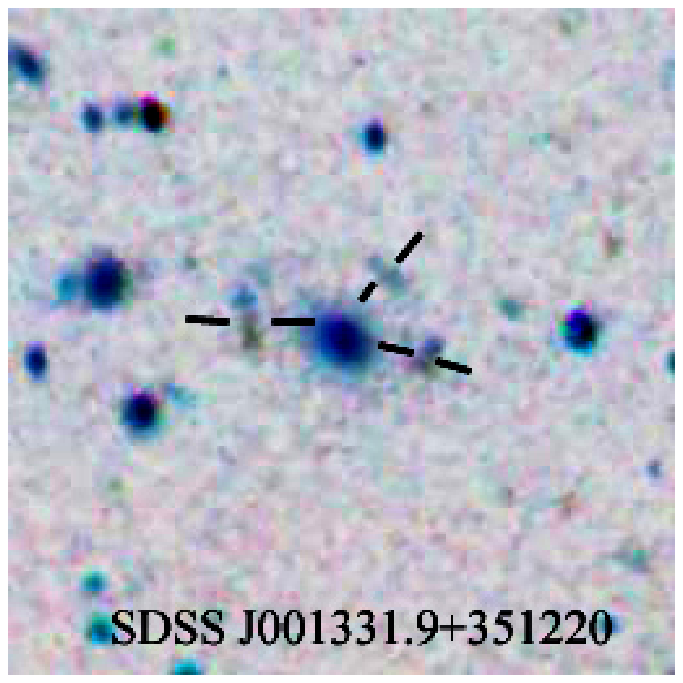}}~%
\resizebox{35mm}{!}{\includegraphics{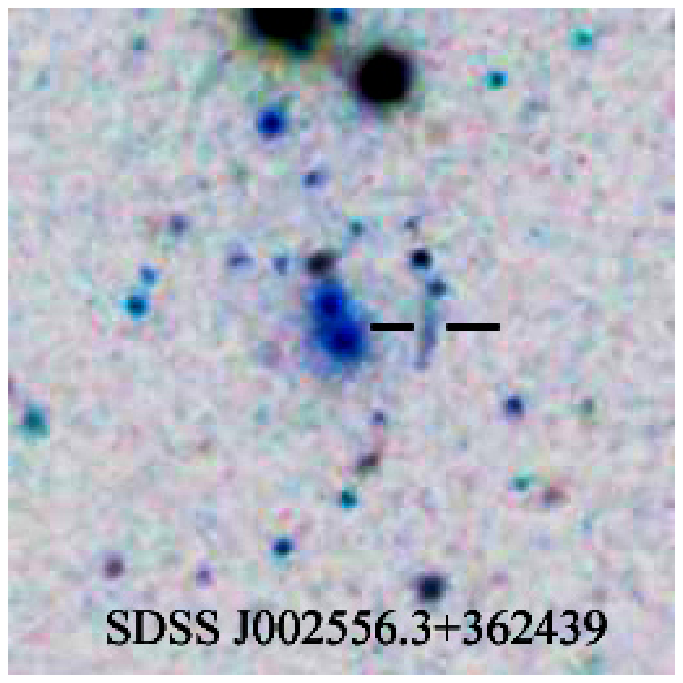}}~%
\resizebox{35mm}{!}{\includegraphics{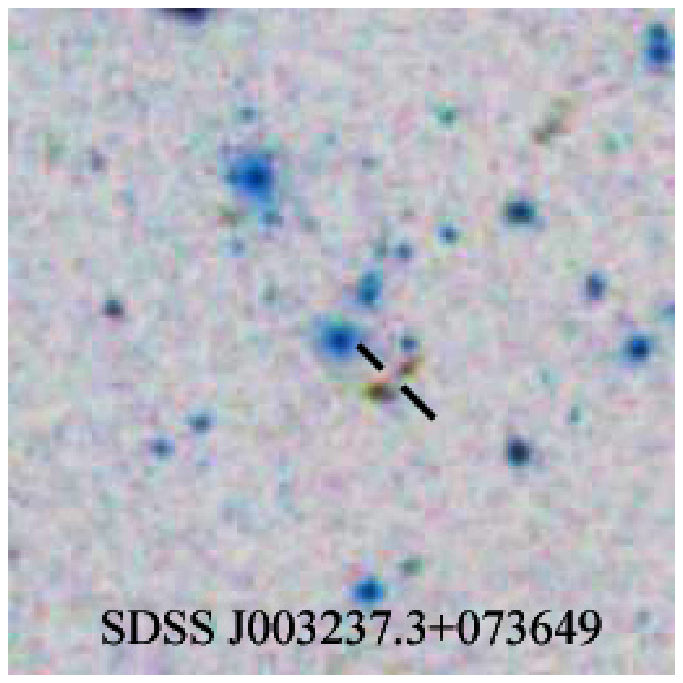}}~%
\resizebox{35mm}{!}{\includegraphics{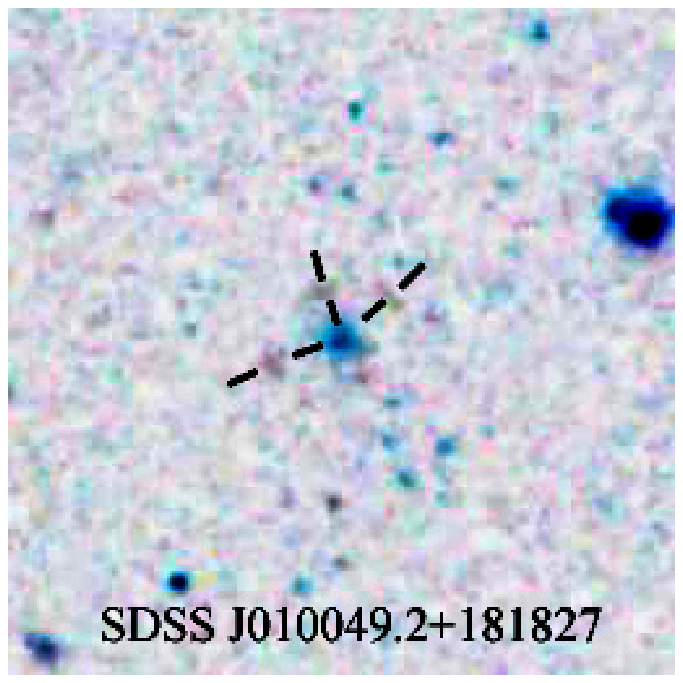}}\\[0.5mm]
\resizebox{35mm}{!}{\includegraphics{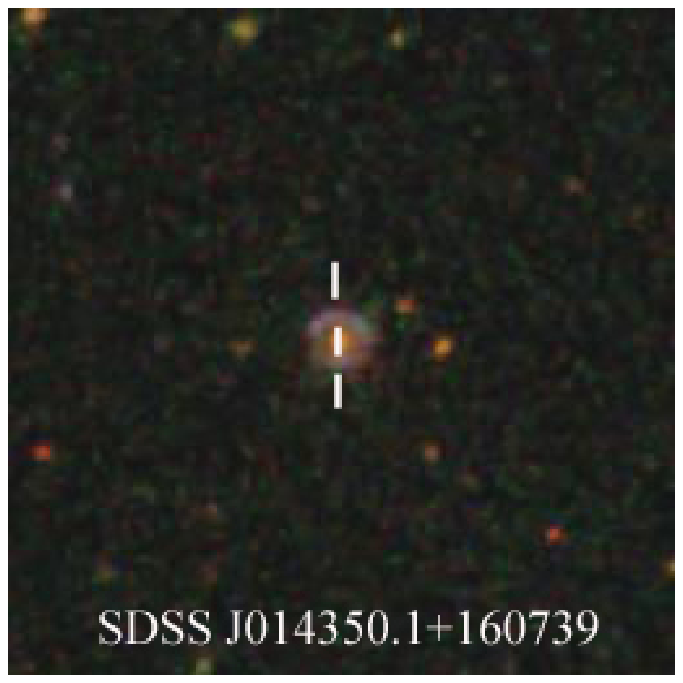}}~%
\resizebox{35mm}{!}{\includegraphics{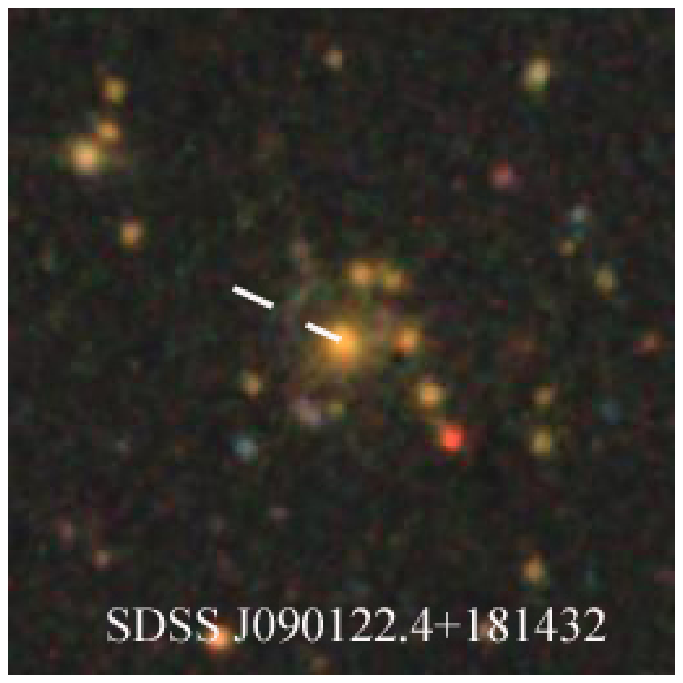}}~%
\resizebox{35mm}{!}{\includegraphics{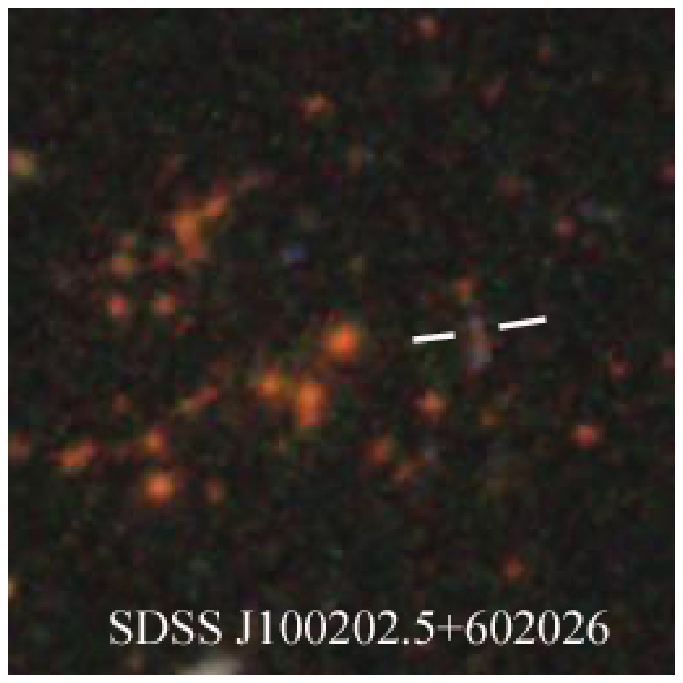}}~%
\resizebox{35mm}{!}{\includegraphics{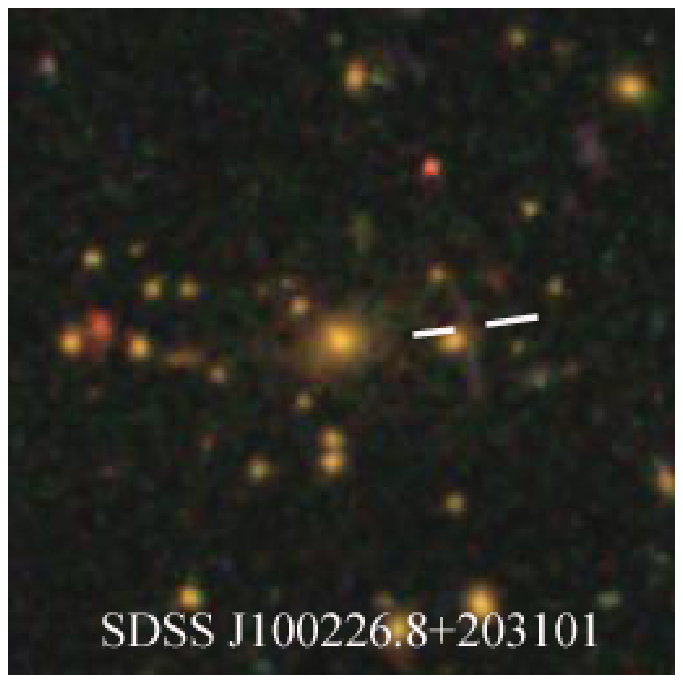}}\\[0.5mm]
\resizebox{35mm}{!}{\includegraphics{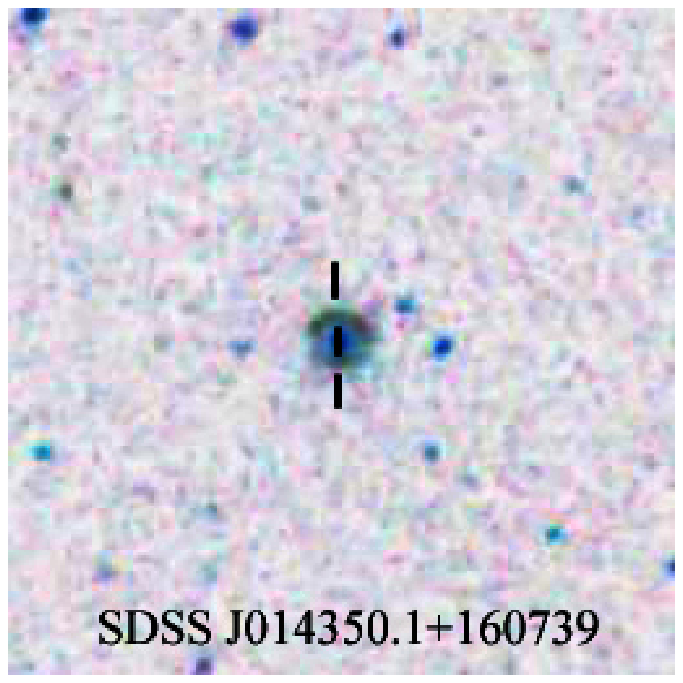}}~%
\resizebox{35mm}{!}{\includegraphics{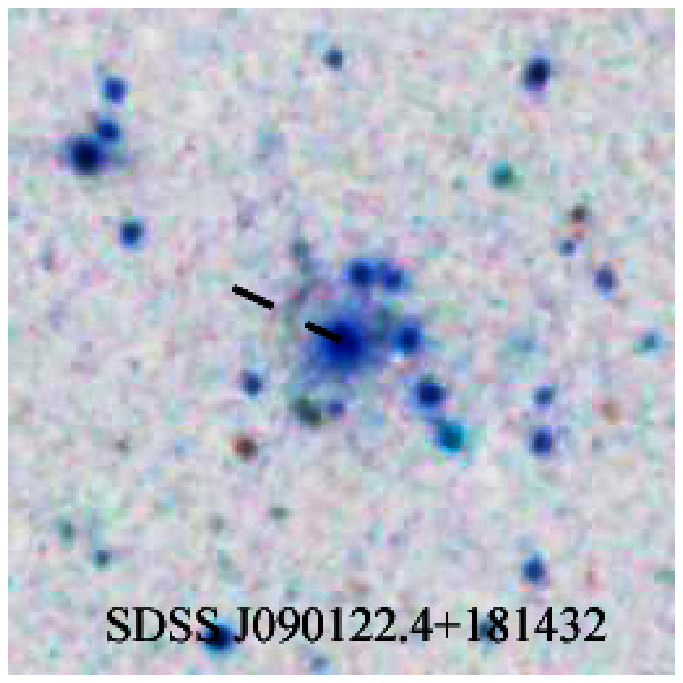}}~%
\resizebox{35mm}{!}{\includegraphics{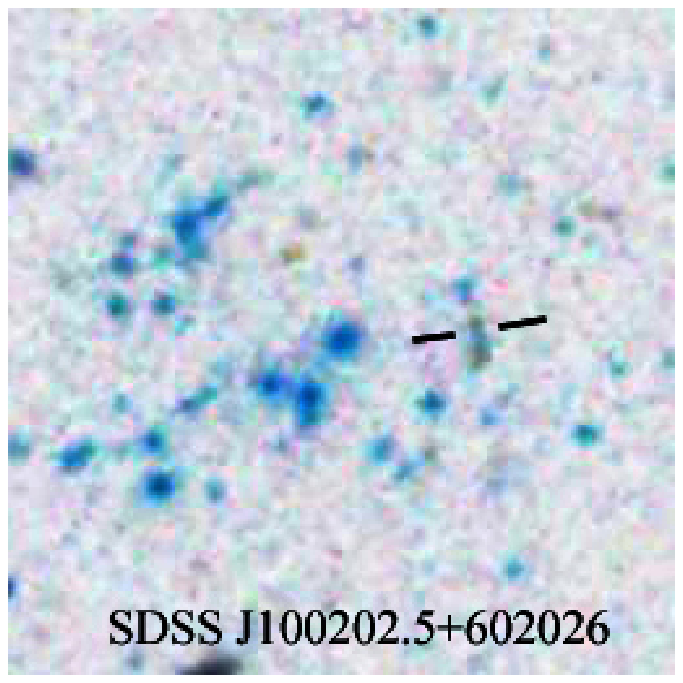}}~%
\resizebox{35mm}{!}{\includegraphics{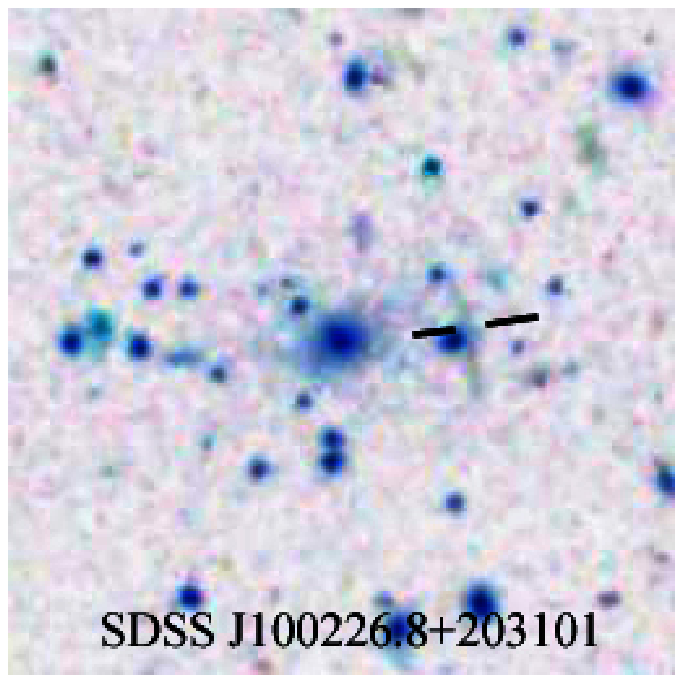}}\\[0.5mm]
\resizebox{35mm}{!}{\includegraphics{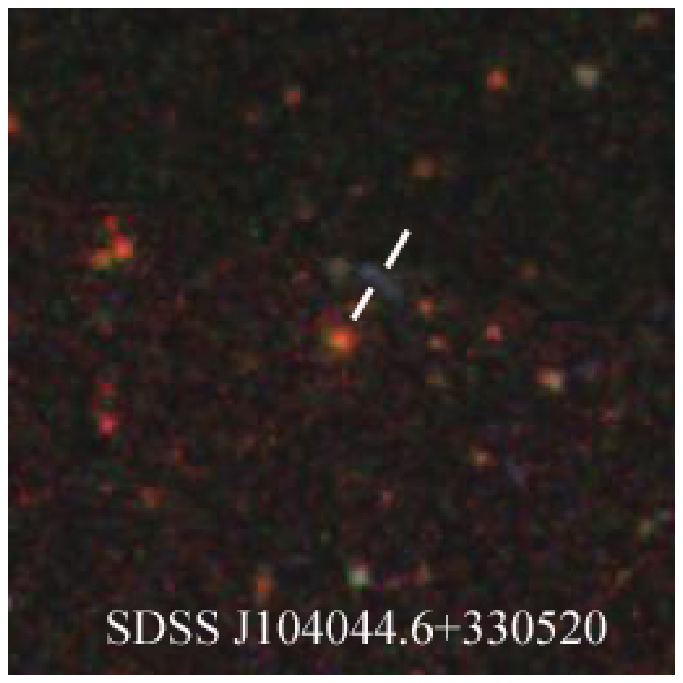}}~%
\resizebox{35mm}{!}{\includegraphics{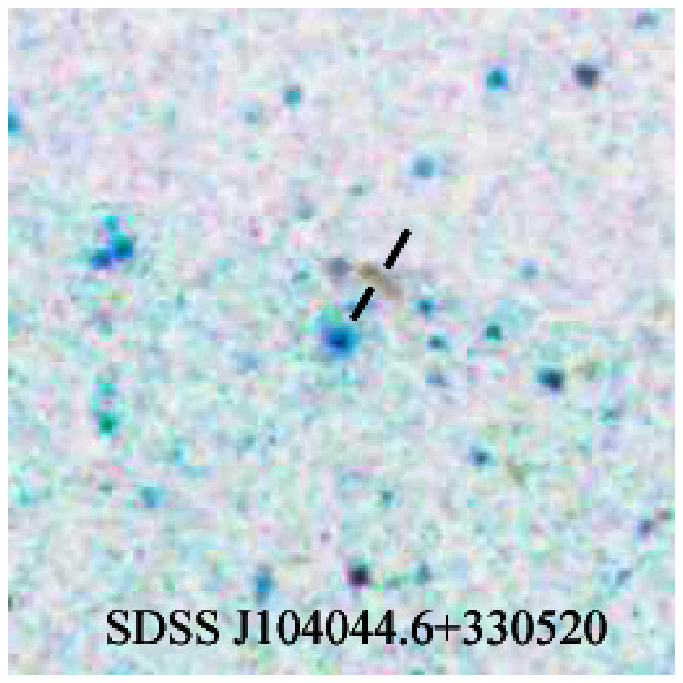}}~%
\resizebox{35mm}{!}{\includegraphics{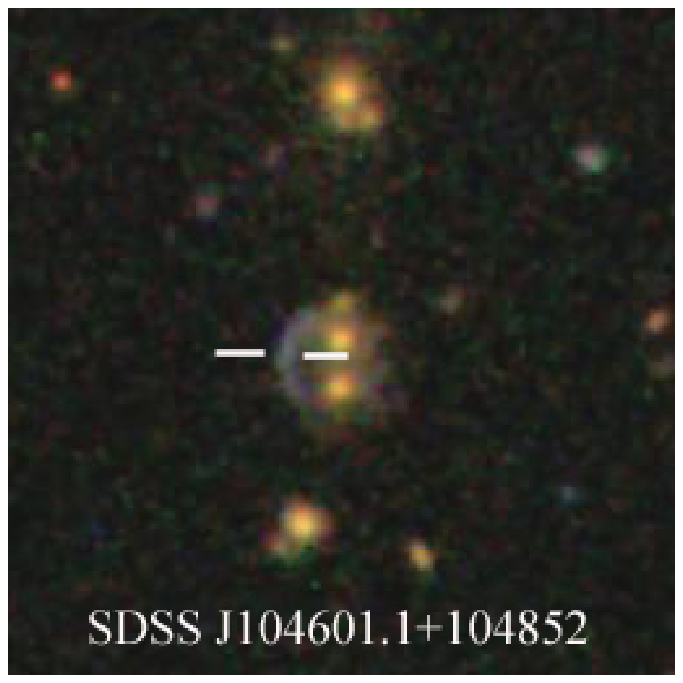}}~%
\resizebox{35mm}{!}{\includegraphics{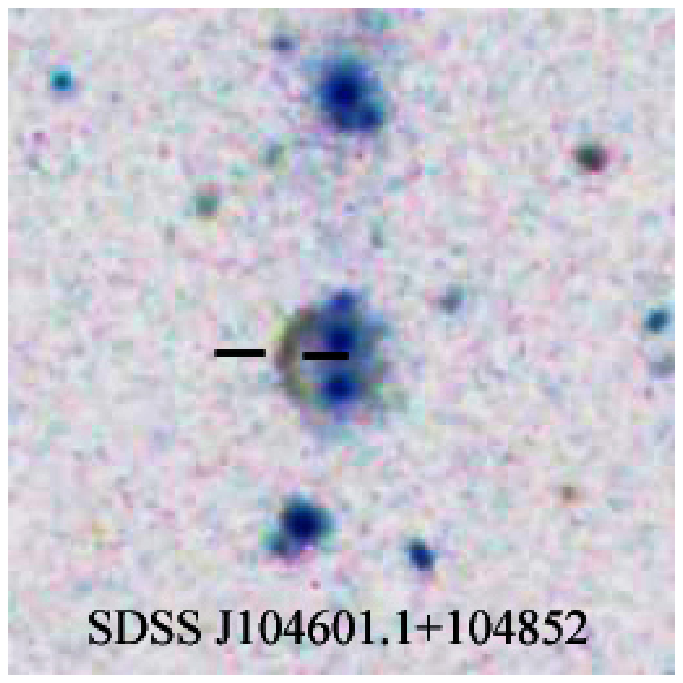}}
\caption{\baselineskip 3.6mm
Same as Fig.~\ref{lens_sure}, but for 22 clusters 
which are {\it probable} lensing systems. 
\label{lens_prob}}
\end{figure}

\begin{figure}[!hpt]
\resizebox{35mm}{!}{\includegraphics{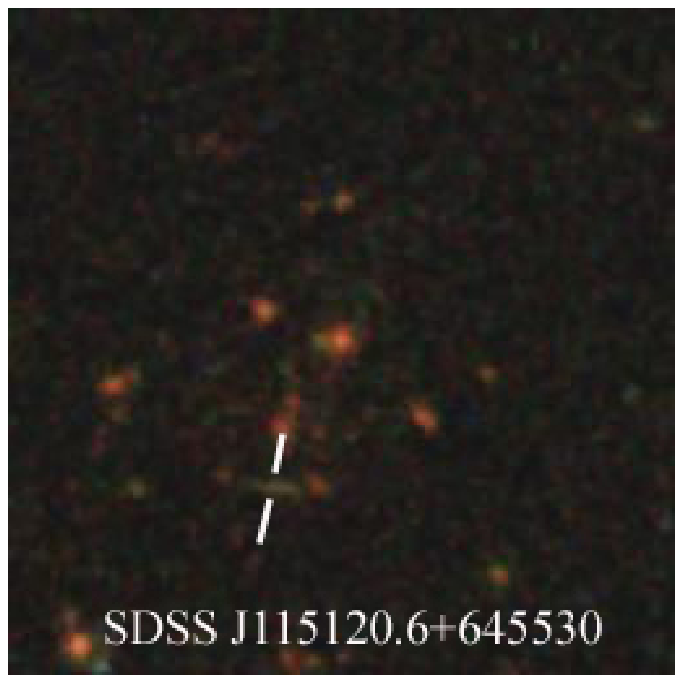}}~%
\resizebox{35mm}{!}{\includegraphics{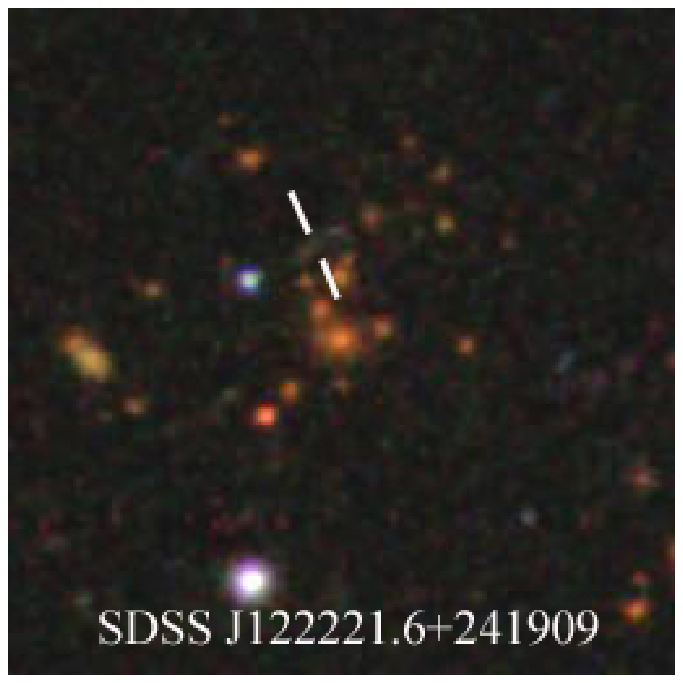}}~%
\resizebox{35mm}{!}{\includegraphics{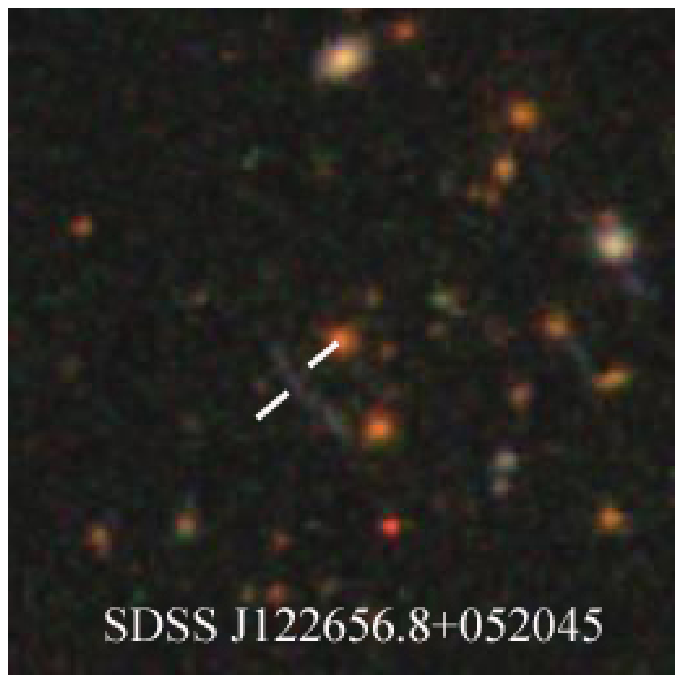}}~%
\resizebox{35mm}{!}{\includegraphics{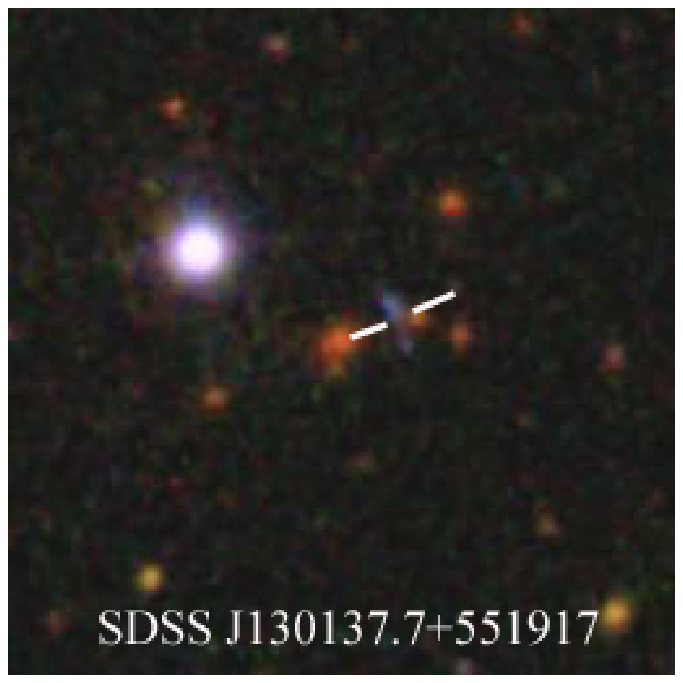}}\\[0.5mm]
\resizebox{35mm}{!}{\includegraphics{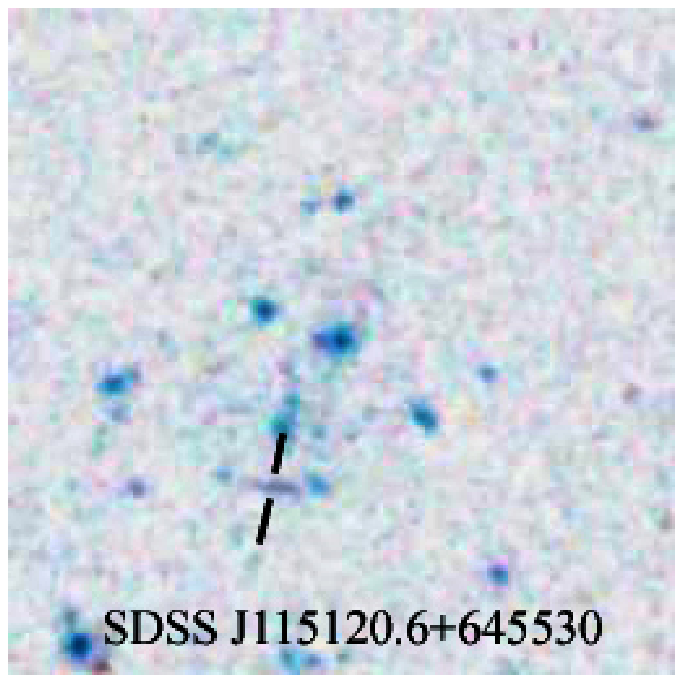}}~%
\resizebox{35mm}{!}{\includegraphics{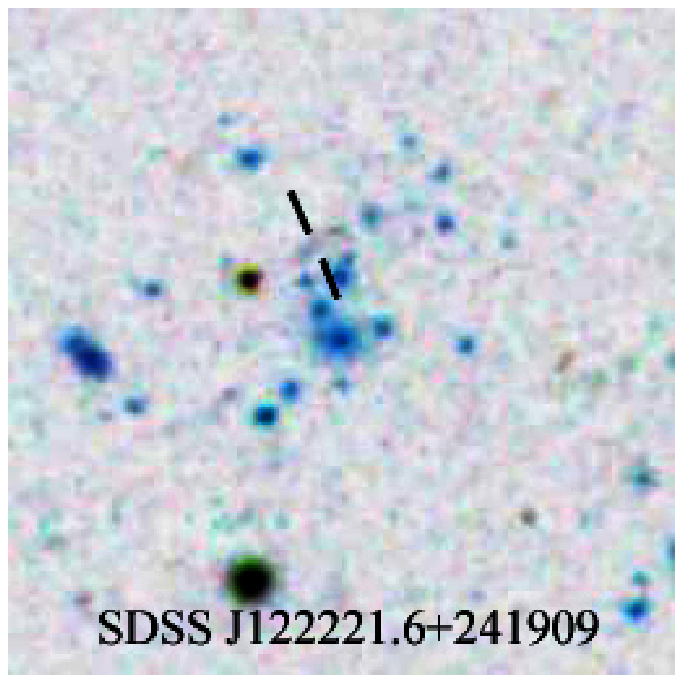}}~%
\resizebox{35mm}{!}{\includegraphics{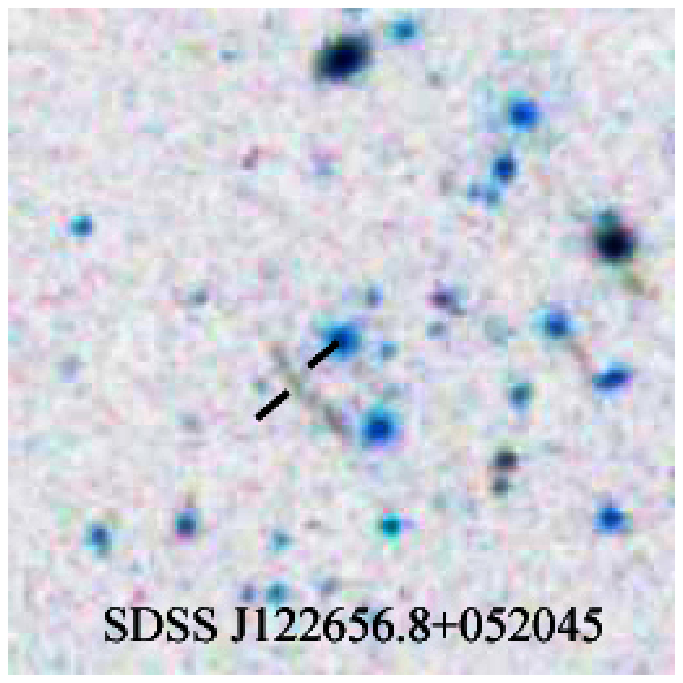}}~%
\resizebox{35mm}{!}{\includegraphics{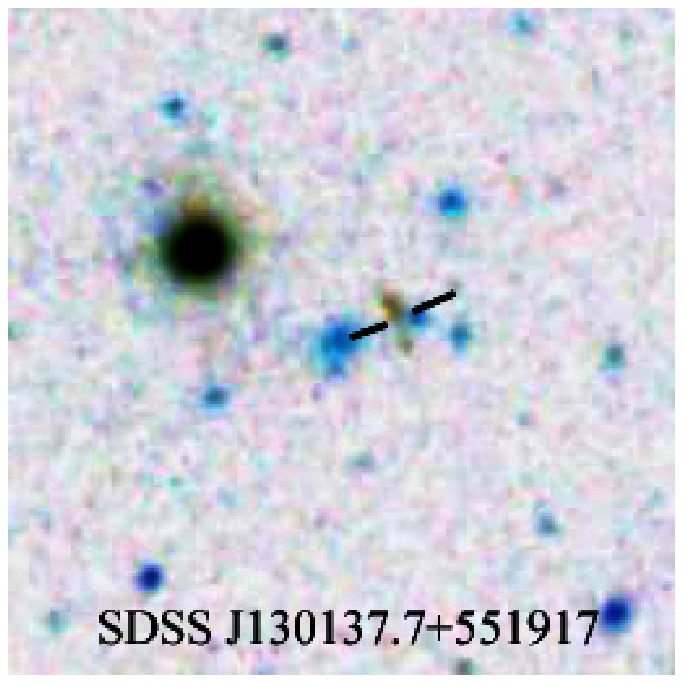}}\\[0.5mm]
\resizebox{35mm}{!}{\includegraphics{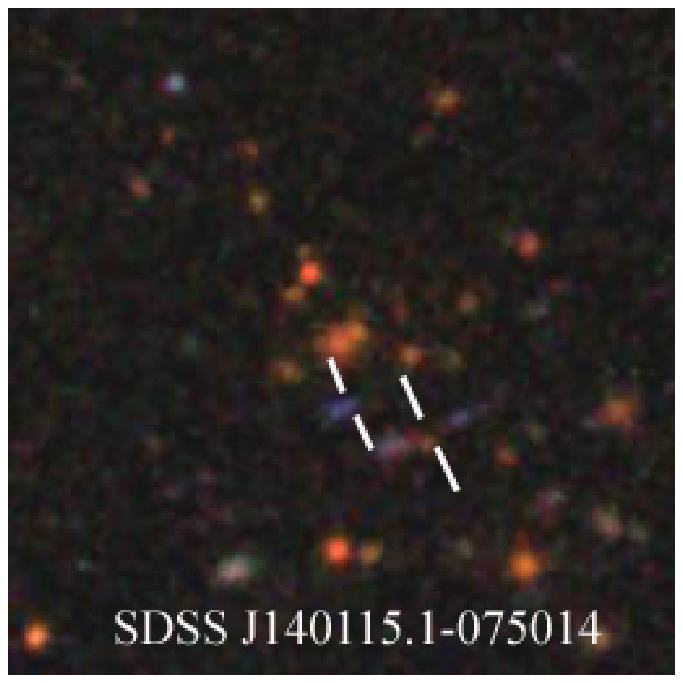}}~%
\resizebox{35mm}{!}{\includegraphics{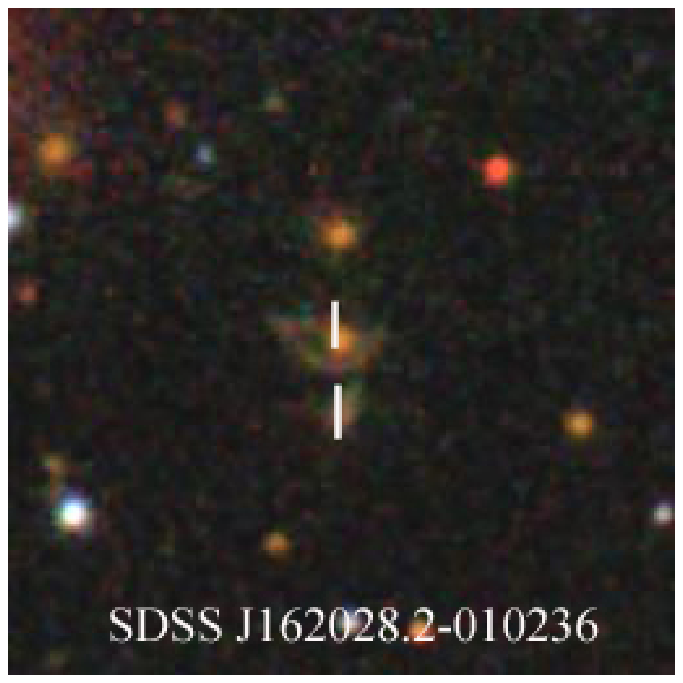}}~%
\resizebox{35mm}{!}{\includegraphics{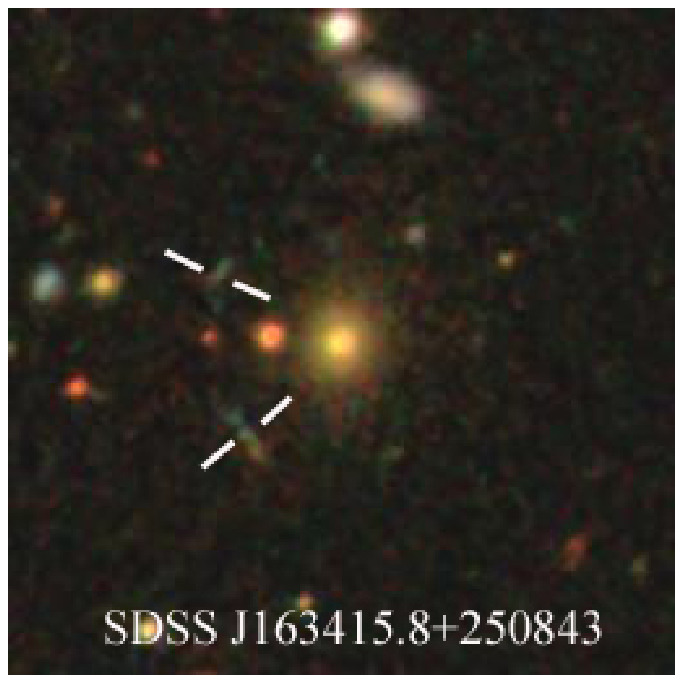}}~%
\resizebox{35mm}{!}{\includegraphics{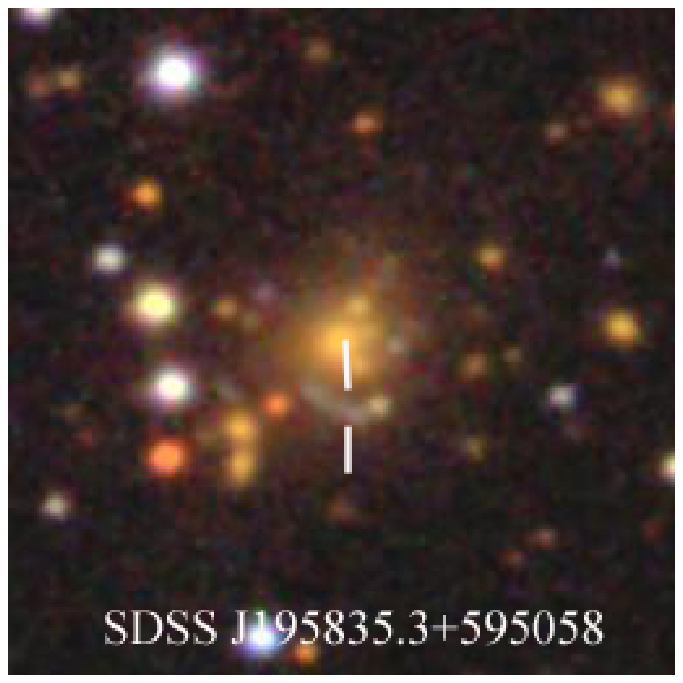}}\\[0.5mm]
\resizebox{35mm}{!}{\includegraphics{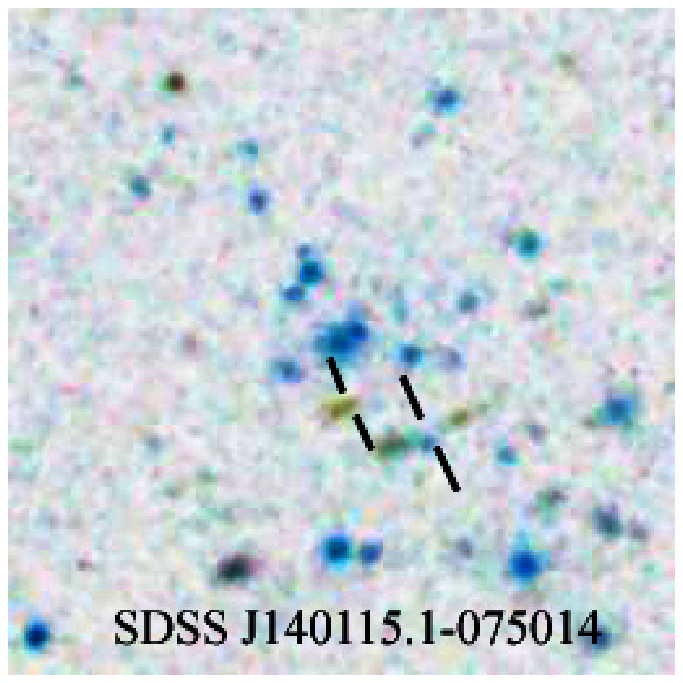}}~%
\resizebox{35mm}{!}{\includegraphics{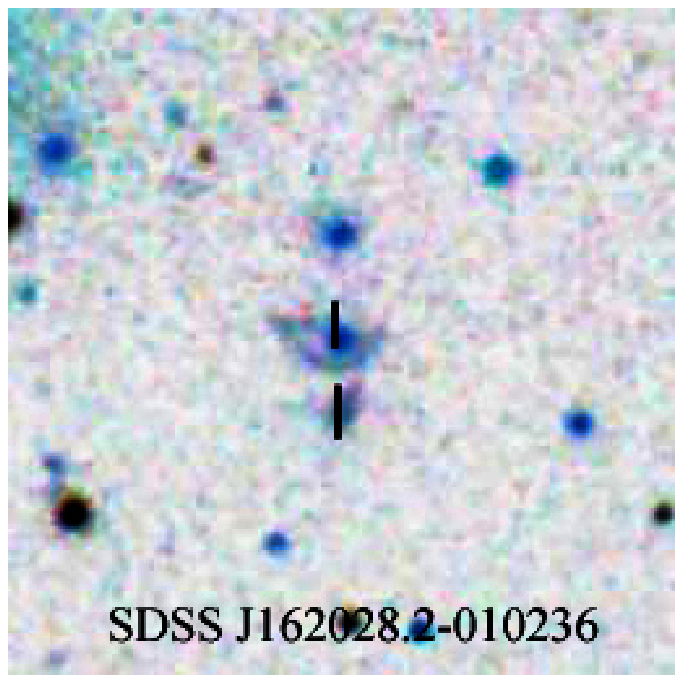}}~%
\resizebox{35mm}{!}{\includegraphics{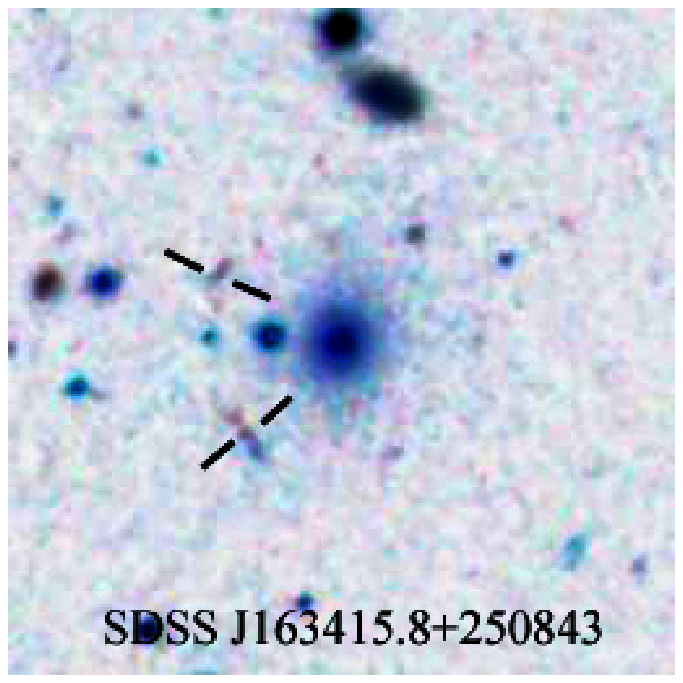}}~%
\resizebox{35mm}{!}{\includegraphics{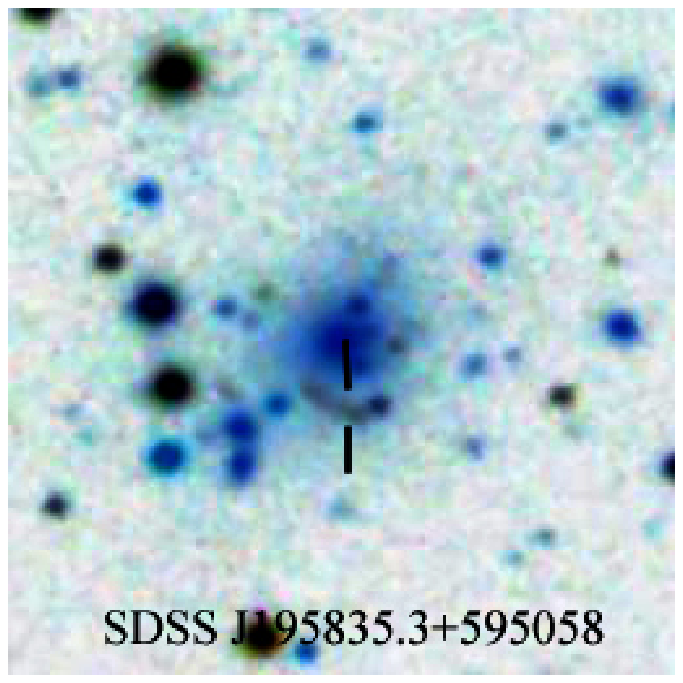}}\\[0.5mm]
\resizebox{35mm}{!}{\includegraphics{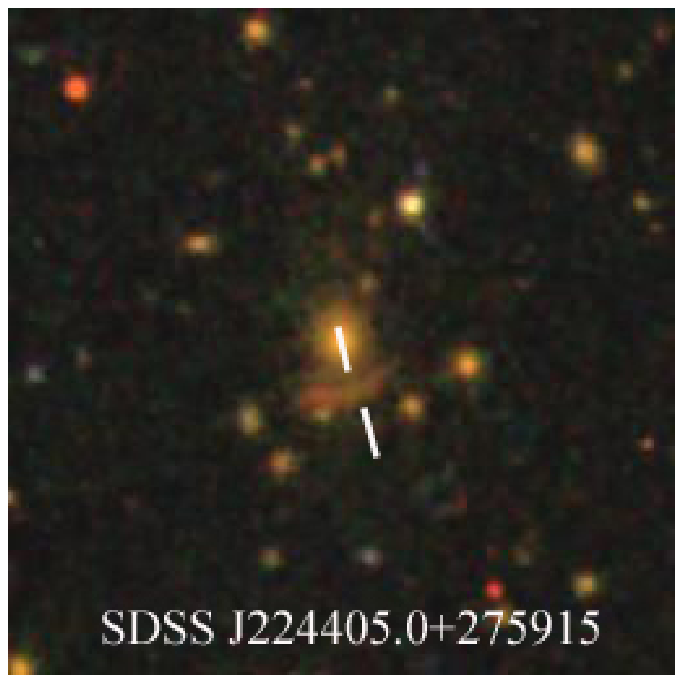}}~%
\resizebox{35mm}{!}{\includegraphics{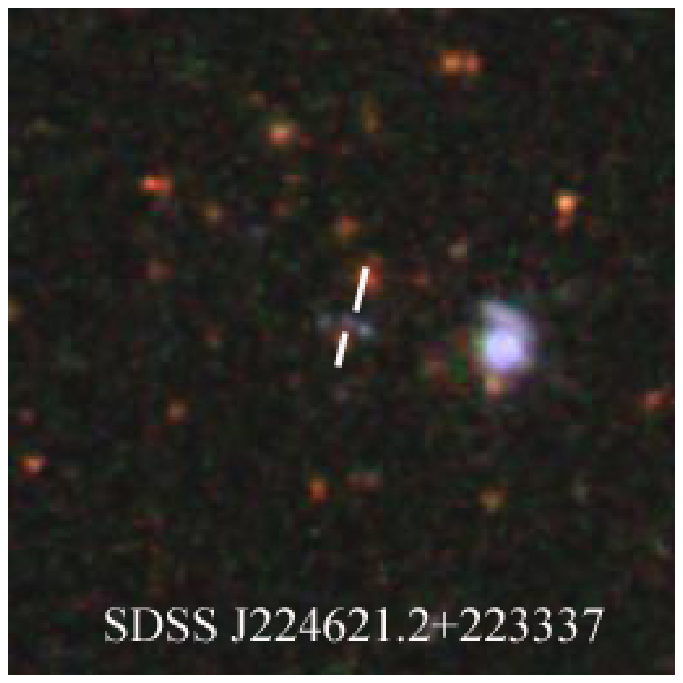}}~%
\resizebox{35mm}{!}{\includegraphics{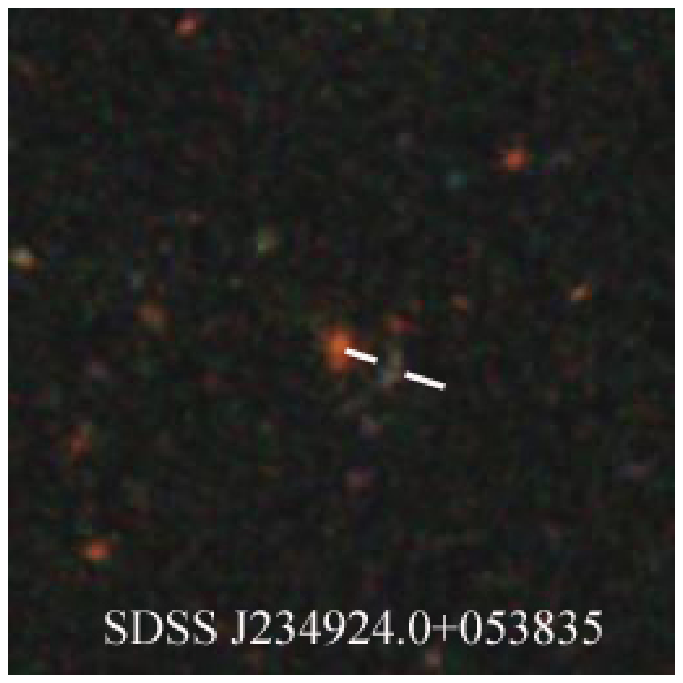}}~%
\resizebox{35mm}{!}{\includegraphics{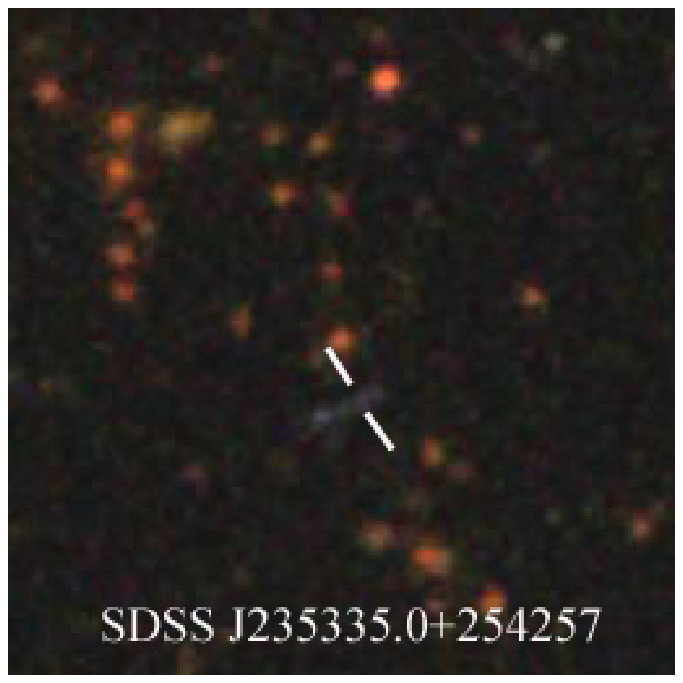}}\\[0.5mm]
\resizebox{35mm}{!}{\includegraphics{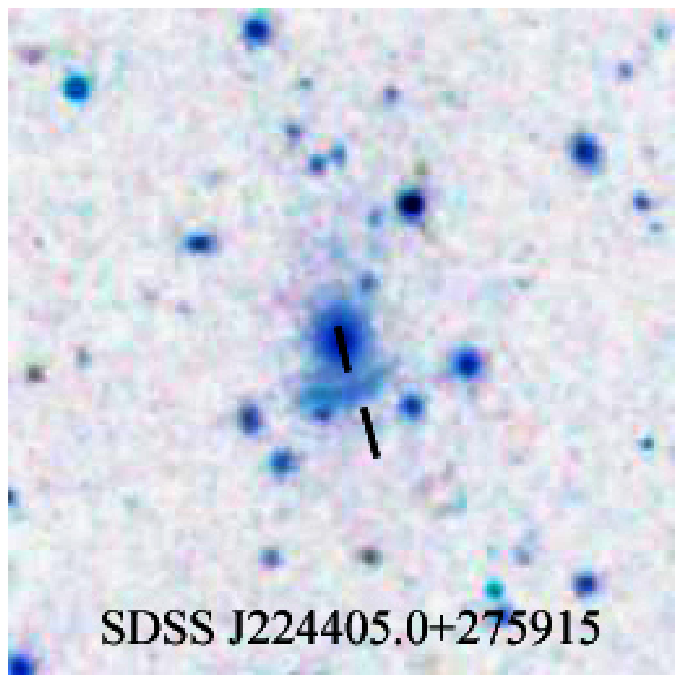}}~%
\resizebox{35mm}{!}{\includegraphics{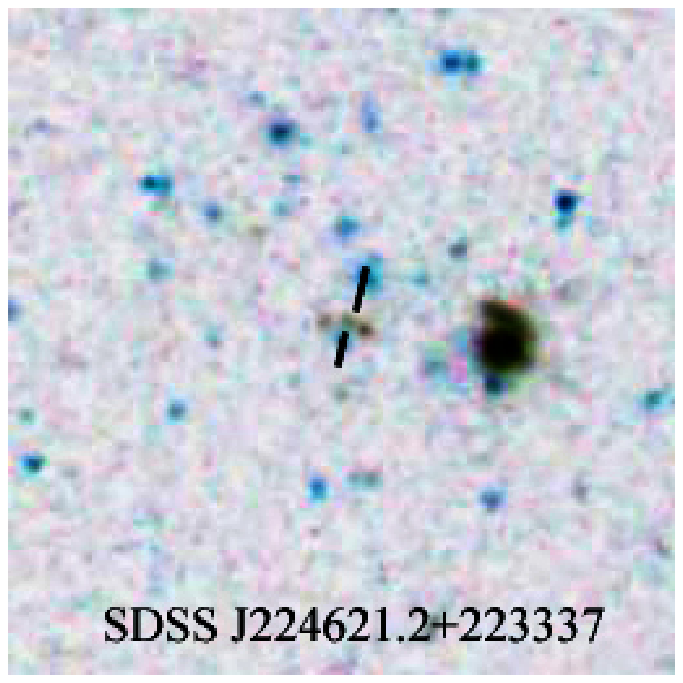}}~%
\resizebox{35mm}{!}{\includegraphics{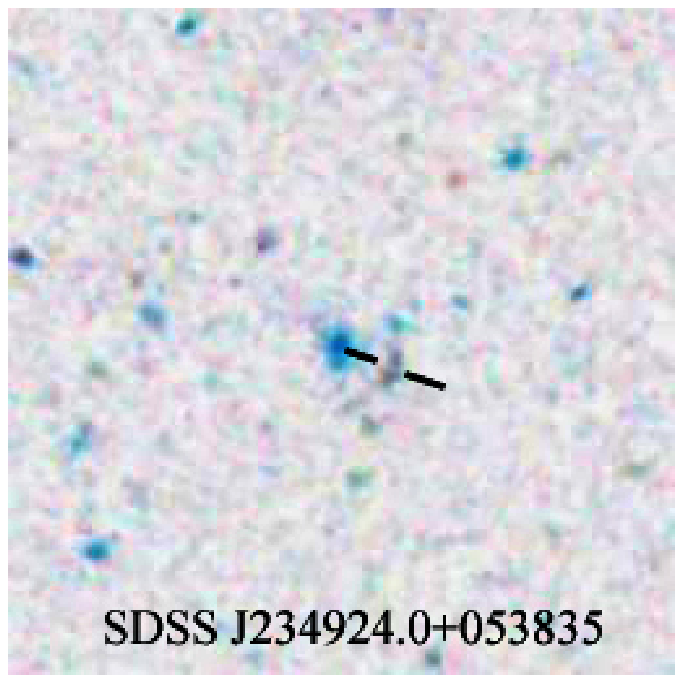}}~%
\resizebox{35mm}{!}{\includegraphics{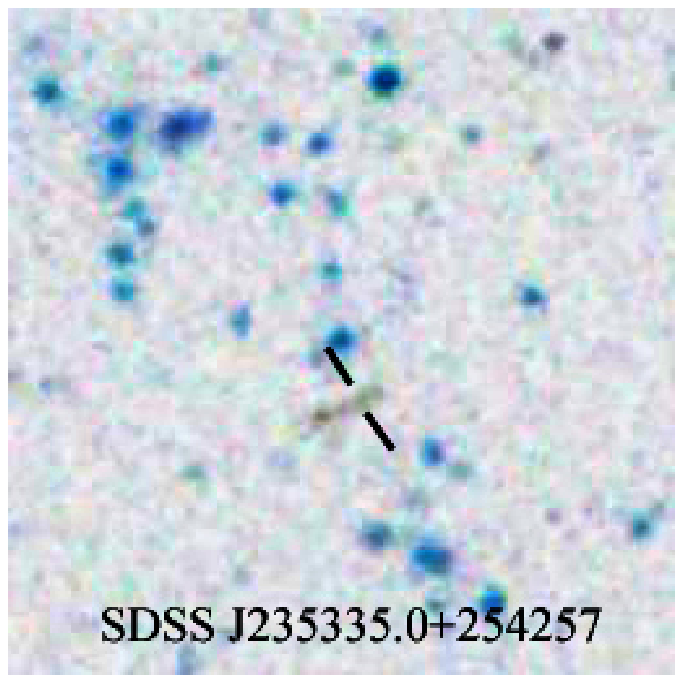}}~%
\setcounter{figure}{1}
\caption{{\it Continued}}
\end{figure}
\begin{figure}[!hpt]
\resizebox{35mm}{!}{\includegraphics{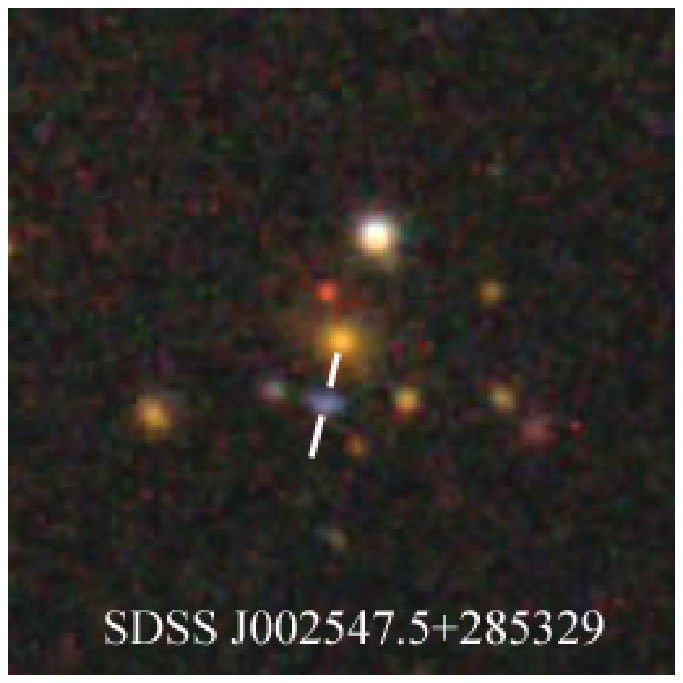}}~%
\resizebox{35mm}{!}{\includegraphics{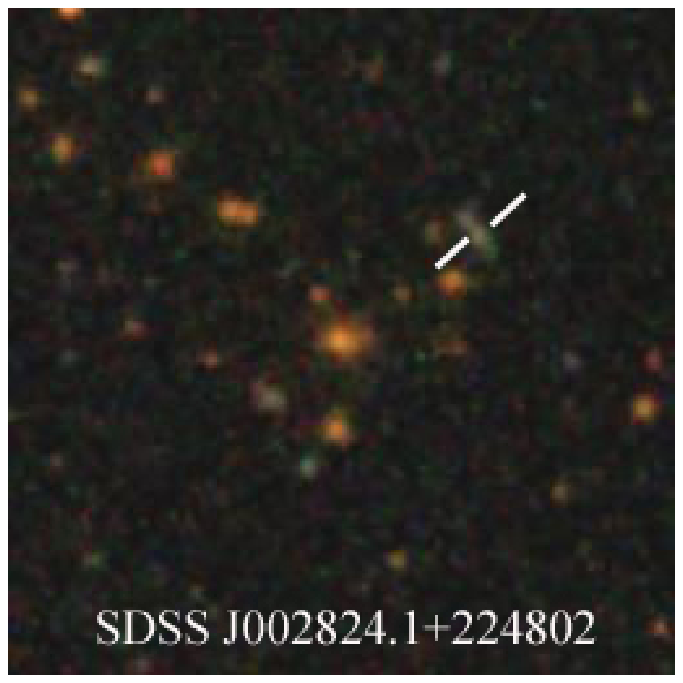}}~%
\resizebox{35mm}{!}{\includegraphics{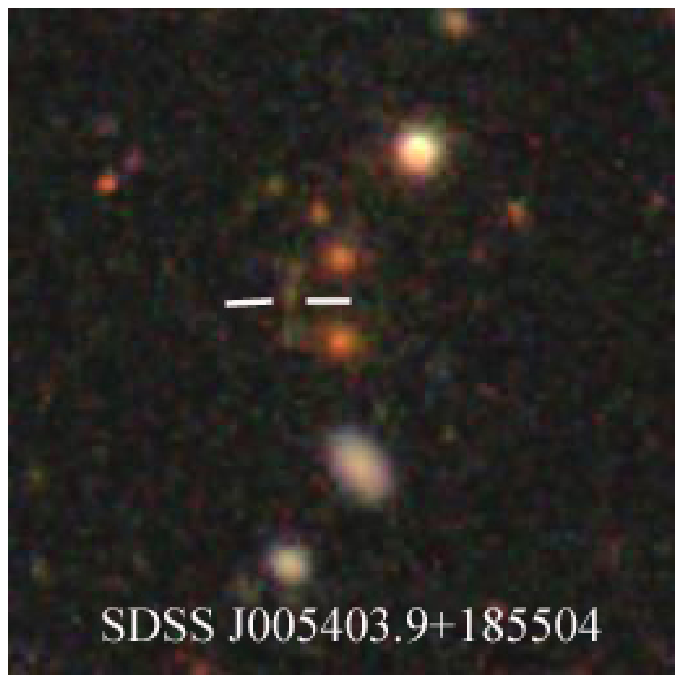}}~%
\resizebox{35mm}{!}{\includegraphics{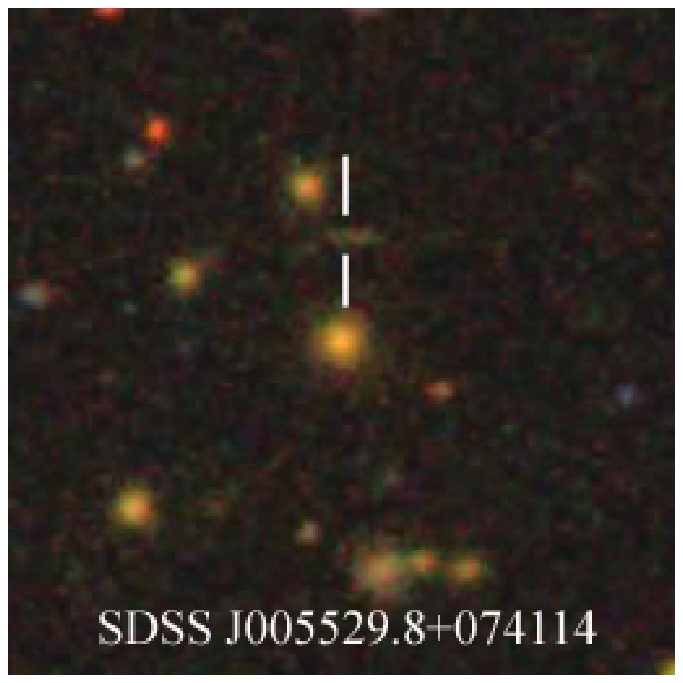}}\\[0.5mm]
\resizebox{35mm}{!}{\includegraphics{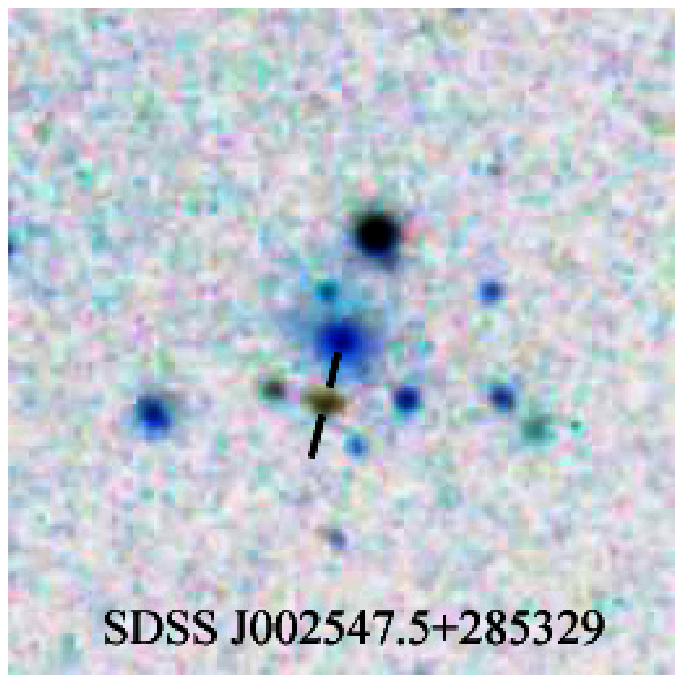}}~%
\resizebox{35mm}{!}{\includegraphics{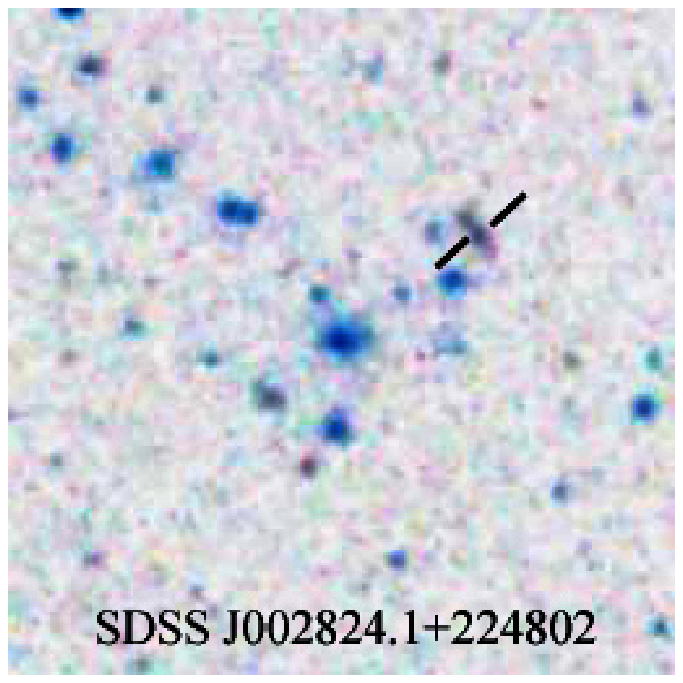}}~%
\resizebox{35mm}{!}{\includegraphics{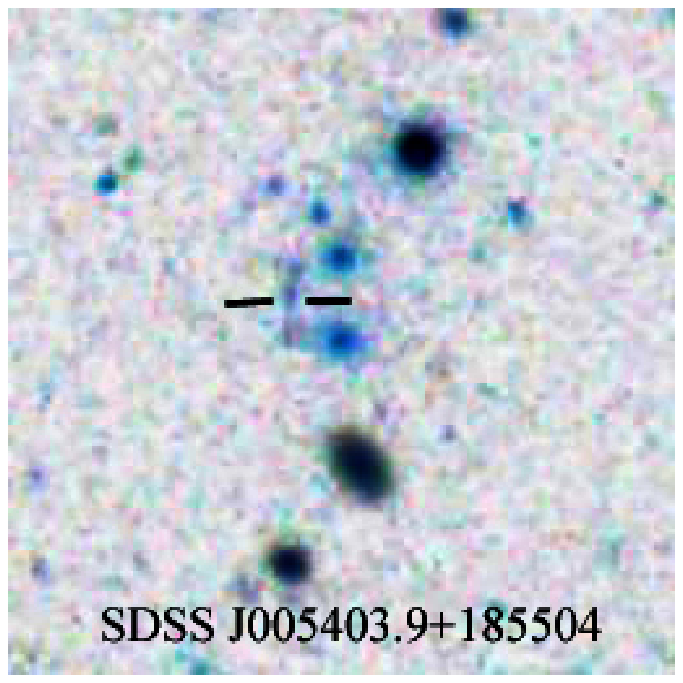}}~%
\resizebox{35mm}{!}{\includegraphics{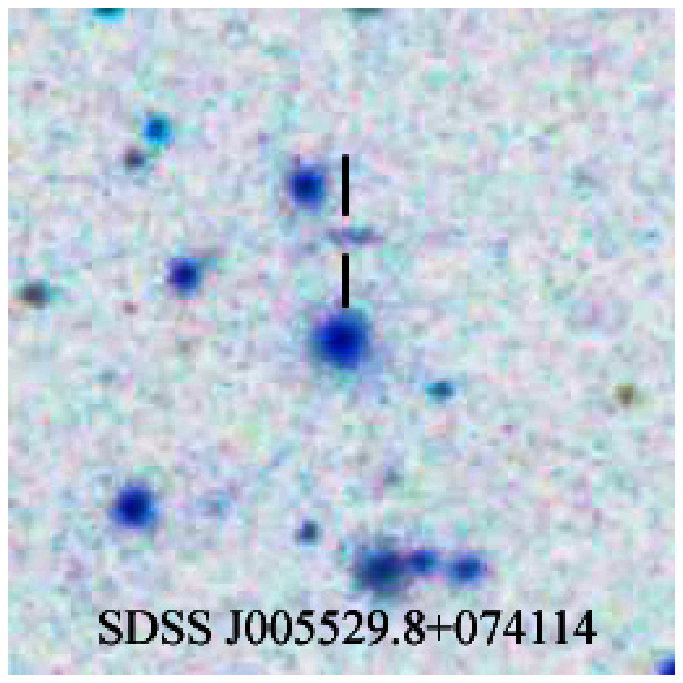}}\\[0.5mm]
\resizebox{35mm}{!}{\includegraphics{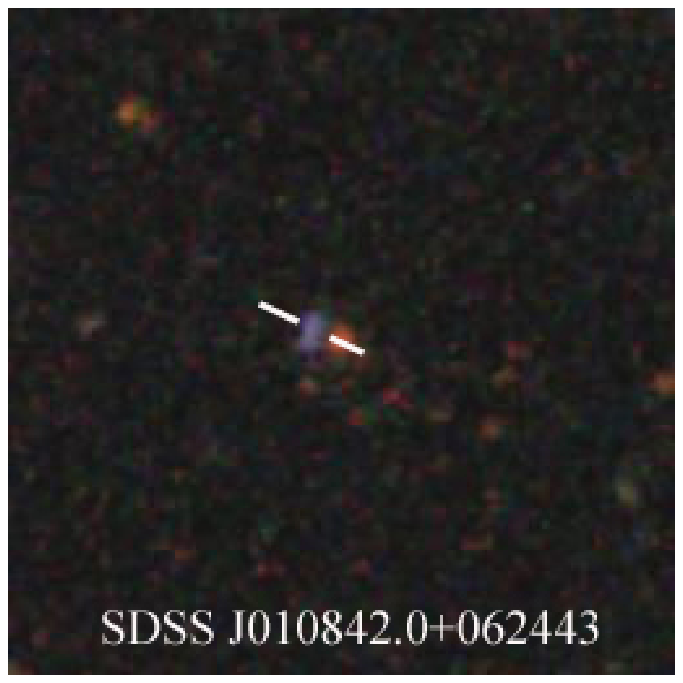}}~%
\resizebox{35mm}{!}{\includegraphics{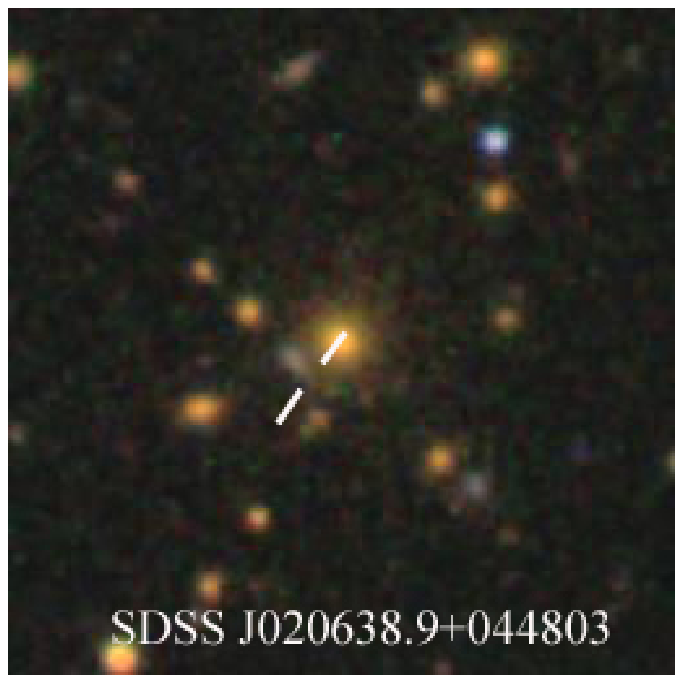}}~%
\resizebox{35mm}{!}{\includegraphics{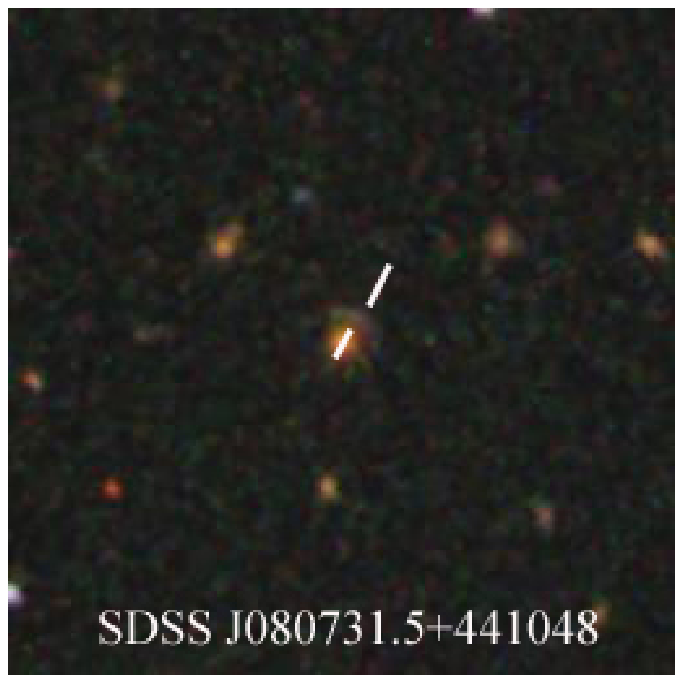}}~%
\resizebox{35mm}{!}{\includegraphics{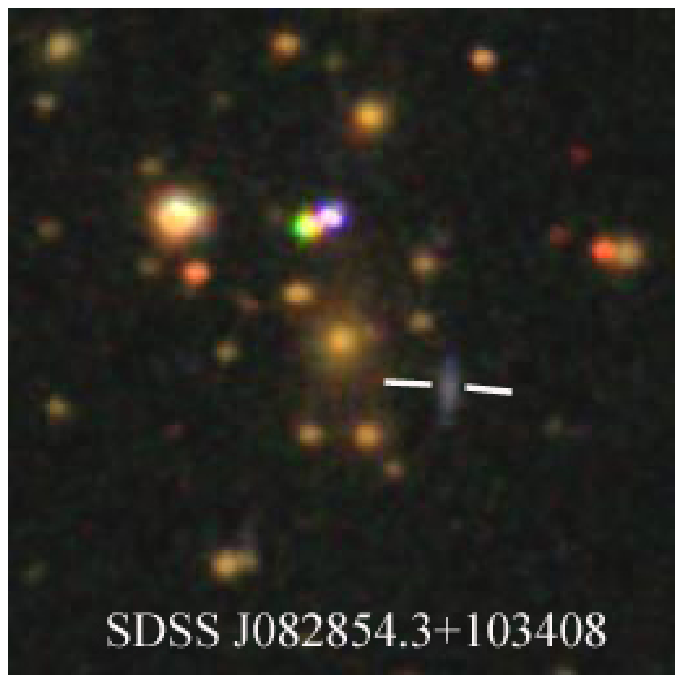}}\\[0.5mm]
\resizebox{35mm}{!}{\includegraphics{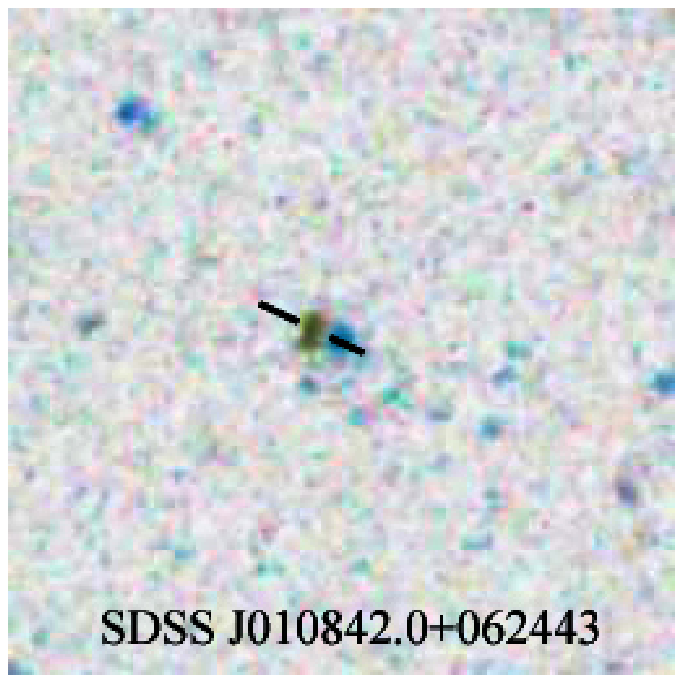}}~%
\resizebox{35mm}{!}{\includegraphics{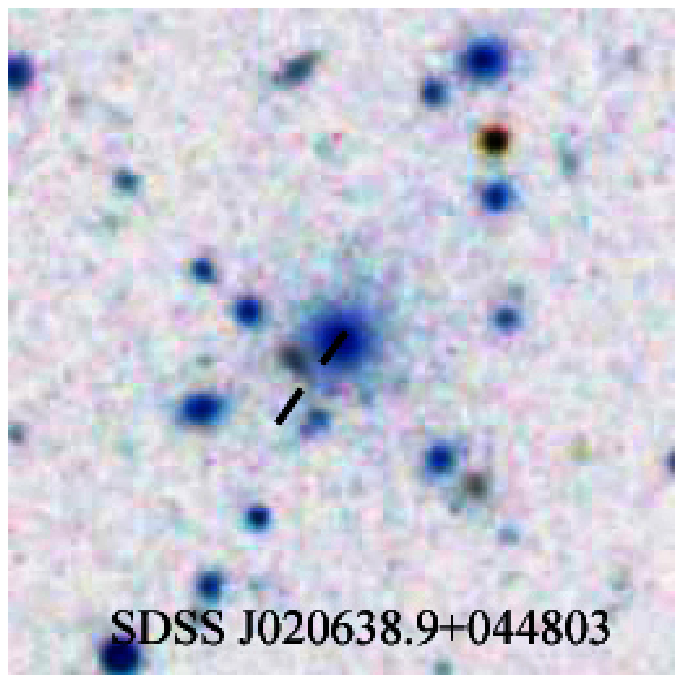}}~%
\resizebox{35mm}{!}{\includegraphics{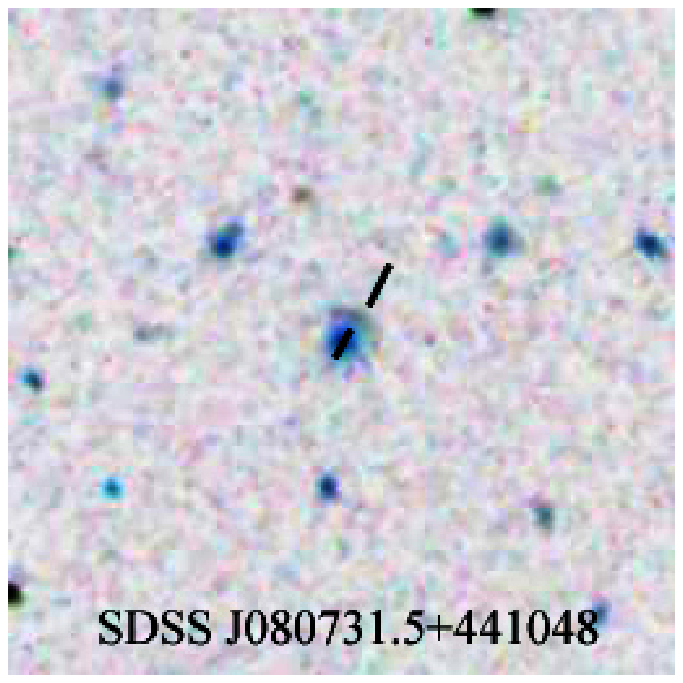}}~%
\resizebox{35mm}{!}{\includegraphics{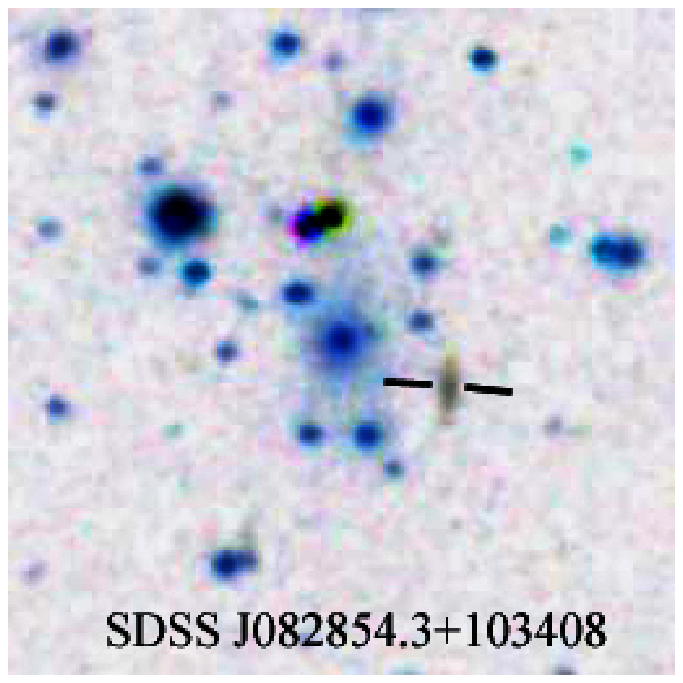}}\\[0.5mm]
\resizebox{35mm}{!}{\includegraphics{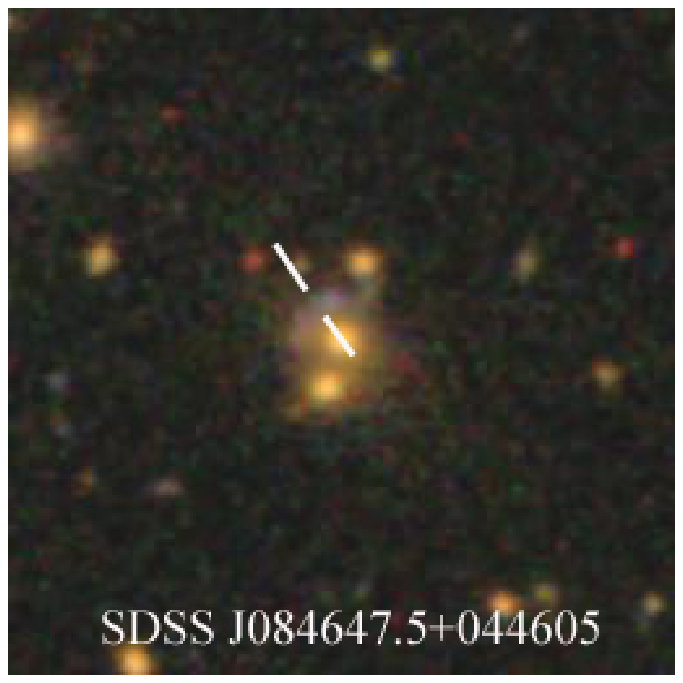}}~%
\resizebox{35mm}{!}{\includegraphics{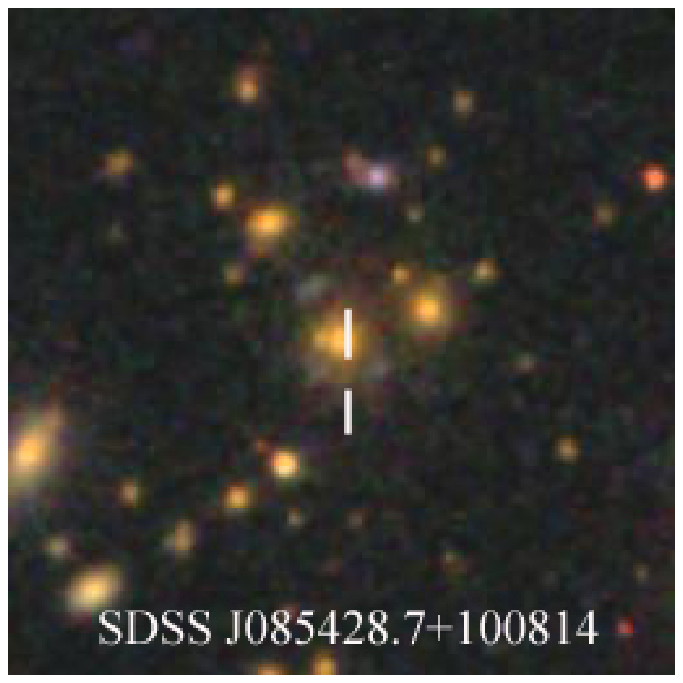}}~%
\resizebox{35mm}{!}{\includegraphics{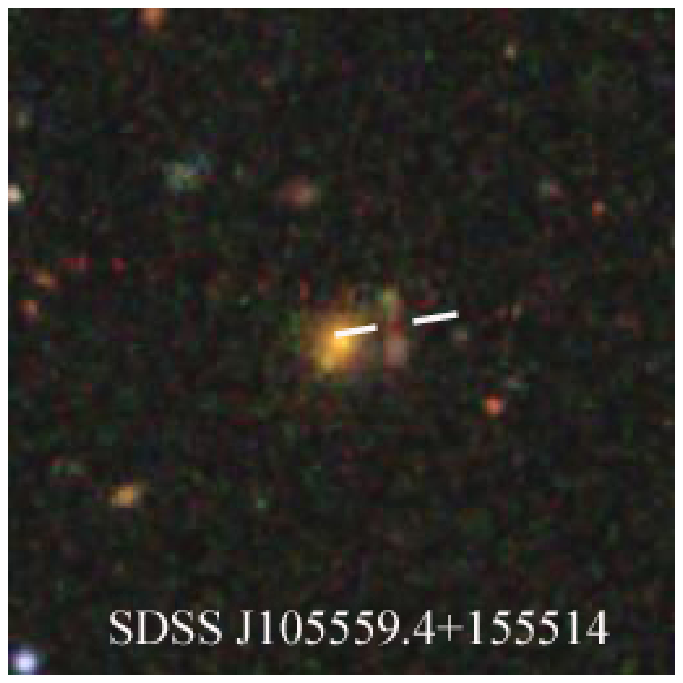}}~%
\resizebox{35mm}{!}{\includegraphics{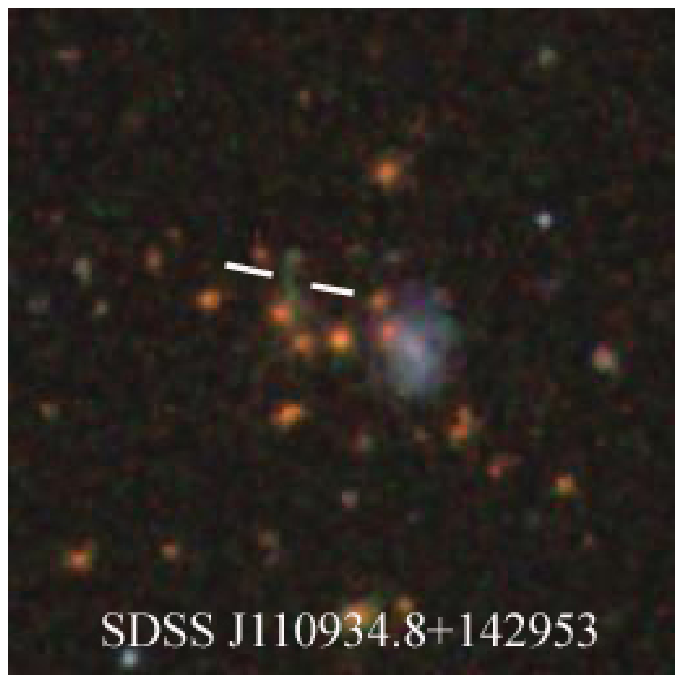}}\\[0.5mm]
\resizebox{35mm}{!}{\includegraphics{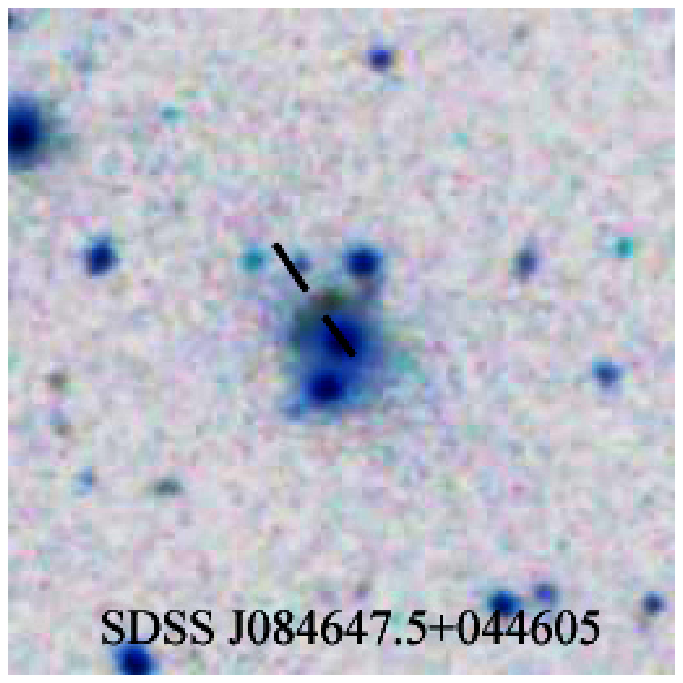}}~%
\resizebox{35mm}{!}{\includegraphics{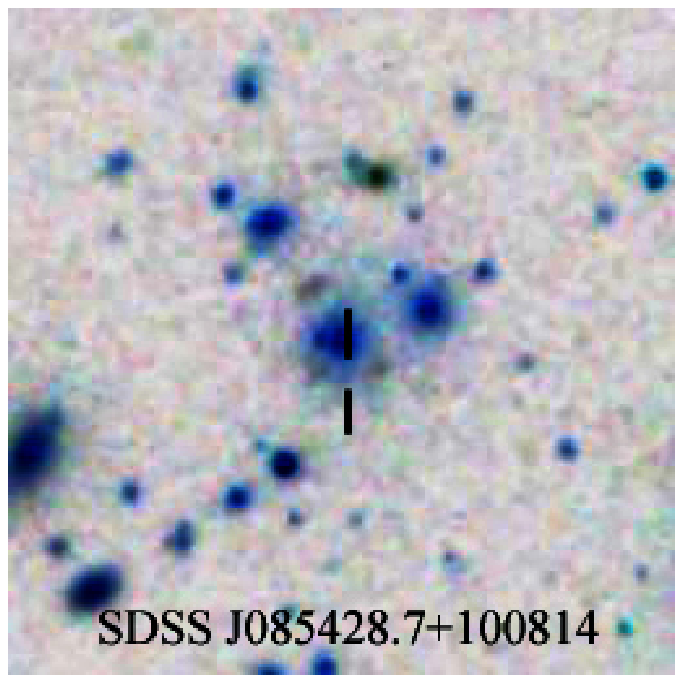}}~%
\resizebox{35mm}{!}{\includegraphics{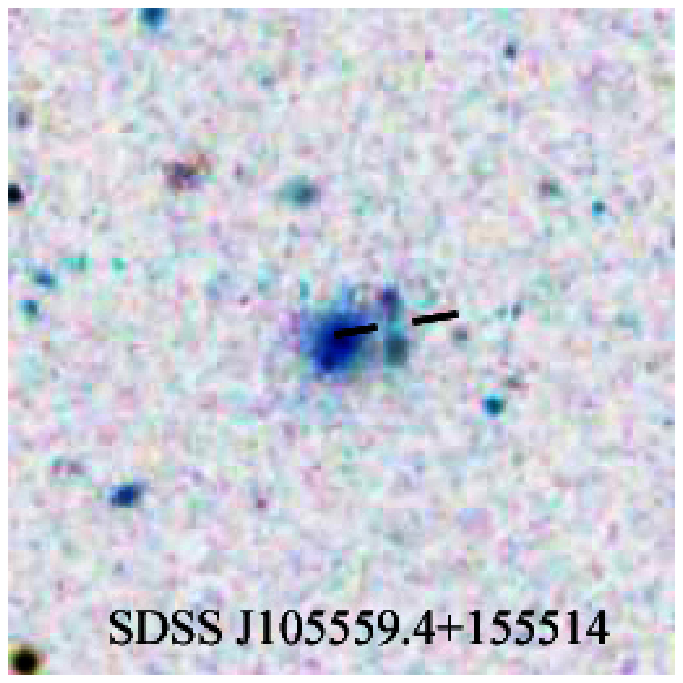}}~%
\resizebox{35mm}{!}{\includegraphics{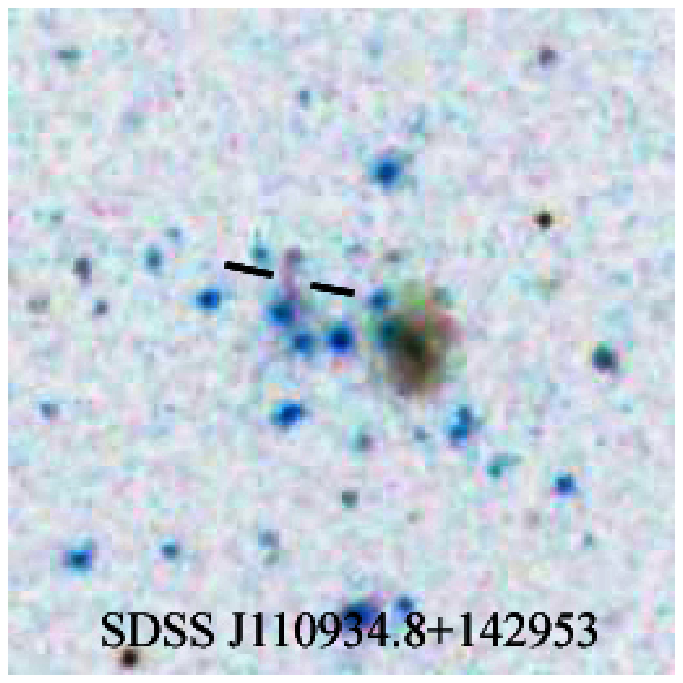}}
\caption{\baselineskip 3.6mm
Same as Fig.~\ref{lens_sure}, but for 31 clusters 
which are {\it possible} lensing systems. 
\label{lens_poss}}
\end{figure}
\begin{figure}[!hpt]
\resizebox{35mm}{!}{\includegraphics{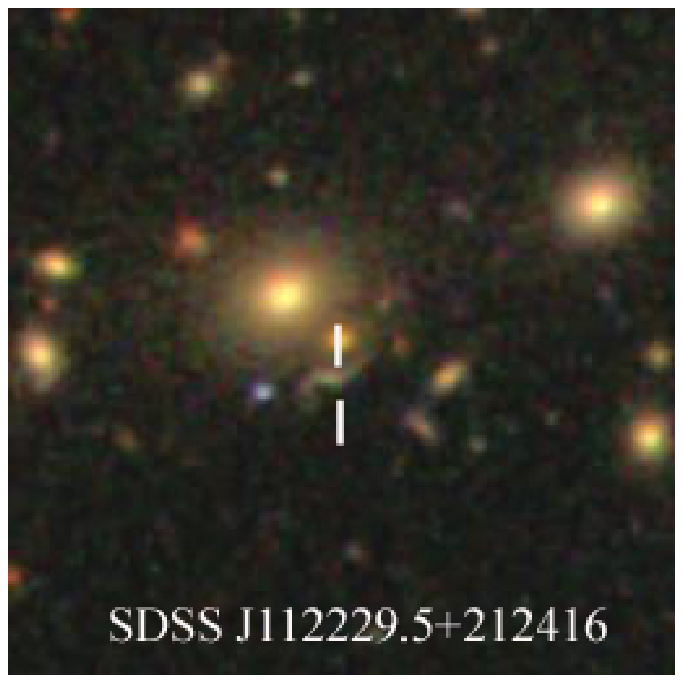}}~%
\resizebox{35mm}{!}{\includegraphics{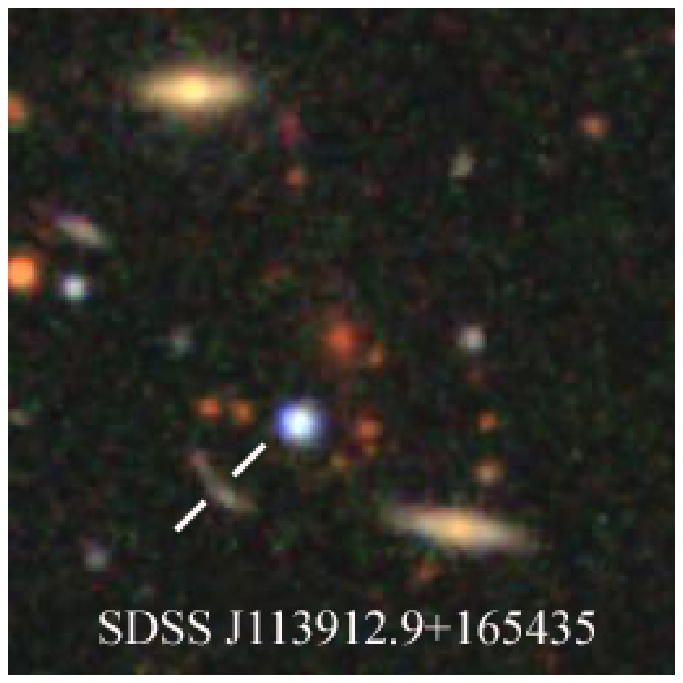}}~%
\resizebox{35mm}{!}{\includegraphics{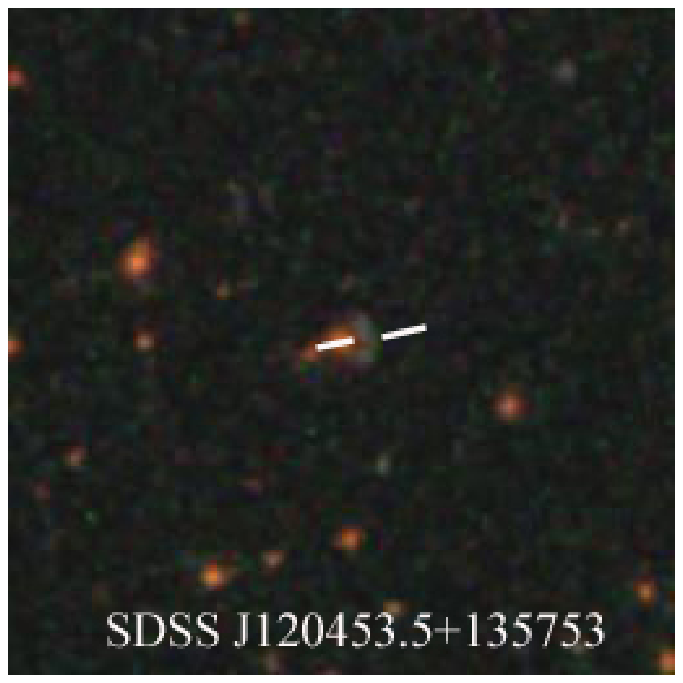}}~%
\resizebox{35mm}{!}{\includegraphics{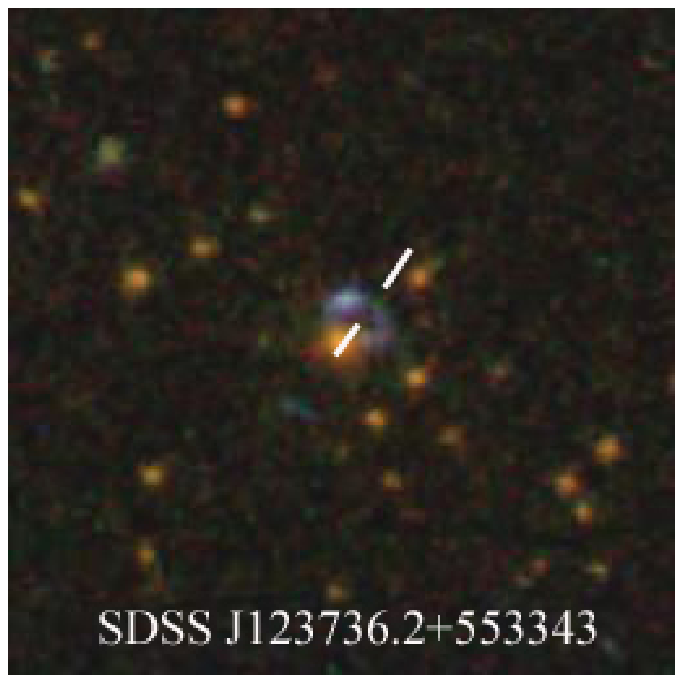}}\\[0.5mm]
\resizebox{35mm}{!}{\includegraphics{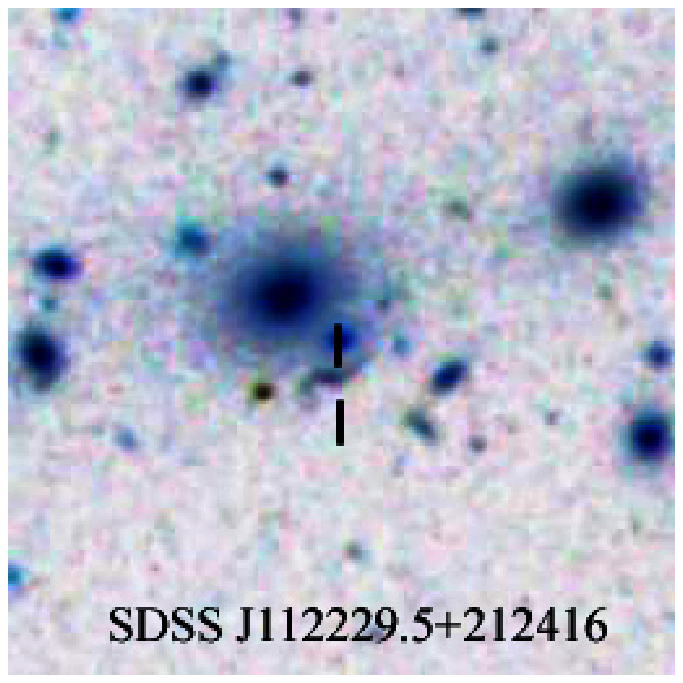}}~%
\resizebox{35mm}{!}{\includegraphics{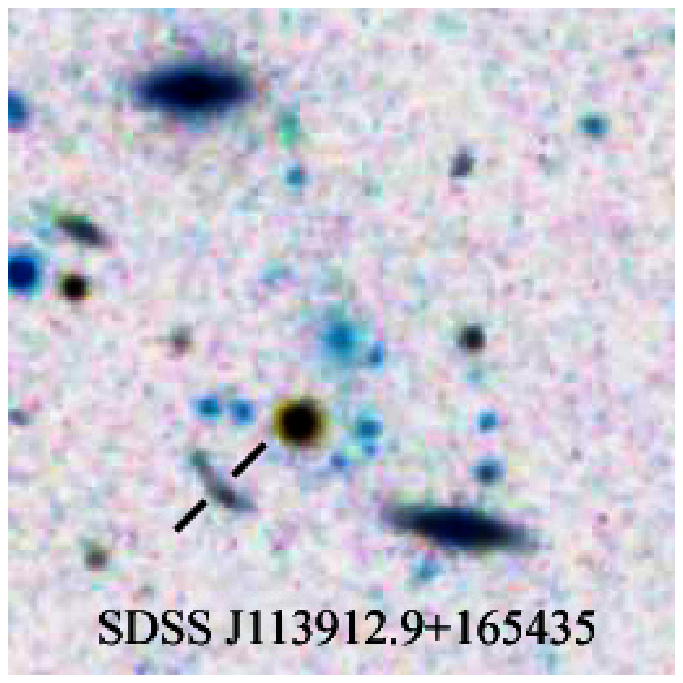}}~%
\resizebox{35mm}{!}{\includegraphics{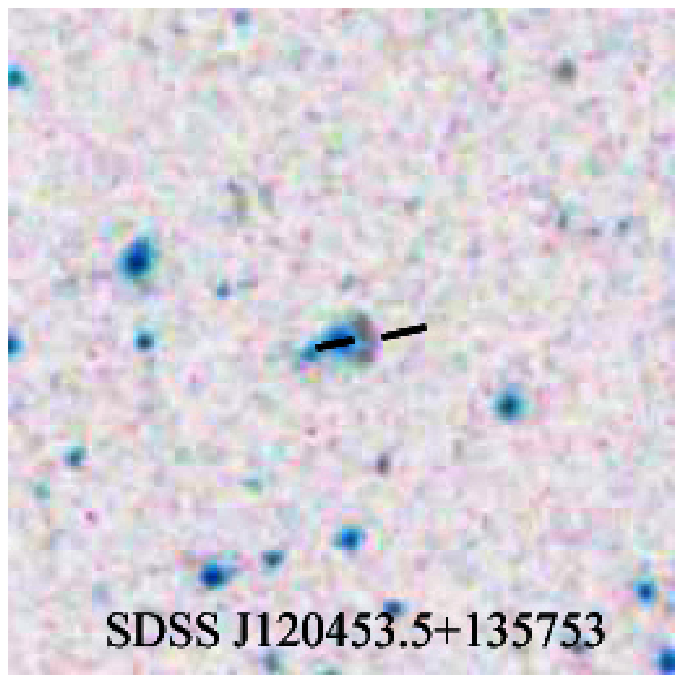}}~%
\resizebox{35mm}{!}{\includegraphics{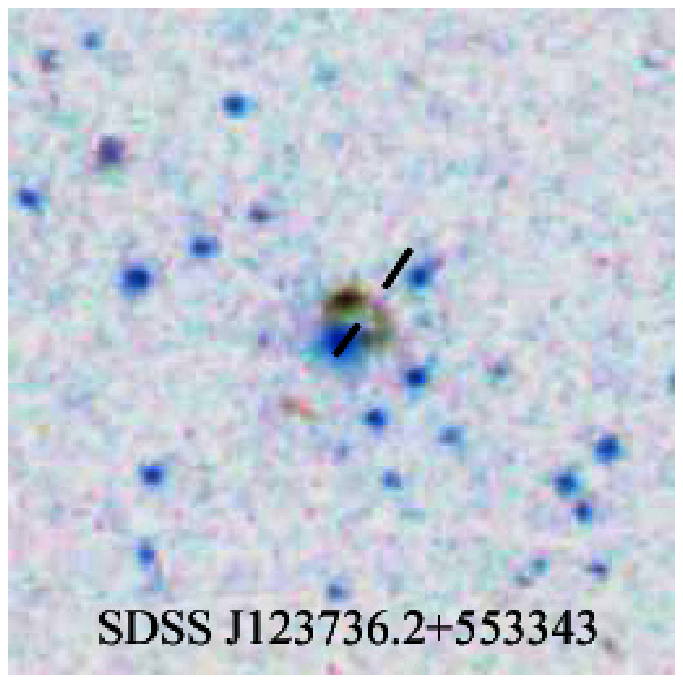}}\\[0.5mm]
\resizebox{35mm}{!}{\includegraphics{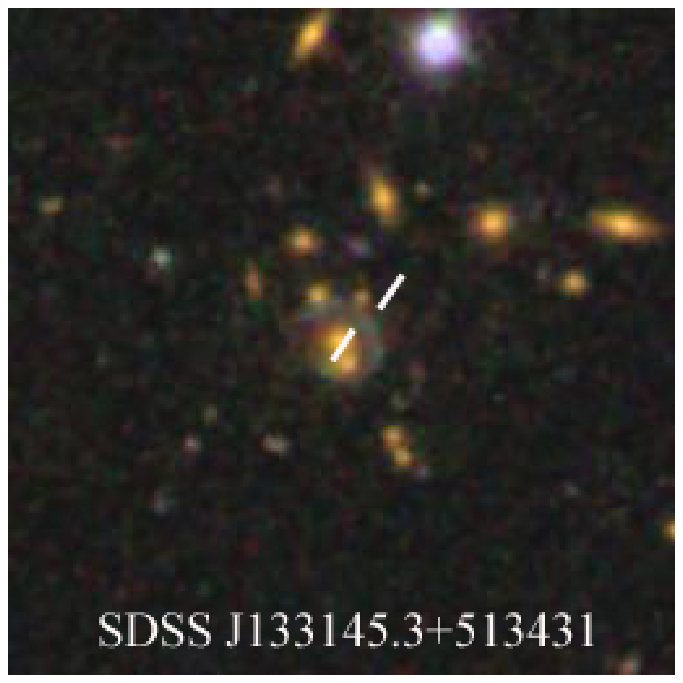}}~%
\resizebox{35mm}{!}{\includegraphics{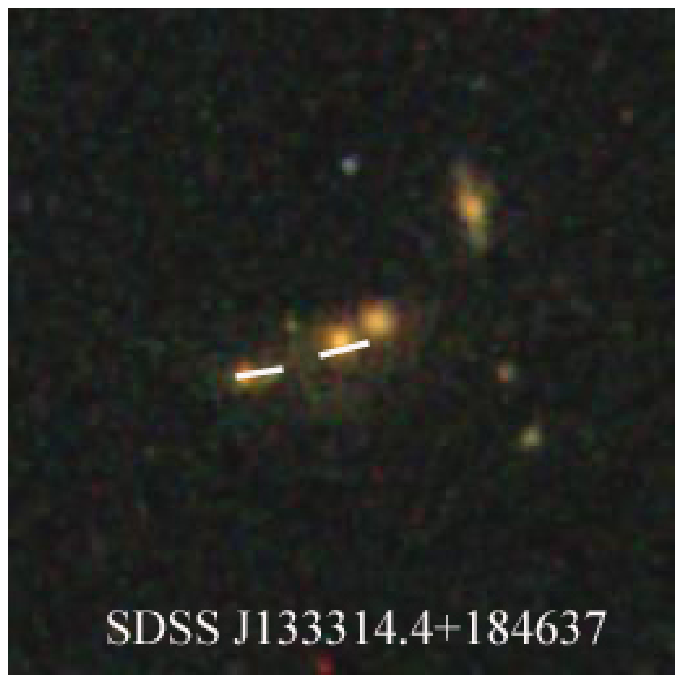}}~%
\resizebox{35mm}{!}{\includegraphics{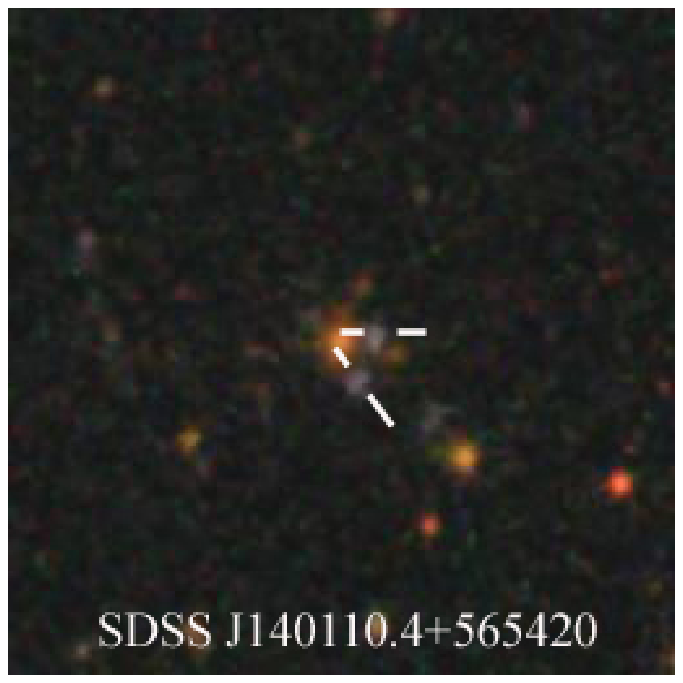}}~%
\resizebox{35mm}{!}{\includegraphics{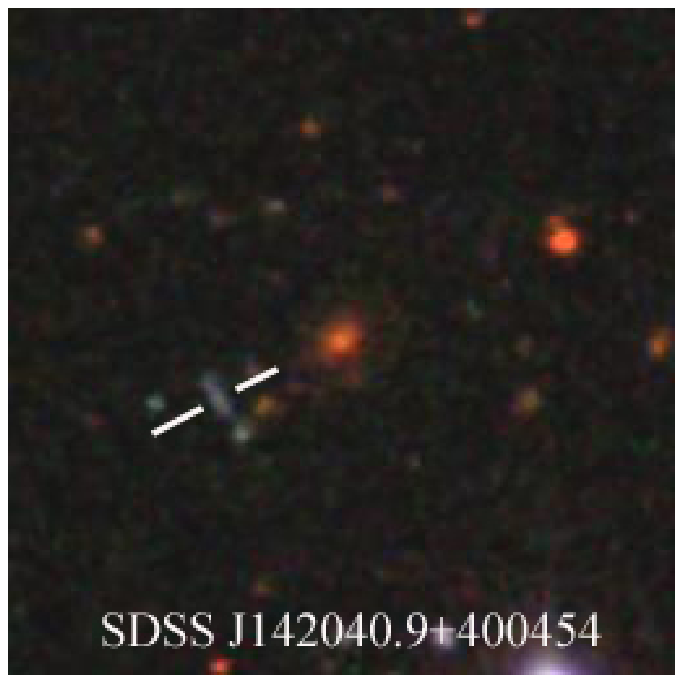}}\\[0.5mm]
\resizebox{35mm}{!}{\includegraphics{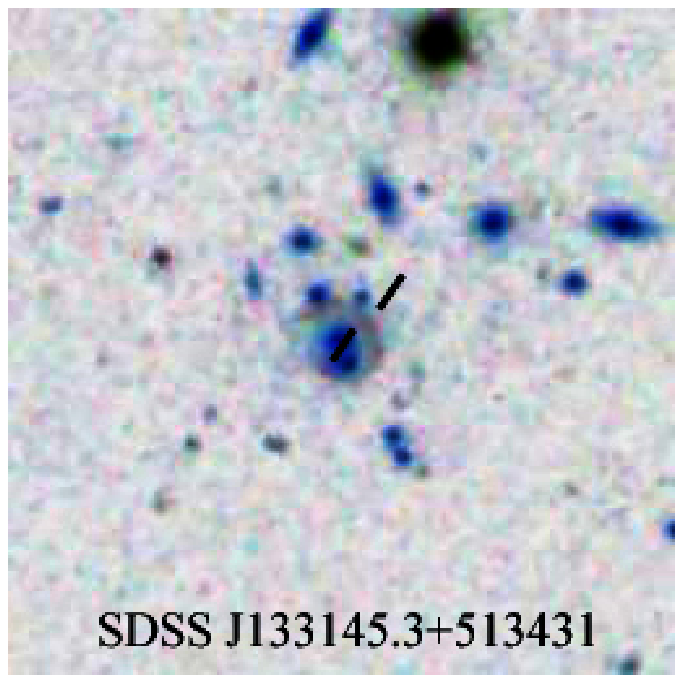}}~%
\resizebox{35mm}{!}{\includegraphics{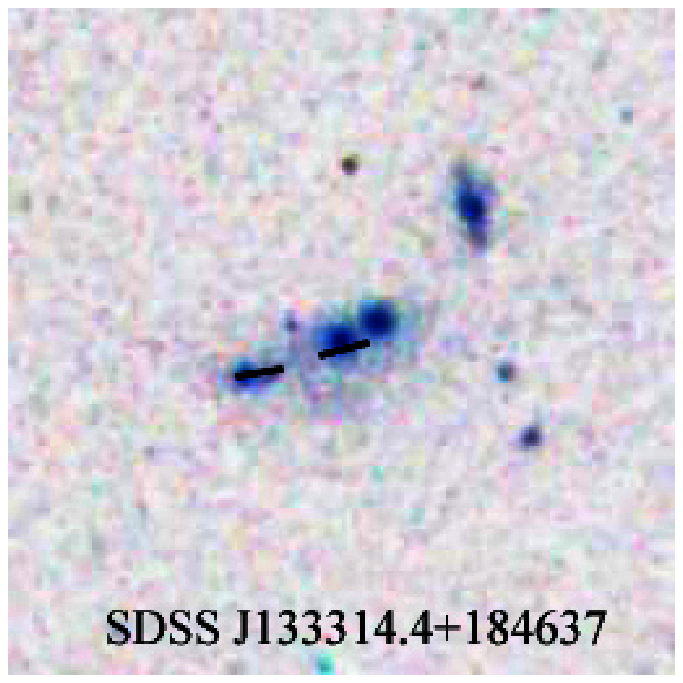}}~%
\resizebox{35mm}{!}{\includegraphics{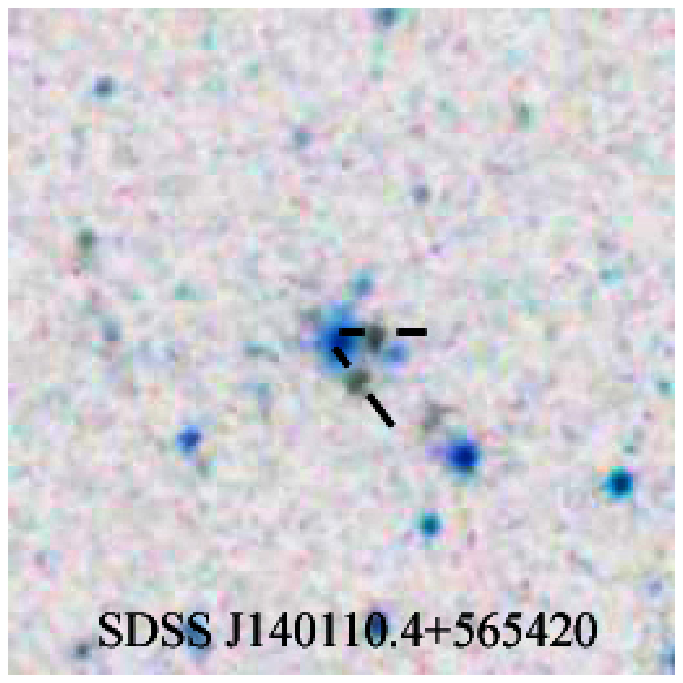}}~%
\resizebox{35mm}{!}{\includegraphics{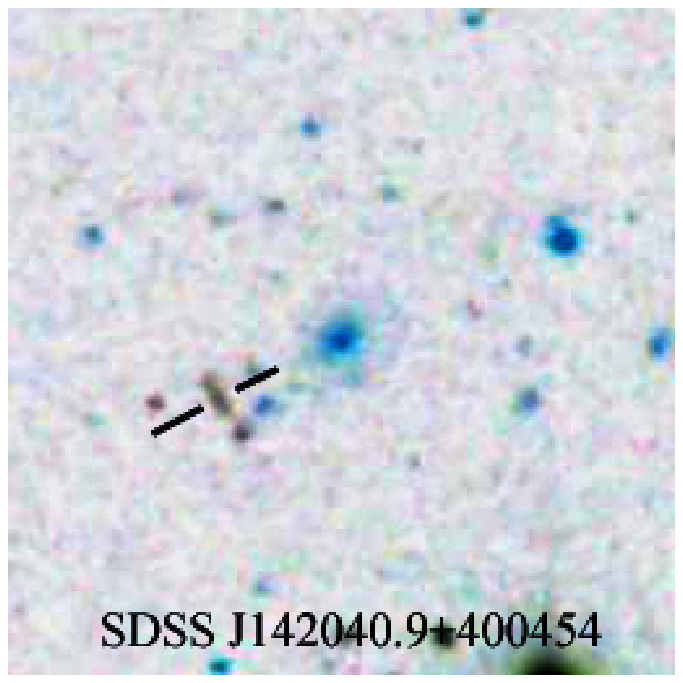}}\\[0.5mm]
\resizebox{35mm}{!}{\includegraphics{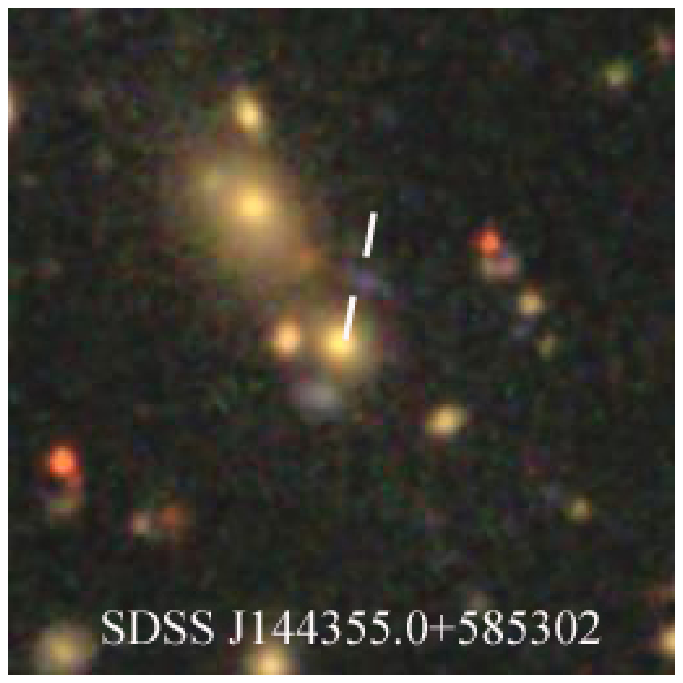}}~%
\resizebox{35mm}{!}{\includegraphics{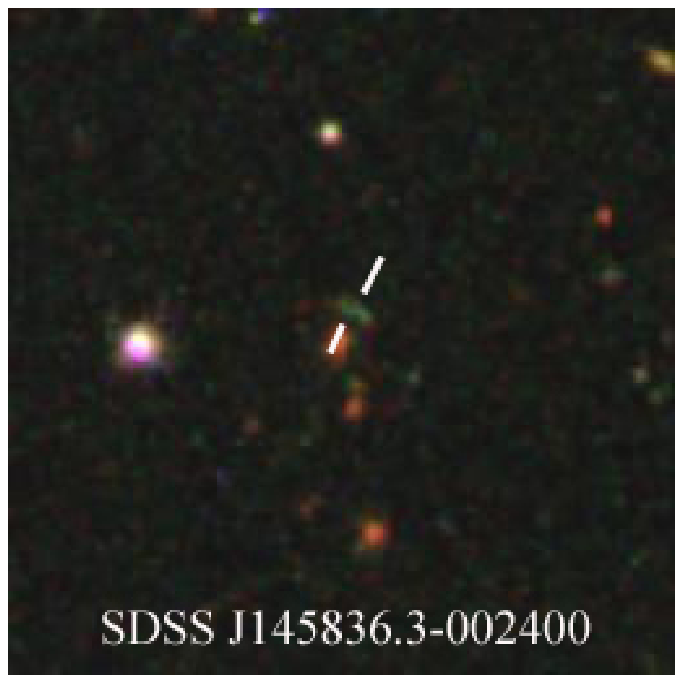}}~%
\resizebox{35mm}{!}{\includegraphics{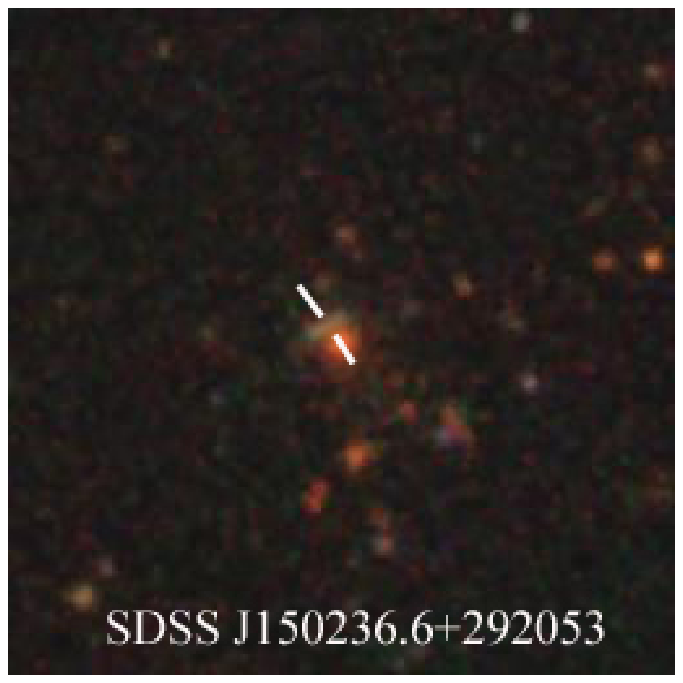}}~%
\resizebox{35mm}{!}{\includegraphics{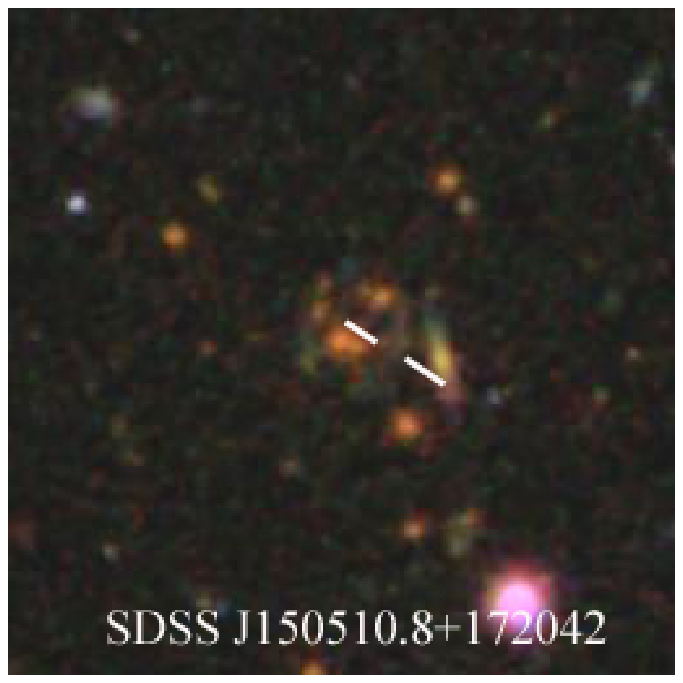}}\\[0.5mm]
\resizebox{35mm}{!}{\includegraphics{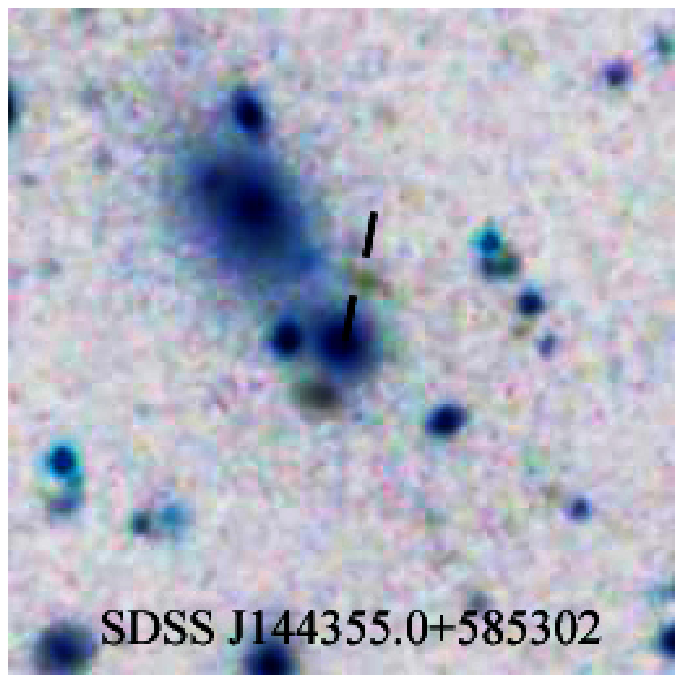}}~%
\resizebox{35mm}{!}{\includegraphics{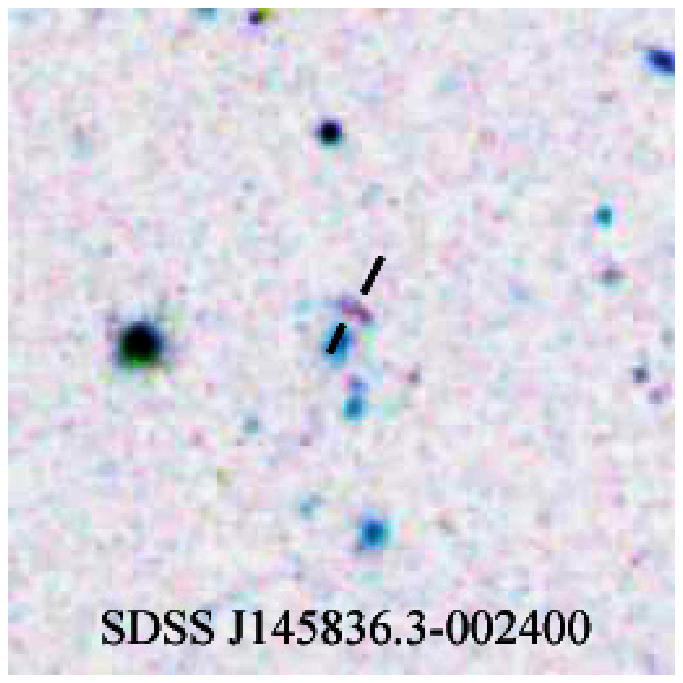}}~%
\resizebox{35mm}{!}{\includegraphics{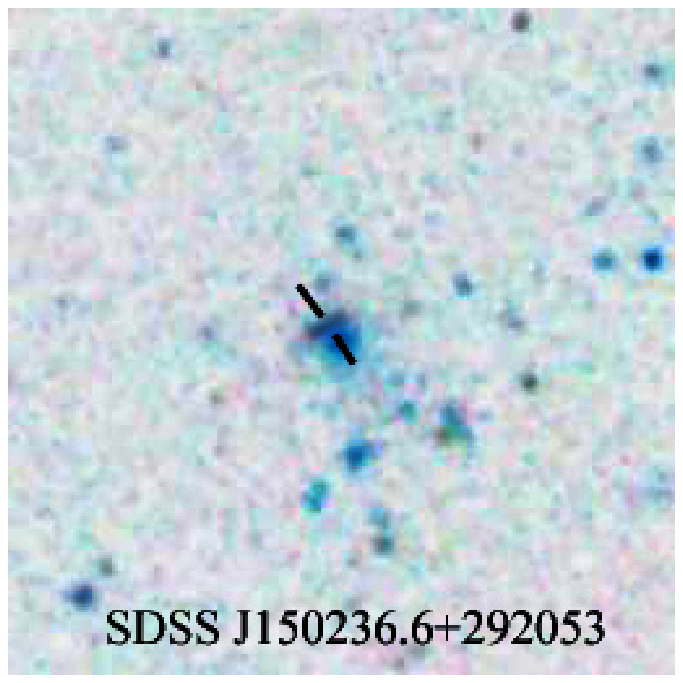}}~%
\resizebox{35mm}{!}{\includegraphics{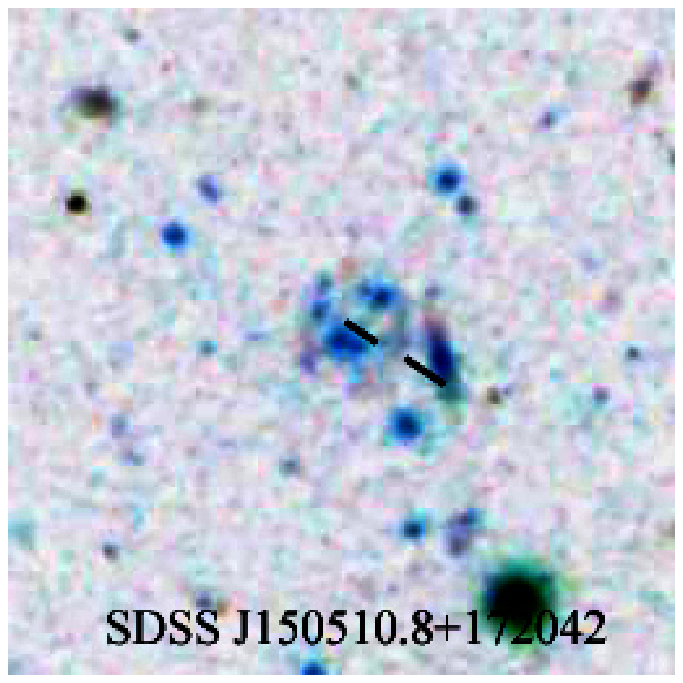}}
\setcounter{figure}{2}
\caption{{\it Continued}}
\end{figure}
\addtocounter{table}{-1}
\begin{figure}[!hpt]
\resizebox{35mm}{!}{\includegraphics{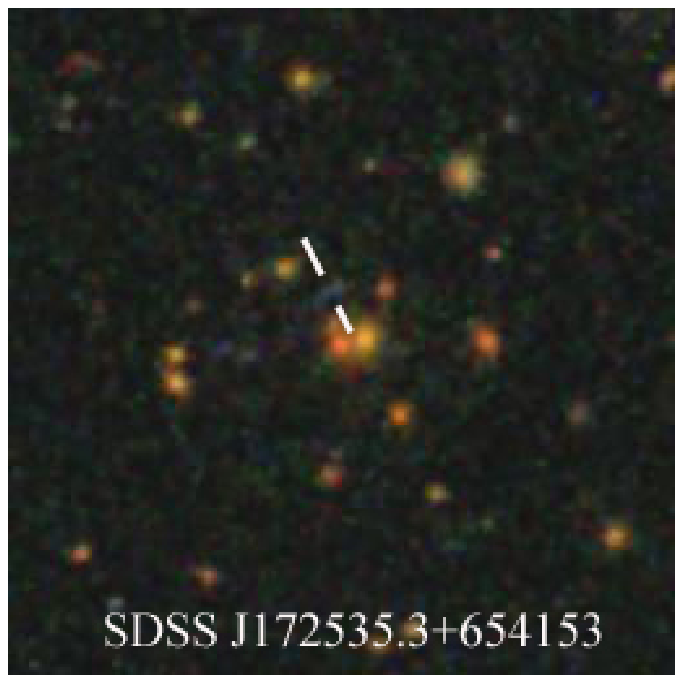}}~%
\resizebox{35mm}{!}{\includegraphics{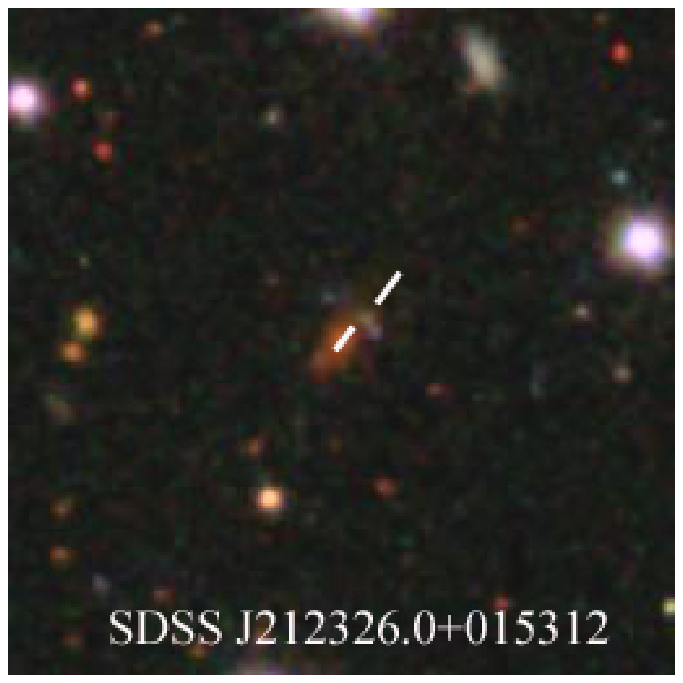}}~%
\resizebox{35mm}{!}{\includegraphics{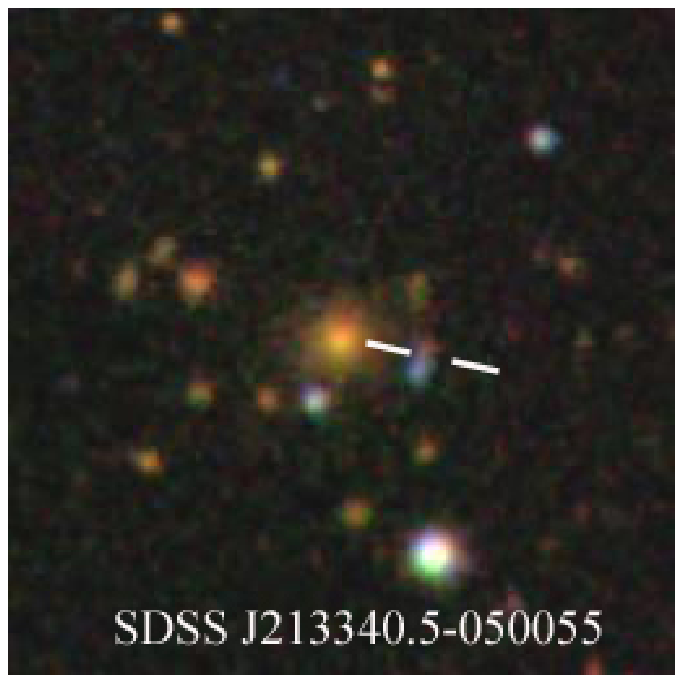}}~%
\resizebox{35mm}{!}{\includegraphics{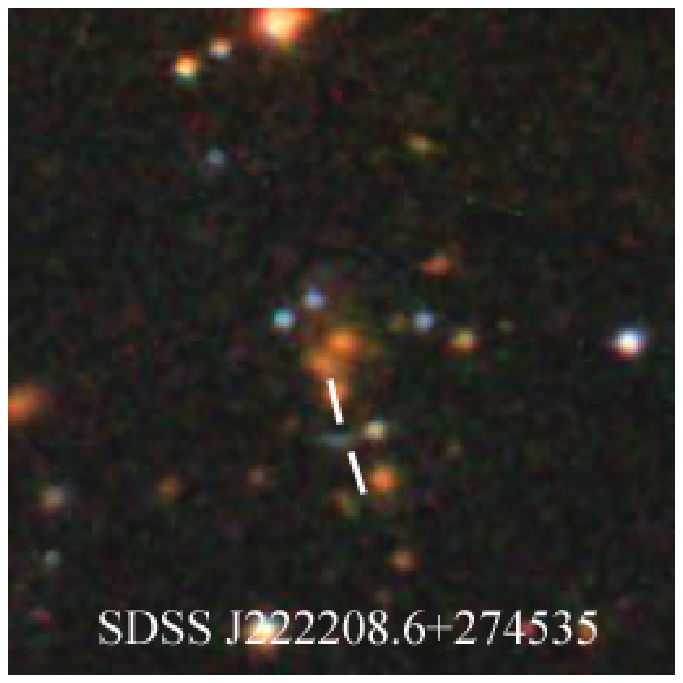}}\\[0.5mm]
\resizebox{35mm}{!}{\includegraphics{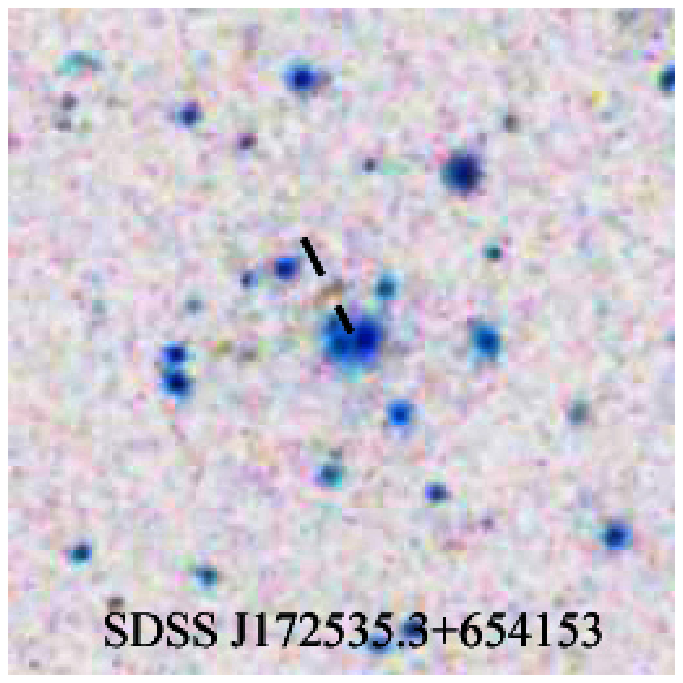}}~%
\resizebox{35mm}{!}{\includegraphics{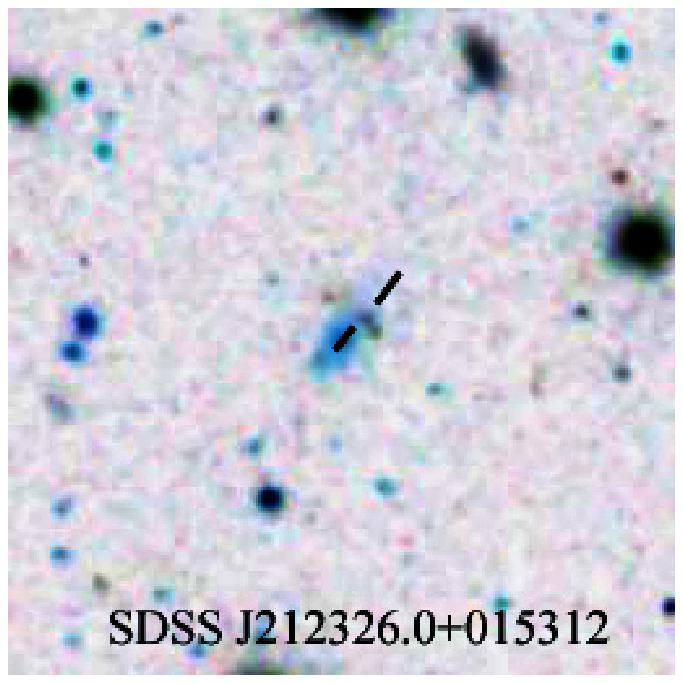}}~%
\resizebox{35mm}{!}{\includegraphics{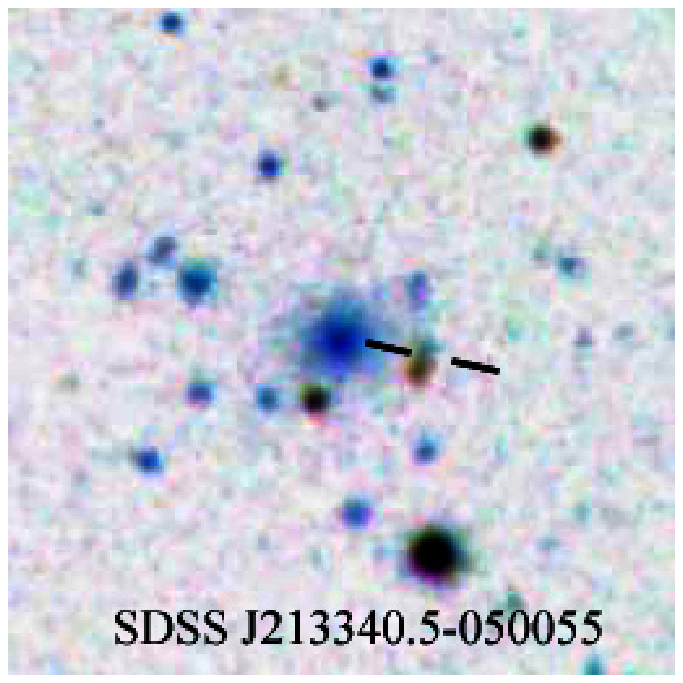}}~%
\resizebox{35mm}{!}{\includegraphics{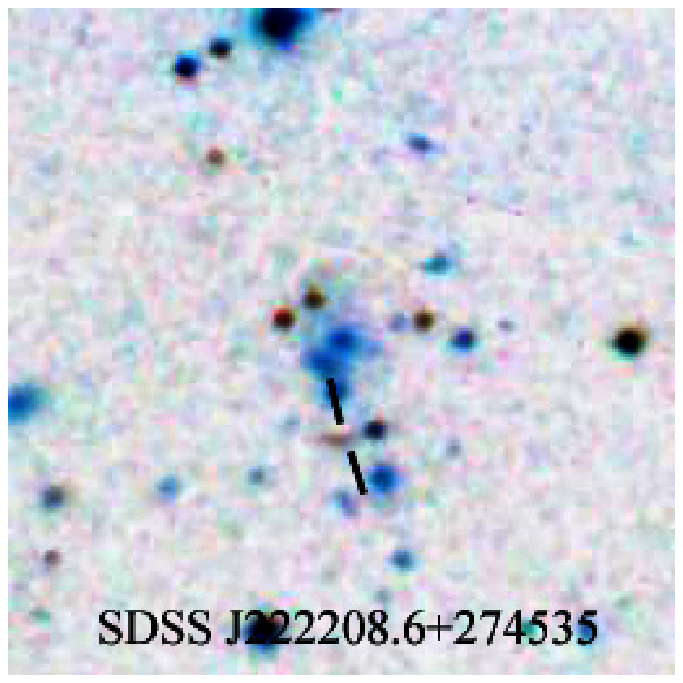}}\\[0.5mm]
\resizebox{35mm}{!}{\includegraphics{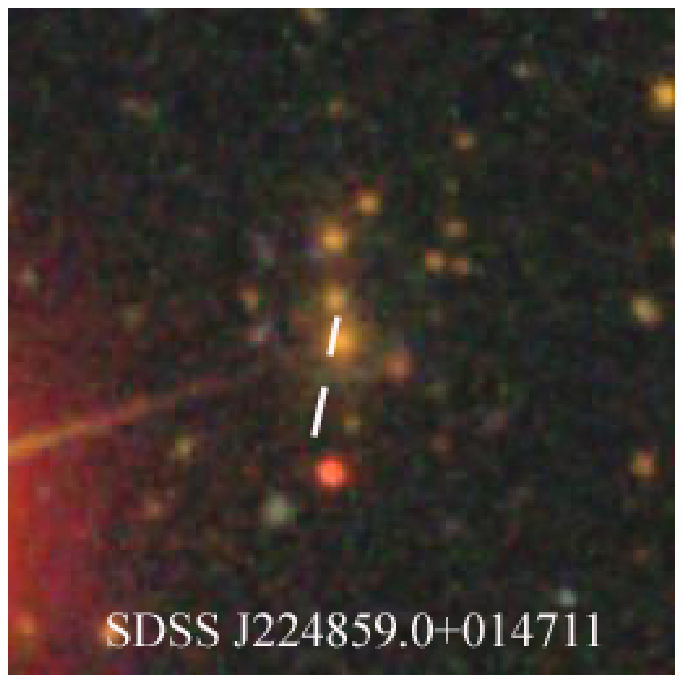}}~%
\resizebox{35mm}{!}{\includegraphics{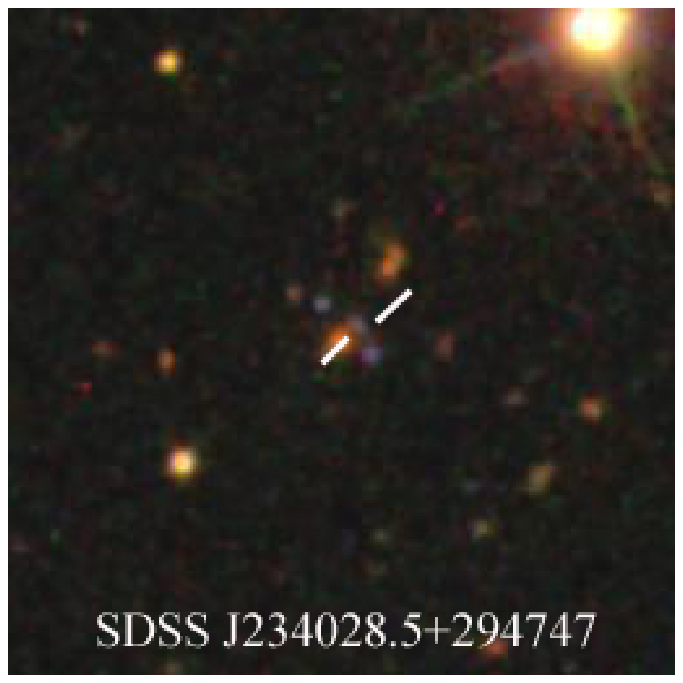}}~%
\resizebox{35mm}{!}{\includegraphics{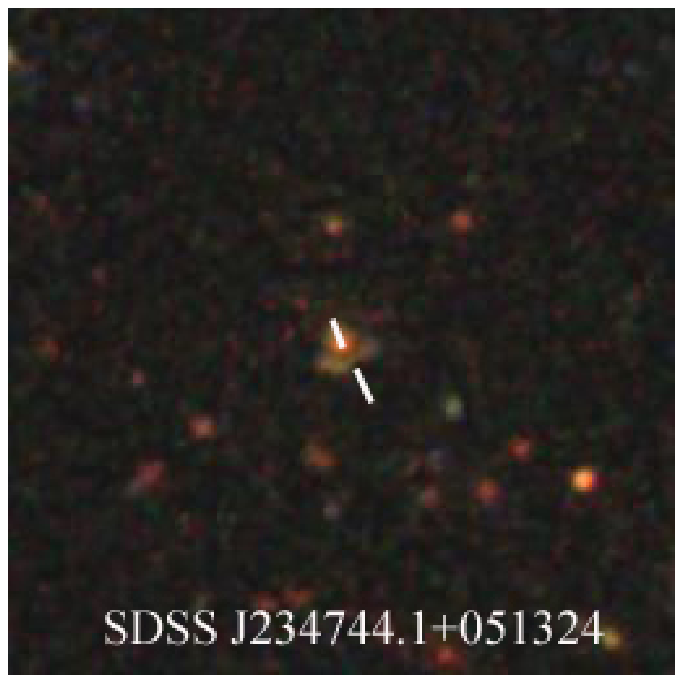}}\\[0.5mm]
\resizebox{35mm}{!}{\includegraphics{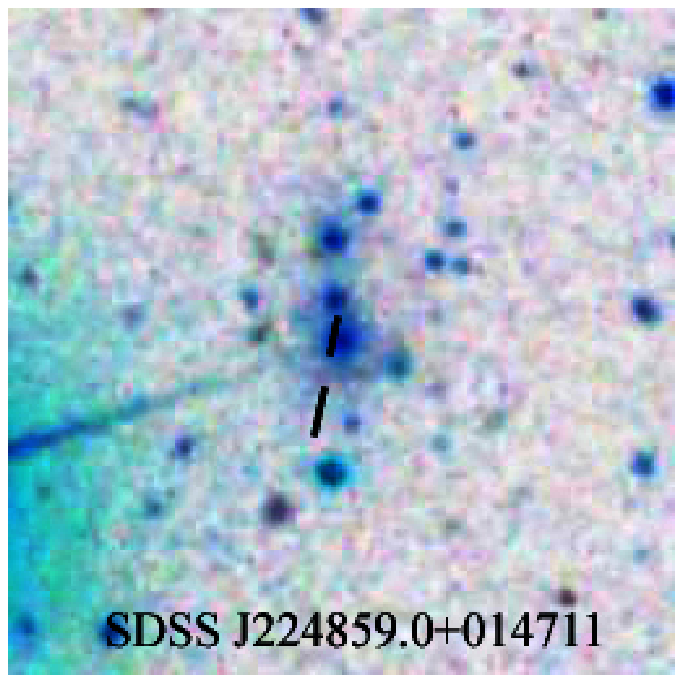}}~%
\resizebox{35mm}{!}{\includegraphics{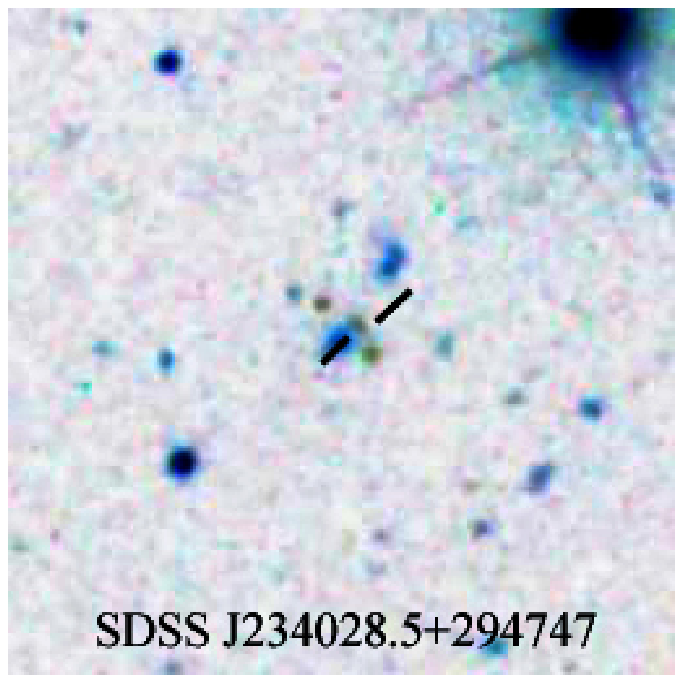}}~%
\resizebox{35mm}{!}{\includegraphics{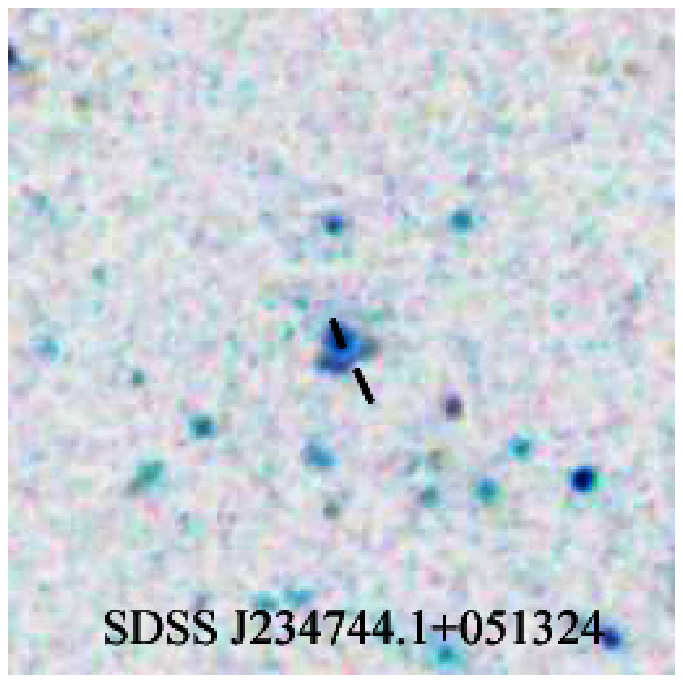}}~%
\setcounter{figure}{2}
\caption{{\it Continued}}
%
\resizebox{35mm}{!}{\includegraphics{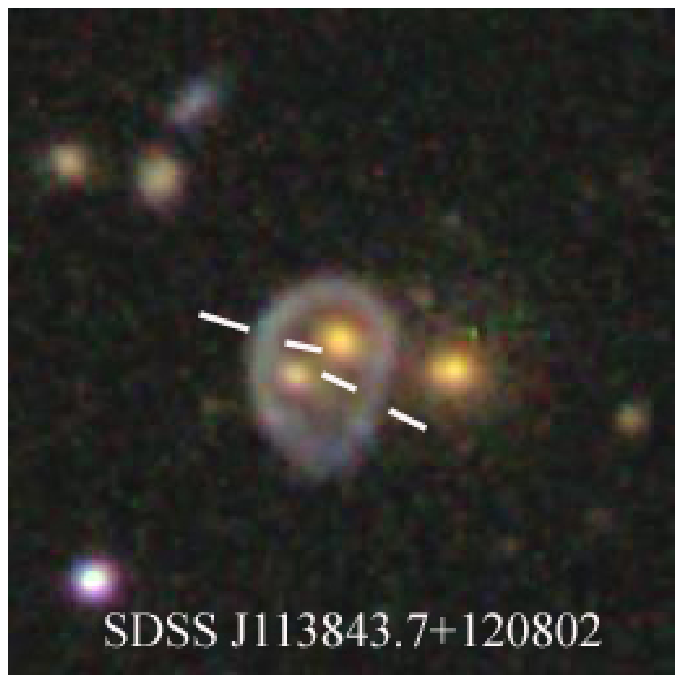}}~%
\resizebox{35mm}{!}{\includegraphics{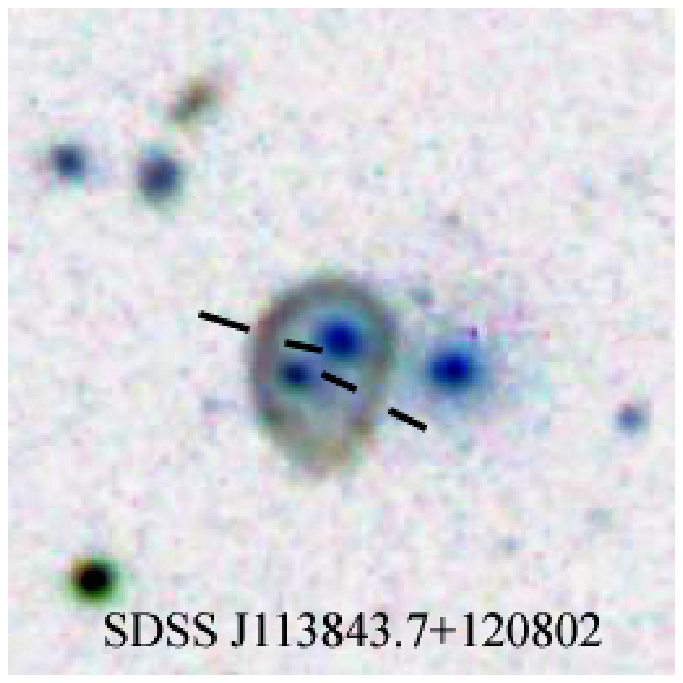}}~%
\resizebox{35mm}{!}{\includegraphics{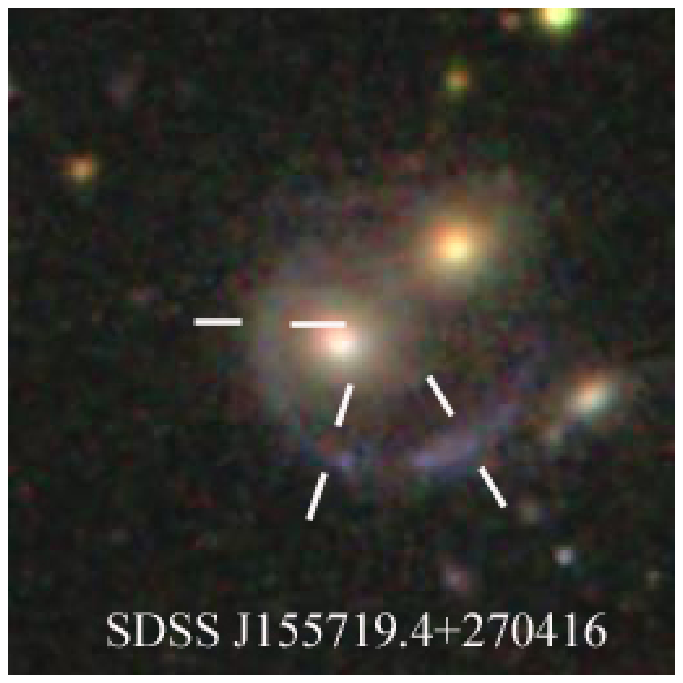}}~%
\resizebox{35mm}{!}{\includegraphics{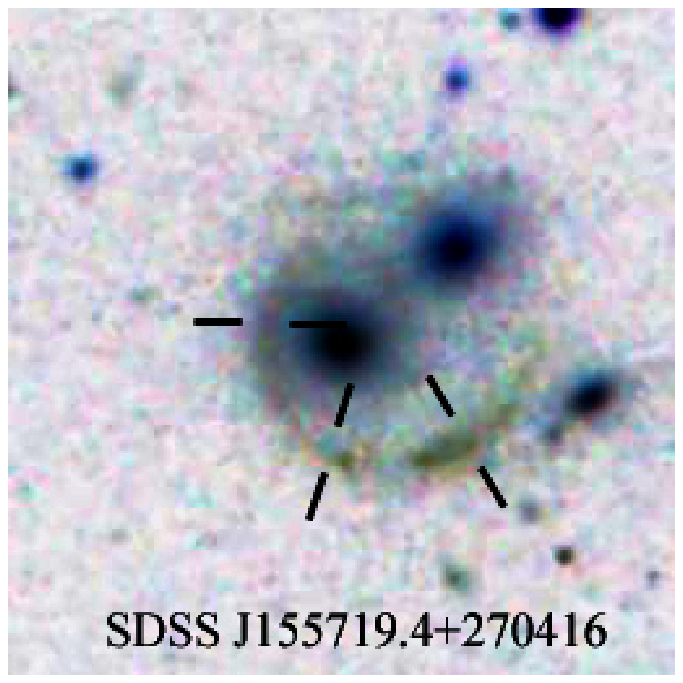}}
\caption{\baselineskip 3.6mm
Same as Fig.~\ref{lens_sure}, but for two systems with exotic rings. 
\label{exotic}}
\end{figure}

We find another 22 clusters which are {\it probable} lensing systems
(see Fig.~\ref{lens_prob}). They also show blue giant arcs, which are
tangential to the bright central galaxies and are distinct in color
from the reddish cluster galaxies. Generally, the arcs in these
clusters are shorter than those in the ``{\it almost certain}''
systems. In some cases, we cannot exclude the possibility that the
arcs are formed by galaxy interaction or the coincidental
superposition of a few faint galaxies.

We also find another 31 clusters which are {\it possible} lensing
systems (see Fig.~\ref{lens_poss}). The arcs in these clusters 
also show blue colors but are shorter than those in the ``{\it almost
  certain}'' and ``{\it probable}'' lensing systems.

We find two clusters showing {\it exotic} rings (see Fig.~\ref{exotic}) 
around two bright cluster member galaxies. In the cluster SDSS
J113843.7$+$120802, the blue ring has a radius of $7.5''$. In this
cluster, a brighter galaxy, SDSS J113842.8$+$120759, is located
$7.0''$ east of the ring, but the ring is not tangential to this
galaxy.
In the case of the cluster SDSS J155719.4$+$270416, a blue giant 
ring surrounds two member galaxies with redshifts $z=0.0678$
and $0.0683$. This giant ring has a radius of $15''$. If it 
is formed by strong lensing, the mass within the arc is
$7.8\times10^{12}~M_{\odot}$ assuming the source redshift $z_s=2$. If
the giant arc is a star forming region, the system is an excellent
example to study how galaxy merging triggers the star formation in
their strange common halo.

\begin{figure}[!t]
\begin{center}
\resizebox{80mm}{!}{\includegraphics{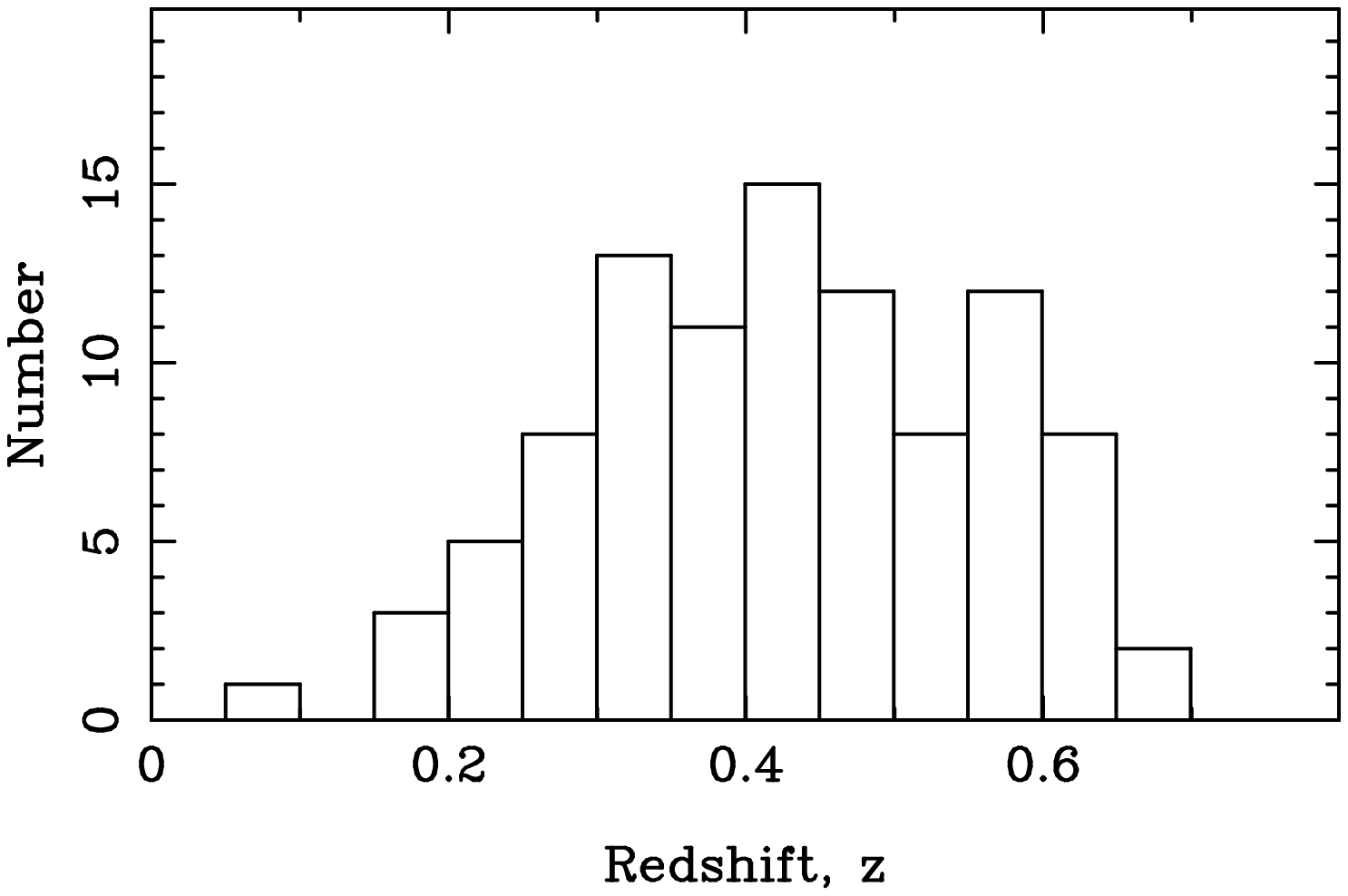}}
\end{center}
\caption{Redshift distribution of 98 lensing clusters identified 
from the SDSS-III images.
\label{lensz}}
\end{figure}

Figure~\ref{lensz} shows the redshift distribution of 98 clusters with
giant arcs (30 previously known lensing systems and 68 new) which we
identify from the SDSS-III images. These clusters are distributed in
the redshift range of $\sim$$0.1<z<0.7$ and have a redshift peak of
$z\sim0.4$.

\section{Distribution of lensing systems}

\begin{figure}[!ht]
\resizebox{65mm}{!}{\includegraphics{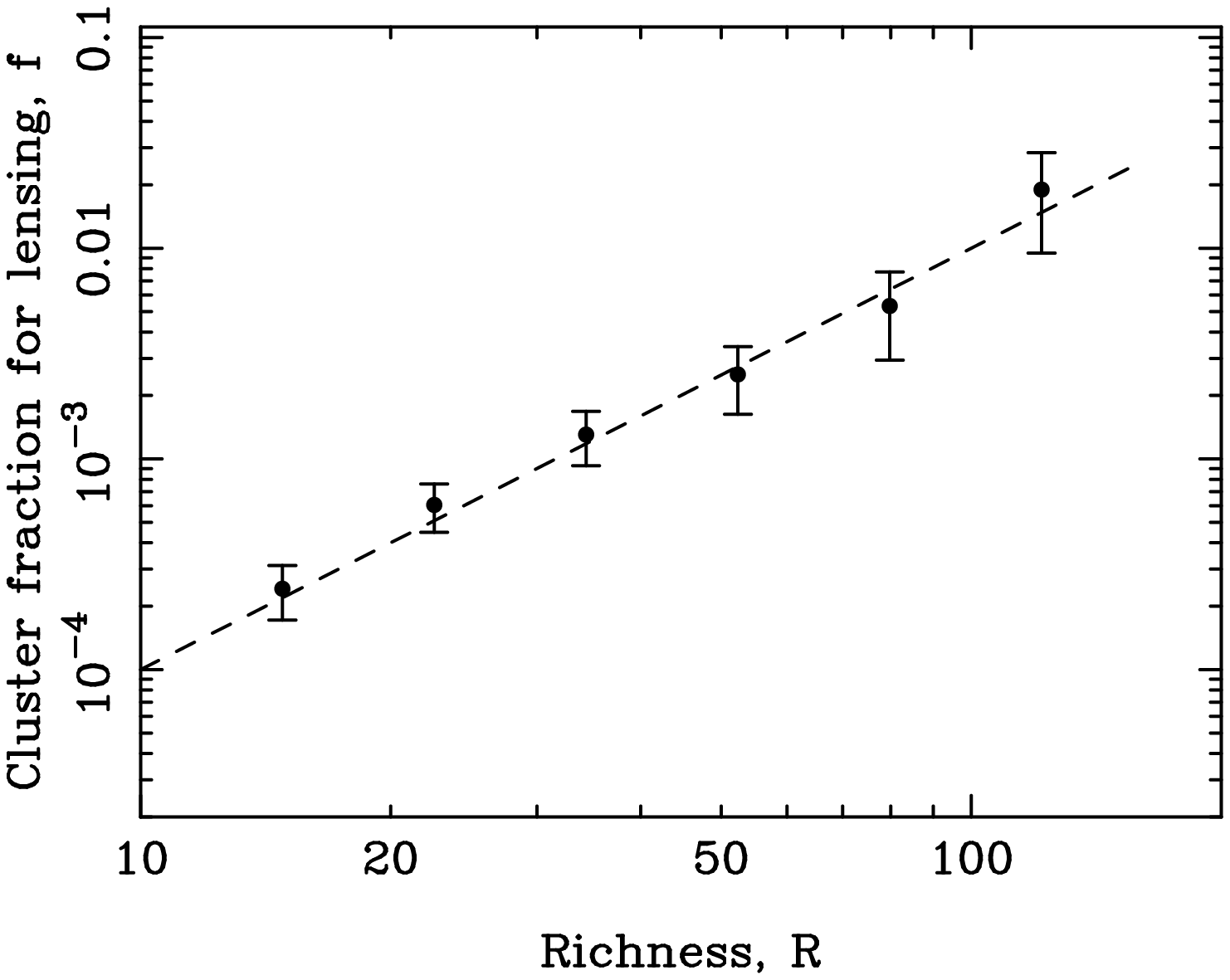}}\hspace{4mm}
\resizebox{65mm}{!}{\includegraphics{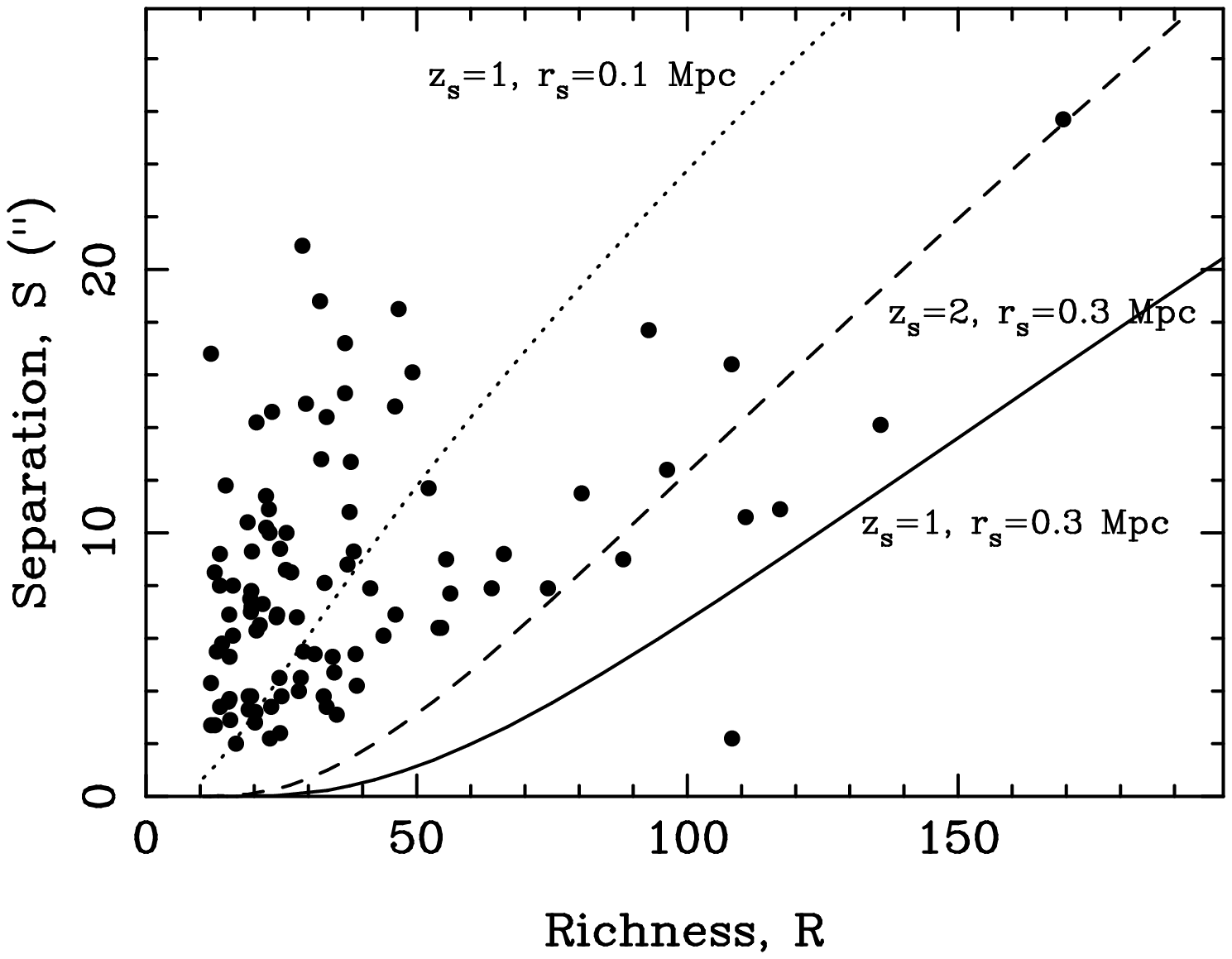}}~%
\caption{{\it Left panel}: the fraction of clusters that act as lensing 
systems against cluster richness. The dashed line is the power law of 
$f=10^{-6}R^2$. {\it Right panel}: the separation between lensed 
arc and central galaxy against cluster richness. The 
lines are the Einstein radius against cluster 
richness for the cluster mass profiles and different redshifts of
 background sources (see text).}
\label{rate}
\end{figure}

We notice that the lensed arcs preferably appear in the images of
massive clusters.  In Figure~\ref{rate}, we show the fraction of
clusters as lensing systems against cluster richness, $R$. Clearly,
the fraction, $f$, strongly depends on cluster richness. This is
expected because richer clusters are more massive and have greater
magnification factors. The fraction $f\sim$$10^{-4}$ is found for
clusters with a richness of $R\sim10$, and it increases to
$\sim$$10^{-3}$ for clusters with $R\sim30$ and $\sim$$10^{-2}$ for
clusters with $R\sim100$. The data are consistent with a power law
\begin{equation}
f=10^{-6}R^2.
\end{equation}

We also show the separation between the lensed arc and the central
galaxy as a function of cluster richness, $R$. Although showing large
scatters, richer lensing clusters tend to have larger separations. The
data scatter may come from the variety of redshifts, locations of
background galaxies and the cluster masses within the lensed arcs. To
explain this general tendency, we draw three lines for the Einstein
radius versus cluster richness with different parameters.
The redshift of the lensing cluster is set at $z_{\rm l}=0.4$ (i.e., the
peak of the redshift distribution in Fig.~\ref{lensz}). The background
source is placed at $z_{\rm s}=1$ or $z_{\rm s}=2$. Considering the
dark matter halo with a profile of \citet{nfw96}
\begin{equation}
\rho=\frac{\rho_s}{(r/r_s)(1+r/r_s)^2},
\end{equation}
where $\rho_{\rm s}$ and $r_{\rm s}$ are the characteristic density
and radius, respectively, we can get the Einstein radius using
Equation~(\ref{lensm}).  The cluster mass, $M_{200}$, is related to the
cluster richness, $R$, by (Wen et al. in preparation)
\begin{equation}
\log(M_{200})=(-1.29\pm0.05)+(1.05\pm0.03)\log(R),
\label{mass}
\end{equation}
where $M_ {200}$ is the total mass within $r_{200}$ in units of
$10^{14}~M_{\odot}$.
For a cluster with a smaller $r_{\rm s}$, the matter is more
concentrated toward the cluster center, so that the cluster has a larger
Einstein radius for the same $z_{\rm l}$ and $z_{\rm s}$.  We take
$r_{\rm s}=0.3$ Mpc or $r_{\rm s}=0.1$ Mpc \citep{sa07} for
calculation. As shown in Figure~\ref{rate}, for clusters with richness
$R>50$, the separations are consistent with the lensing systems with
parameters between $z_{\rm s}=1$, $r_{\rm s}=0.3$ Mpc and $z_{\rm
  s}=1$, $r_{\rm s}=0.1$ Mpc. We notice that one rich cluster, SDSS
J082728.4$+$223245 with a richness of $R=108.26$, has an arc with a
separation of $2.2''$, in which the arc is lensed by one of the member
galaxies \citep{sso+08}. For clusters with lower richnesses of $R<50$,
the separations have large scatters and tend to be consistent with the
calculation with a small $r_{\rm s}$. This indicates that the lensing
is obviously dominated by the bright central galaxy rather than the
massive halo of a cluster.

\section{Summary}

By inspecting color images of a large sample of clusters of galaxies
identified from the SDSS-III (DR8), we find 13 lensing clusters which
are almost certain, together with 22 probable and 31 possible
lensing clusters and two exotic systems. Together with 30 previously 
known lensing clusters, at least 43 lensing systems have been
identified from the SDSS. The lensing clusters have a redshift in the
range of $\sim$$0.1<z<0.7$. The arcs in the cluster images have
angular separations of $2.0''-25.7''$ from the bright central galaxies
and show bluer colors compared with the reddish cluster galaxies.
We find that the fraction of clusters that act as lensing systems
strongly depends on cluster richness. It increases from
$\sim$$10^{-4}$ for clusters of richness $R=10$ to $\sim$$10^{-2}$ for
clusters of $R=100$.  The separations between the arcs and the central
galaxies tend to increase with cluster richness.

\normalem
\begin{acknowledgements}

%
The authors are supported by the National Natural Science Foundation
of China (10821061 and 10833003) and Young Researcher Grant of
National Astronomical Observatories, Chinese Academy of Sciences.
Funding for SDSS-III has been provided by the Alfred P. Sloan
Foundation, the Participating Institutions, the National Science
Foundation, and the U.S. Department of Energy. The SDSS-III web site
is http://www.sdss3.org/.
SDSS-III is managed by the Astrophysical Research Consortium for the
Participating Institutions of the SDSS-III Collaboration including the
University of Arizona, the Brazilian Participation Group, Brookhaven
National Laboratory, University of Cambridge, University of Florida,
the French Participation Group, the German Participation Group, the
Instituto de Astrofisica de Canarias, the Michigan State/Notre
Dame/JINA Participation Group, Johns Hopkins University, Lawrence
Berkeley National Laboratory, Max Planck Institute for Astrophysics,
New Mexico State University, New York University, Ohio State
University, Pennsylvania State University, University of Portsmouth,
Princeton University, the Spanish Participation Group, University of
Tokyo, University of Utah, Vanderbilt University, University of
Virginia, University of Washington, and Yale University.

\end{acknowledgements}

\bibliographystyle{raa}
\bibliography{journals,lens}

\end{document}